\documentclass{svjour3}                     
\smartqed

\usepackage{graphicx}
\usepackage{setspace}
\usepackage{amsmath}
\usepackage{graphics}
\usepackage{float}
\usepackage{amsfonts}
\usepackage{multicol}
\usepackage{multirow}
\usepackage{enumerate}
\usepackage{url}
\usepackage{bm}
\usepackage{booktabs}
\usepackage{siunitx}
\usepackage{subcaption}
\usepackage{algorithm,algorithmic}
\usepackage{color}
\usepackage[table,xcdraw]{xcolor}
\usepackage{array}
\usepackage{placeins}
\usepackage[toc,page]{appendix}
\usepackage{amsfonts}
\usepackage{grffile}
\usepackage{geometry}
\usepackage[misc]{ifsym}
\definecolor{arsenic}{rgb}{0.23, 0.27, 0.29}
\definecolor{charcoal}{rgb}{0.21, 0.27, 0.31}
\definecolor{hanblue}{rgb}{0.27, 0.42, 0.81}
\definecolor{blue-ncs}{rgb}{0.0, 0.53, 0.74}
\definecolor{awesome}{rgb}{1.0, 0.13,0.32}
\definecolor{darkgreen}{rgb}{0, .4,0}
\definecolor{purple}{rgb}{.55, .2,.9}

\begin{document}

\title{Using high-fidelity discrete element simulation to calibrate an expeditious terramechanics model in a multibody dynamics framework}
	
\author{
		Yuemin Zhang \and
		Junpeng Dai \and
		Wei Hu \and
		Dan Negrut
}
	
\institute{Yuemin Zhang, Junpeng Dai, \Letter~Wei Hu \at
		School of Ocean and Civil Engineering, Shanghai Jiao Tong University, Shanghai 200240, China \\
		\email{weihu@sjtu.edu.cn}
		\and
		Dan Negrut \at
		Department of Mechanical Engineering, University of Wisconsin-Madison, Madison, WI 53706, USA \\
		\email{negrut@wisc.edu}
}
	
\date{Received: XXX / Accepted: XXX}
	
\maketitle
	
\begin{abstract}
		The wheel-soil interaction has great impact on the dynamics of off-road vehicles in terramechanics applications. The Soil Contact Model (SCM), which anchors an empirical method to characterize the frictional contact between a wheel and soil, has been widely used in off-road vehicle dynamics simulations because it quickly produces adequate results for many terramechanics applications. The SCM approach calls for a set of model parameters that are obtained via a bevameter test. This test is expensive and time consuming to carry out, and in some cases difficult to set up, e.g., in extraterrestrial applications. We propose an approach to address these concerns by conducting the bevameter test in simulation, using a model that captures the physics of the actual experiment with high fidelity. To that end, we model the bevameter test rig as a multibody system, while the dynamics of the soil is captured using a discrete element model (DEM). The multibody dynamics--soil dynamics co-simulation is used to replicate the bevameter test, producing high-fidelity ground truth test data that is subsequently used to calibrate the SCM parameters within a Bayesian inference framework. To test the accuracy of the resulting SCM terramechanics, we run single wheel and full rover simulations using both DEM and SCM terrains. The SCM results match well with those produced by the DEM solution, and the simulation time for SCM is two to three orders of magnitude lower than that of DEM. All simulations in this work are performed using Chrono, an open-source, publicly available simulator. The scripts and models used are available in a public repository for reproducibility studies and further research.

\end{abstract}

\section{Introduction}
\label{section:intro}
The interaction between a vehicle's wheels and the terrain has a significant impact on the vehicle's dynamics, particularly when the terrain is highly deformable. Over the past decades, several computational strategies for modeling wheel-soil interaction have been proposed in terramechanics \cite{bekker56,bekker60,bekker69,wong83,wong2001,janosi61,Krenn2008SCM,Krenn2011,chauchat2014three,ionescu2015viscoplastic,sulsky1994particle,bardenhagen2000material,cundall1979discrete,iwashita1999mechanics,jensen1999simulation}; for a review, see \cite{Taheri2015}. These models aim to improve various off-road mobility metrics through simulation before the production and deployment of physical prototypes. Based on their accuracy and efficiency, the terramechanics models that capture the terrain-vehicle interaction can be conceptually divided into three categories: ($i$) empirical models \cite{bekker56,bekker60,wong83,wong2001,janosi61,Krenn2008SCM,Krenn2011}; ($ii$) continuous representation models (CRMs) \cite{chauchat2014three,ionescu2015viscoplastic,sulsky1994particle,bardenhagen2000material,chenSPH3DgranMat2012,nguyenSPHgranFlows2017,hurley2017continuum}; and ($iii$) discrete element models (DEMs)  \cite{cundall1979discrete,iwashita1999mechanics,jensen1999simulation,antonioVehicleTireGranMatSim2017}. The empirical terramechanics models, which use relatively simple phenomenological equations to characterize the interaction between the wheel and soil, are the most efficient. Their accuracy is adequate but limited to a relatively narrow set of wheel--terrain interaction regimes. When either the geometry of the wheel changes or the terrain properties changes, the formulas producing the resistive forces and torques need to be recalibrated. This is due to the fact that the model is not physics-based, and scenarios out of the training/calibration distribution cannot be captured well in simulation. At the other end of the accuracy spectrum, the DEM-based terramechanics is physics-based, as it treats each individual grain in the deformable soil as a small body that engages in frictional contact events with neighboring bodies, which might be other grains or implements, e.g., a rover wheel, a tire, a plow \cite{bertrandDEMmixing2005,longmore-GPU-DEM2013,houDEM-ScrewFeeder2014,ganDEM-GPU2016,he-powderGPU2018,toson-PowdMxr2018}. The manifest drawback of a DEM approach is its poor efficiency -- because the method keeps track of the motion of each element of the terrain, the DEM solution requires high simulation times even for relatively small problems \cite{antonioVehicleTireGranMatSim2017}.The CRM approach is physics-based and strikes a balance between simulation speed and accuracy, making it an attractive terramechanics method for many practical engineering problems  \cite{chauchat2014three,ionescu2015viscoplastic,sulsky1994particle,bardenhagen2000material,chenSPH3DgranMat2012,nguyenSPHgranFlows2017,hurley2017continuum,bui2008lagrangian,weiGranularSPH2021,weiTracCtrl2022}. 

Empirical terramechanics models have been improved for more than a century, and nowadays are the most widely used approaches for handling wheel-soil interaction problems in off-road mobility. Bernstein was the first to study the relationship between the wheel load and sinkage, and to establish an early formula by introducing a parameter called sinkage modulus \cite{Bernstein1913}. By collecting and analyzing data from a large number of experimental tests, Bekker and Wong established a more accurate pressure sinkage formula, in which the sinkage modulus was divided into two parts -- the friction modulus and the cohesive modulus \cite{bekker60,wong67a,wong67b}. Their empirical formula, extensively validated with experimental data and subsequently refined, is the most widely used for characterizing wheel-soil interaction. Janosi and Hanamoto \cite{janosi61}, further improved the shear stress model initially proposed by Bekker to address the frictional force between the wheel and soil. Recently, a more versatile empirical approach called the soil contact model (SCM), which incorporates the Bekker-Wong and Janosi-Hanamoto formulations, has been proposed in \cite{Krenn2008SCM,Krenn2011}. This model has achieved real-time performance in many complex problems. The SCM approach yields good results in cases in which the wheel sinkage is small, wheel-soil slip is low to moderate, and the shape of the wheel is close to a cylinder \cite{Smith2014,meirion2011modified}. The parameters associated with the SCM model need to be identified using the so-called bevameter test rig \cite{Apfelbeck2011Systematic,edwards2017bevameter,mason2020overview}, which makes the approach cumbersome, especially in problems when the soil properties change frequently. 

In CRM-based terramechanics, the soil is treated as a continuum, thus reducing the degree-of-freedom (DOF) count relative to the one encountered in the DEM approach. In the CRM approach, one solves a set of partial differential equations that include the mass and momentum balance, in addition to a closure condition such as the Jaumann equation describing the rate of change of the Cauchy stress tensor \cite{weiGranularSPH2021,KamrinFluidMechanics2015}. These partial differential equations can be spatially discretized using the finite element method (FEM), thus capturing both the stress distribution and soil deformation field \cite{chauchat2014three,ionescu2015viscoplastic,chiroux05,guo2016parallel,zhao2020multiscale}. However, in problems involving large soil deformations, grid-based FEM methods often require prohibitively expensive re-meshing operations or fail to converge to a satisfactory numerical solution. Meshless methods, e.g., the Material Point Method (MPM) or Smoothed Particle Hydrodynamics (SPH), which use Lagrangian particles instead of Eulerian grids in discretizing the field equations, are well equipped to deal with large deformation artifacts -- for implementation approaches and sample applications, see \cite{sulsky1994particle,bardenhagen2000material,chenSPH3DgranMat2012,nguyenSPHgranFlows2017,hurley2017continuum,bui2008lagrangian,weiTracCtrl2022,KamrinFluidMechanics2015,bandara2015coupling,kenichiMPM2016,kamrin2019,kularathna2017implicit}. Due to a marked reduction in the DOF count, the CRM approach typically results in a one to two orders of magnitude reduction in simulation time compared to the high-fidelity DEM approach \cite{weiGranularSPH2021,chen2020gpu,xu2019analysis}. 

Since the DEM approach factors in the shape and material properties of each individual grain of soil, it is more expressive and thus better positioned to capture the wheel-soil interactions in terramechanics applications. For this reason, the DEM approach has been widely used in various terramechanics applications especially in cases with a complex mechanical system, e.g., \cite{antonioVehicleTireGranMatSim2017,iagnema2015,ucgul-EDEM-tillage2015,zhao-FEM-DEMtire-terrain2017,negrut2017sand}. However, the number of grains in real-world terramechanics applications can be extremely large; for example, a cubic meter of typical sand contains one to two billion elements. Thus, it is not common to carry out full vehicle terramechanics studies using fully-resolved DEM granular simulations. By the same token, it is worth pointing out that a high-fidelity terramechanics model is not always necessary. For instance, when vehicle-terrain interaction simulations are utilized to validate traction control algorithms, test path planning strategies, or investigate human-in-the-loop scenarios, computational efficiency takes precedence over the accuracy of the terramechanics solution. Under such conditions, where computational efficiency is paramount, an empirical terramechanics model like SCM becomes highly attractive. 

Physics-based terramechanics models, e.g., the ones using DEM or CRM, are simpler to set up -- parameters such as bulk density, friction angle, bulk modulus, particle size, etc. can be lifted from the literature, see, for instance, the case of lunar regolith \cite{heRegolithProperties2010}. For empirical approaches, such as SCM, the unknown model parameters typically lack direct physical interpretations and are usually identified through bevameter test rig experiments  \cite{Apfelbeck2011Systematic,edwards2017bevameter,mason2020overview,mahonen2021portable,kruger2023experimental,weiVirtualBevameter2024}. The test calls for both plate sinkage and annulus shear experiments. Unfortunately, once the soil changes, e.g., from location to location, or when the gravitational pull changes, the SCM parameters will change as well. As a consequence, re-performing bevameter tests is necessary, but this process can be time-consuming and sometimes impractical, especially for extraterrestrial applications. While online estimation is a viable option \cite{iagnemma2004}, it requires the existence of an actual vehicle, which may not be available—particularly during the optimal design phase of a rover before a mission is launched. Against this backdrop, we address the issue of calibrating SCM by proposing the use of virtual bevameter tests conducted within a multibody dynamics framework. Thus, the plate sinkage and annulus shear test rigs are modeled as multibody systems interacting with deformable soil within a DEM terramechanics framework. The data collected from the DEM simulation via a virtual bevameter digital test is treated as the ground truth subsequently used to calibrate the SCM parameters. The calibration takes place in a Bayesian inference framework \cite{gelmanBayesian1995,robertMH2015} using a third-party package \cite{pymc3} that produces the posterior distribution of the SCM model parameters. The chosen set of SCM parameters maximizes the probability of generating the ground truth data; i.e., we employ maximum likelihood estimation. Simulation results obtained with the estimated set of parameters are compared against Bekker-Wong \cite{bekker60,wong2001} and Janosi-Hanamoto \cite{janosi61} results in several benchmark cases, and subsequently validated against DEM results for single wheel and full vehicle simulations.

The cornerstone of this contribution is the idea of producing high-fidelity and expensive-to-obtain data in sufficient quantities to enable the calibration of the expeditious model. Properly calibrated, SCM will captures well the terramechanics of the ``expensive'' model, in this case DEM. The question is whether DEM can be replaced by CRM, as done in \cite{weiVirtualBevameter2024}. The CRM approach employs a homogenization mechanism to represent the soil as a continuum \cite{chiroux05,kenichiMPM2016,kamrin2019,dai2017sph,he2018study,abdelrazek2016simulation}. In real world applications, the assumptions anchoring the homogenization mechanism might not hold true, e.g., the grain size in deformable soil is not always uniform and small, and the shape of the terrain particles might not be as trivial as that associated with collections of spheres. This is the motivation for considering here the DEM approach rather than the CRM one as the source of the ground truth data used for SCM calibration.

This contribution is organized as follows. Section \ref{sec:methods} outlines the SCM and DEM terramechanics solutions, and summarizes the Bayesian calibration approach employed. Section \ref{sec:calibration} introduces a virtual bevameter test rig, which draws on the high-fidelity DEM approach and a multibody simulation framework. The section also details how the high-fidelity data produced via DEM terramechanics is used as ground truth to calibrate the parameters associated with the lower-fidelity SCM model. In Section \ref{sec:validation}, the proposed approach is validated using single wheel and full rover simulation experiments. The mechanics of the comparison are as follows: both single wheel and full rover simulations are conducted using DEM and calibrated SCM terramechanics, with the results subsequently compared to evaluate how effectively SCM terramechanics serves as a proxy for DEM terramechanics. Section \ref{sec:conclusion} summarizes our findings and provides directions of future work. All simulations carried out in this contribution have been performed in an open source software called Chrono with the Chrono::GPU module \cite{chronoOverview2016}. The scripts to reproduce the results reported in this paper are provided in \cite{dem2scm_scripts}.

\section{Numerical methods preamble}\label{sec:methods}

\subsection{Overview of the DEM method}\label{subsec:demMethod}
In DEM, the frictional contact problem can be solved either in a complementarity-based framework, or in a penalty-based one. For a comparison of these two approaches, see \cite{armanDEMP-DEMC2017}. Therein, the authors indicate that for granular problems, the penalty approach is faster and more expressive; i.e., it can be adjusted better to replicate experimental data. Consequently, this study employs the penalty approach to produce the ground truth data used to calibrate the SCM approach. Figure~\ref{fig:dem_sketch} presents a schematic of the contact between two sphere particles using the penalty-based DEM approach. The model captures the normal contact force, the tangential friction force, and the rolling-resistance torque.

\subsubsection{The Newton-Euler equations of motion}%

\begin{figure}[htp]
	\centering
	\includegraphics[width=5in]{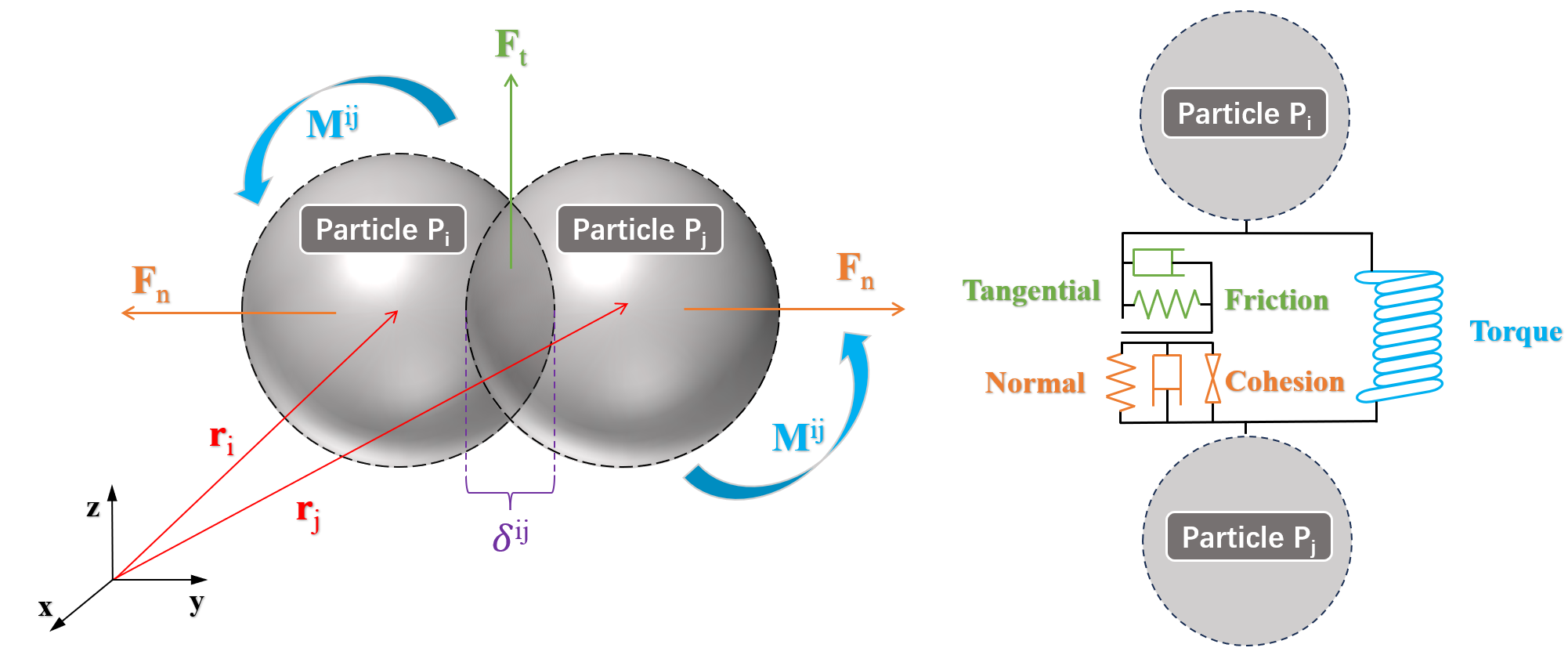}
	\caption{Schematic of the contact between two spherical particles using the penalty-based DEM approach.}
	\label{fig:dem_sketch}
\end{figure}

We consider all terrain particles to be monodisperse spheres. For a particle $P_i$, its mass and mass moment of inertia are denoted by $m^i$ and $I^i$, respectively. At time $t$, the collection of particles in collision with $P_i$ is defined as $C(i;t)$. The time evolution of particle $P_i$ can be described by solving the Newton–Euler equations of motion:
\begin{subequations}
\begin{equation}\label{equ:Newton-Euler equation(1)}
	m^i \frac{d\mathbf{v}^i}{dt} = m^i\mathbf{g} +\sum_{j\in C(i;t)}(\mathbf{F}_n^{ij} + \mathbf{F}_t^{ij}) \; ,
\end{equation}

\begin{equation}\label{equ:Newton-Euler equation(2)}
	I^i \frac{d{\boldsymbol{\omega}} ^i}{dt} =\sum_{j\in C(i;t)}\mathbf{M}^{ij} =\sum_{j\in C(i;t)}\Delta \mathbf{r}^{ij} \times \mathbf{F}_t^{ij} + \mathbf{M}_{rr}^{ij} \; ,
\end{equation}
\end{subequations}
where $\mathbf{v}^i$ and ${\boldsymbol{\omega} }^i$ are the translational and angular velocities of particle $P_i$; for $P_j \in C(i;t)$, $\mathbf{F}_n^{ij}$ is the normal contact force between $P_i$ and $P_j$, and $\mathbf{F}_t^{ij}$ is the friction contact force between the two particles (see Fig.~\ref{fig:dem_sketch}); $\Delta \mathbf{r}^{ij}$ is the vector that starts from the center of mass of $P_i$ and ends at the contact point where $\mathbf{F}_t^{ij}$ is applied; and $\mathbf{M}_{rr}^{ij}$ is the rolling resistance torque between particles $P_i$ and $P_j$.

\subsubsection{The Hertz-Mindlin contact model}%
We employ the Hertz-Mindlin contact model to calculate the normal contact force $\mathbf{F}_n^{ij}$ and the tangential friction force $\mathbf{F}_t^{ij}$ \cite{johnson1987contact}:

\begin{subequations}
\begin{equation}\label{equ:HM_FN}
	\mathbf{F}_n^{ij}=\sqrt{\frac{ \delta^{ij}  }{2R^{ij}}}(k_n \delta^{ij}\mathbf{n}^{ij}-\gamma_nm^{ij}\mathbf{v}_n^{ij})+\mathbf{f}_c \; ,
\end{equation}
\begin{equation}\label{equ:HM_FT}
	\mathbf{F}_t^{ij}=\sqrt{\frac{ \delta^{ij}  }{2R^{ij}}}(-k_t\mathbf{u}_t^{ij}-\gamma_tm^{ij}\mathbf{v}_t^{ij}) \; ,
\end{equation}
\end{subequations}
where $\mathbf{f}_c$ is the constant cohesion force; and $k_n$  and $k_t$ are the normal and tangential elastic stiffness coefficients, while $\gamma_n$ and $\gamma_t$ are the corresponding normal and tangential damping coefficients. In Eqs. (\ref{equ:HM_FN}) and (\ref{equ:HM_FT}), $R^{ij}$ is defined as the effective radius of the two particles, and $m^{ij}$ is the effective mass:
\begin{subequations}
	\begin{equation}
		\frac{1}{R^{ij}}=\frac{1}{R^i}+\frac{1}{R^j} \; ,
	\end{equation}
	\begin{equation}
		\frac{1}{m^{ij}}=\frac{1}{m^i}+\frac{1}{m^j} \; .
	\end{equation}
\end{subequations}
The relative normal velocity $\mathbf{v}_n^{ij}$ and the relative tangential velocity $\mathbf{v}_t^{ij}$ are calculated using the unit vector $\mathbf{n}^{ij}$ from the center of mass of $P_i$ to the center of mass of $P_j$, i.e.:
\begin{subequations}
	\begin{equation}
		\mathbf{v}^{ij}=(\mathbf{v}^j+\boldsymbol{\omega} ^j \times \Delta \mathbf{r}^{ji}) - (\mathbf{v}^i+\boldsymbol{\omega} ^i \times \Delta \mathbf{r}^{ij}) \; ,
	\end{equation}
	\begin{equation}
		\mathbf{v}_n^{ij}= (\mathbf{v}^{ij} \cdot \mathbf{n}^{ij})  \mathbf{n}^{ij} \; ,
	\end{equation}
	\begin{equation}
		\mathbf{v}_t^{ij}= \mathbf{v}^{ij}-\mathbf{v}_n^{ij}\; .
	\end{equation}
\end{subequations}
The penetration of two particles in mutual contact is defined as $\delta^{ij}$, the geometrical overlap,
\begin{equation}
\delta^{ij}=R^i+R^j-\| \Delta \mathbf{r}^{ij} \|-\| \Delta \mathbf{r}^{ji} \| \; .
\end{equation}
The tangential friction force in Eq.~(\ref{equ:HM_FT}) depends on the tangential displacement $\mathbf{u}_t^{ij}$. Since the latter is an accumulated value monitored from the initiation of the contact event, a history tracking procedure is employed and associated with each contact event. If $\mathbf{u}_{t,k-1}^{ij}$ is the tangential displacement between the pairwise particles at time step $k-1$, the tangential displacement at step $k$ can be obtained via a two-step algorithm:
\begin{subequations}
	\begin{equation}
		\hat{\mathbf{u}}_{t,k}^{ij} = \mathbf{u}_{t,k-1}^{ij} + \mathbf{v}_{t,k}^{ij}  \Delta t \; ,
	\end{equation}
	\begin{equation}
		\mathbf{u}_{t,k}^{ij} = \hat{\mathbf{u}}_{t,k}^{ij} - (\mathbf{n}_k^{ij} \cdot \hat{\mathbf{u}}_{t,k}^{ij} ) \mathbf{n}_k^{ij} \; .
	\end{equation}
\end{subequations}
It is noted that: ($i$) the tangential displacement histories of all neighbor particles in $C(i;t)$ should be maintained for each $P_i$ to achieve an accumulated friction force \cite{jonJCND2015}; and ($ii$) to update the value of tangential displacement at the current step, the Coulomb capping of the friction force should also be enforced, which herein is accomplished as
\begin{equation}
	\mathbf{u}_{t}^{ij} = \text{min} \left(\mathbf{u}_{t,k}^{ij}, \mathbf{u}_{t,k}^{ij} \frac{\mu_s k_n \delta^{ij}}{k_t | \mathbf{u}_{t,k}^{ij} |}\right) \; ,
\end{equation}
where $\mu_s$ is the static friction coefficient between particles. Finally, the rolling resistance torque $\mathbf{M}_{rr}^{ij}$ applied on the particles is obtained as:
\begin{equation}
	{\mathbf M}_{rr}^{ij}=\frac{\boldsymbol \omega^{ij}}{\left|{\boldsymbol \omega}^{ij} \right| }\mu_r R^{ij}\left| {\mathbf F}_n^{ij}\right| \; ,
\end{equation}
where ${\boldsymbol \omega}^{ij} \equiv {\boldsymbol \omega}^{i}-{\boldsymbol \omega}^{j}$, and $\mu_r$ is the rolling resistance coefficient \cite{AiAssessRolling2011}.

\subsection{Overview of the SCM method}\label{subsec:scmMethod}
\begin{figure}[htp]
	\centering
	\includegraphics[width=6in]{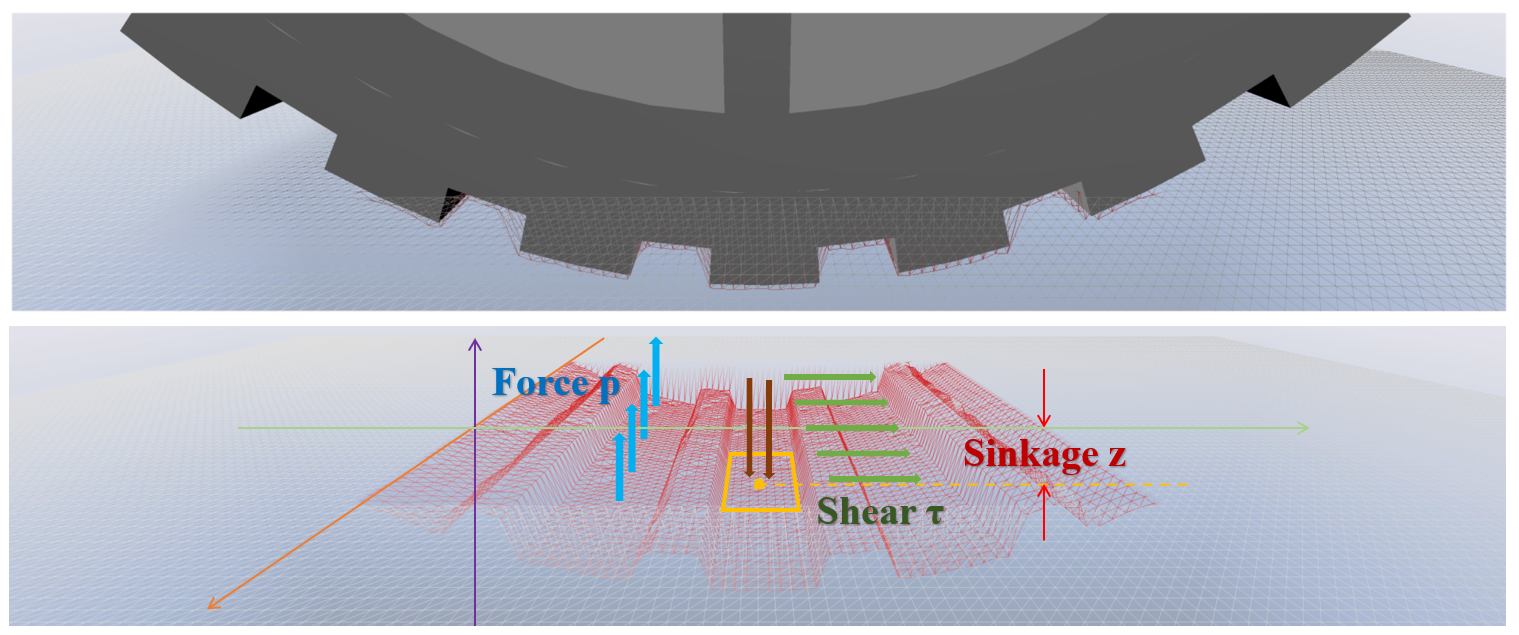}
	\caption{Schematic view of the SCM model for wheel-soil interaction; wheel can have arbitrary shape.}
	\label{fig:scm_sketch}
\end{figure}

The SCM approach employed in this study was first proposed in~\cite{Krenn2008SCM,Krenn2011}. It is an empirical terramechanics model that draws on the Bekker-Wong and Janosi-Hanamoto formulas. Thus, the normal pressure $p$ between the soil and the cylindrical wheel is expressed as~\cite{bekker56}:
\begin{subequations}
\begin{equation}\label{equ:bekker}
	p =  \left(  \frac{K_c}{b} + K_\phi  \right)   z^n \, ,
\end{equation}
where $z$ is the wheel sinkage and $b$ is the wheel width. The parameters $K_c$, $K_\phi$, and $n$ need to be identified using experimental data and a curve fitting algorithm. Since the original Bekker-Wong formula assumes that the wheel is cylindrical, it doesn't apply to problems with general wheel geometries. The SCM approach generalizes the original formula and extends it to problems with arbitrary wheel shapes, e.g., the rover wheel with grousers shown in Fig.~\ref{fig:scm_sketch}. The wheel can be any shape described using a mesh, while the terrain can be non-flat and described using a terrain height map or a mesh. Once the wheel mesh makes contact with the terrain mesh, a ray casting approach is employed to compute the contact patch, see \cite{chronoSCM2022}. The contact patch is then used to evaluate its perimeter $L_{\text{patch}}$ and area $A_{\text{patch}}$. The characteristic length $b$ in Eq.~(\ref{equ:bekker}) is approximated using $ b \approx 2 A_{\text{patch}} / L_{\text{patch}} $ and subsequently used to evaluate the normal contact as in  Eq.~(\ref{equ:bekker}). 

To evaluate the  shear stress between the wheel and terrain, the Janosi-Hanamoto formula \cite{janosi61} was employed in conjunction with the normal pressure evaluated by the Bekker-Wong formula:
	\begin{equation}\label{equ:janosi}
		\tau =  \tau_{\text{max}} ( 1 - e^{-J_s/K_s}) \; , \qquad \tau_{\text{max}} \equiv  c + p \tan \varphi
	\end{equation}
\end{subequations}
where $c$ is the cohesion coefficient; $\varphi$ is the internal friction angle; $J_s$ is the accumulated shear displacement between the contact patches; and $K_s$ is the so-called Janosi parameter. Equations.~(\ref{equ:bekker}) and (\ref{equ:janosi}) can be used to apply both normal contact and tangential friction forces onto a wheel with arbitrary shape. Soil deformation can also be tracked, with the caveat that this deformation is assumed to occur solely in the direction normal to the surface of the terrain. The DOF count for SCM depends on the resolution of the terrain as dictated by the density of the height-map used. A typical resolution value for both the X and Y directions is 0.01 meters. At moderate height-map resolutions, and if the area of terramechanics interest is not very large, the SCM implementation is expected to provide real-time performance. More details regarding the SCM methodology can be found in \cite{Krenn2008SCM,Krenn2011,chronoSCM2022}.

\subsection{Overview of the Bayesian inference framework}\label{subsec:bayesianMethod}
The Bayesian inference approach employed in this work aims to use a statistical framework to provide estimates of the parameters associated with the SCM model. In a statistical sense, a measured (or observed) value $ \mathcal{Y} $ can be expressed as the sum of two components:
\begin{align}
	\label{eq:gauss}
	\mathcal{Y}=\mathcal{G}(\mathcal{X})+\varepsilon, \quad \varepsilon \sim N(0,\Gamma) \;,
\end{align}
where $\mathcal{X}$ is the set of unknown parameters of a known computer model $\mathcal{G}$, and $\varepsilon$ is the statistical term that follows a normal distribution with zero-mean and a covariance matrix $\Gamma$ and which is meant to capture small and random deviations of the model from the observed data. The goal of the approach is to estimate the posterior distribution $p(\mathcal{X}|\mathcal{Y})$ using Bayesian inference:
\begin{align}
	\label{eq:like}
	p(\mathcal{X}|\mathcal{Y}) \propto \exp\left(-\frac{1}{2}\|\mathcal{Y}-\mathcal{G}(\mathcal{X})\|_\Gamma^2 \right) p(\mathcal{X}) \; ,
\end{align}
where $p(\mathcal{X})$ is the prior distribution of the unknown parameters stored in $\mathcal{X}$, a distribution that in our case encapsulates existing knowledge about the SCM parameters. To draw samples out of the posterior distribution $p(\mathcal{X}|\mathcal{Y})$, we employ the Metropolis-Hastings algorithm, see \cite{gelmanBayesian1995,robertMH2015,andrieu2003introduction} for technical details. The calibration algorithm, which was available in a third party software called PyMC \cite{pymc3}, anchors a two-step methodology. In Step 1, the higher-fidelity DEM terramechanics is employed in a virtual bevameter test simulation to generate ground truth data used to calibrate the lower-fidelity SCM terramechanics. A plate sinkage test and an annulus shear test are conducted using the virtual bevameter test rig. The goal is to find the model parameters $\mathcal{X}$ that makes SCM's terramechanics performance close to $\mathcal{Y}$; i.e., that of the DEM approach. 

Finally, Step 2 is tied to the process of obtaining sample values of the parameter array $\mathcal{X}$ according to the posterior distribution $p(\mathcal{X}|\mathcal{Y})$. The iterative Bayesian process produces this statistical distribution that characterizes how well candidate sets of SCM parameters explain the ground truth data. If enough iterations are carried out, a smooth posterior distribution emerges, out of which one can extract the set of parameters $\mathcal{X}$ that has the highest likelihood of explaining the ground truth $\mathcal{Y}$. This was the approach embraced in this contribution, i.e., we used the maximum likelihood estimator for subsequent SCM analysis. To increase the confidence that the posterior distribution has converged, we ran four Markov Chains in parallel and compared the results from each chain.

\section{Calibration of the SCM model}\label{sec:calibration}

This section describes ($i$) the process of ground truth data generation by means of a virtual bevameter test rig, and ($ii$) how the SCM parameters are calibrated via Bayesian inference using the ground truth data produced via DEM terramechanics. For ($i$), two groups of virtual bevameter simulations were involved -- the plate sinkage test and the annulus shear test, both simulated in a multibody system dynamics framework. The dynamics of the granular test bed was modeled using the high-fidelity DEM approach. The dynamics of the granular material and the bevameter test rig were concurrently simulated within a DEM-multibody dynamics co-simulation framework. The dynamics of the multibody system was handled on a multi-core CPU; the time evolution of the DEM particles was determined using GPU computing. At each time step, the position and velocity of the plate or annulus were transferred to the DEM solver where they were used to impose moving boundary conditions. Conversely, the contact force  and resulting torque applied on the plate or annulus were transferred back to the multibody solver as external loads. A terrain settling simulation is first carried out to produce a stable particle distribution as a pre-processing step for the actual virtual bevameter test. The initial particle distribution was the outcome of a settling simulation under a given gravity for about 10-15 s, which is long enough to obtain a good initial configuration for the material. Note that this settling process would lead to different outcomes in lunar and Earth conditions. Screenshots from the terrain settling simulation are shown in Fig.~\ref{fig:terrain}. The full animation of this simulation is provided in the supplementary information. All simulations concerning the plate sinkage and the annulus shear tests started from this particular settled configuration, which is shown in Fig.~\ref{fig:terrain}(b).

\begin{figure}[h]
	\centering
	\begin{subfigure}{0.49\textwidth}
		\centering
		\includegraphics[width=3in]{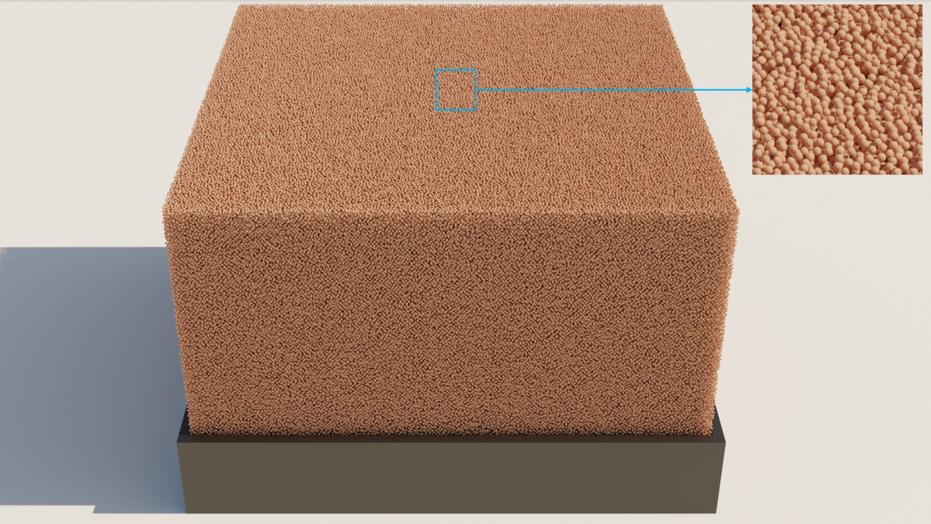}
		\caption{Initial particle distribution.}
	\end{subfigure}
	\begin{subfigure}{0.49\textwidth}
		\centering
		\includegraphics[width=3in]{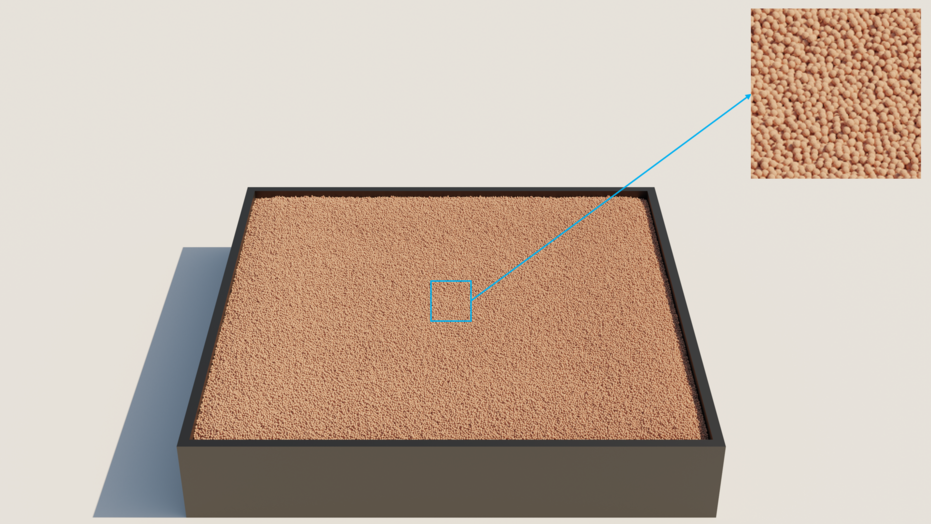}
		\caption{Final particle distribution.}
	\end{subfigure}
	\caption{Screenshots of the terrain settling simulation with DEM particles. Animation of the settling simulation is provided in the supplementary materials.} 
	\label{fig:terrain}
\end{figure}

\subsection{The plate sinkage virtual test, and the $K_c$, $K_\phi$, and $n$  calibration}
\begin{figure}[htp]
	\centering
	\includegraphics[width=6in]{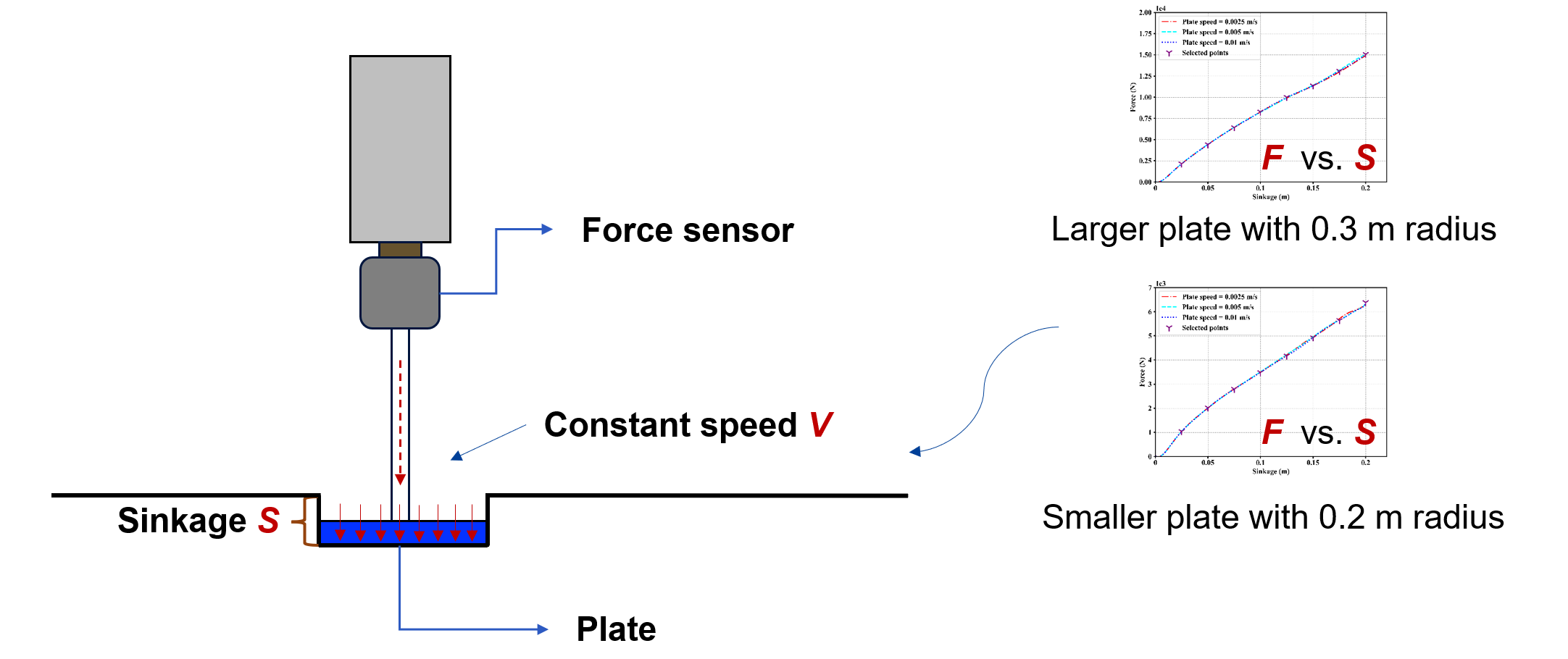}
	\caption{Schematic of the plate sinkage test.}
	\label{fig:sinkage}
\end{figure}

\begin{figure}[h]
	\centering
	\begin{subfigure}{0.49\textwidth}
		\centering
		\includegraphics[width=3in]{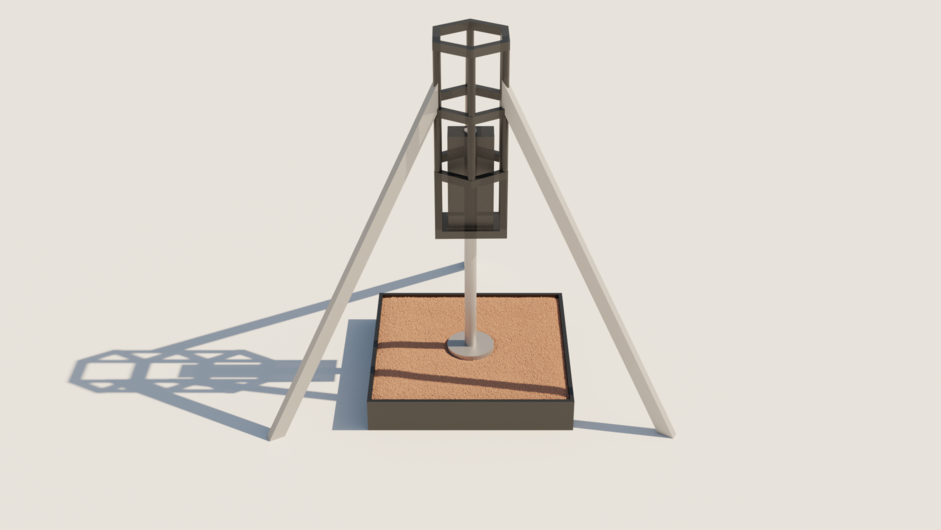}
		\caption{Overall view.} 
	\end{subfigure}
	\begin{subfigure}{0.49\textwidth}
		\centering
		\includegraphics[width=3in]{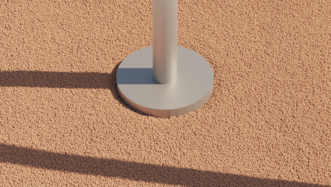}
		\caption{Zoomed-in perspective.}
	\end{subfigure}
	\caption{Screenshot of the plate sinkage test with 0.2 m radius plate. Animation of the plate sinkage test simulation is provided in the supplementary materials.} 
	\label{fig:plate_sinkage}
\end{figure}

\begin{figure}[h]
	\centering
	\begin{subfigure}{0.49\textwidth}
		\centering
		\includegraphics[width=3in]{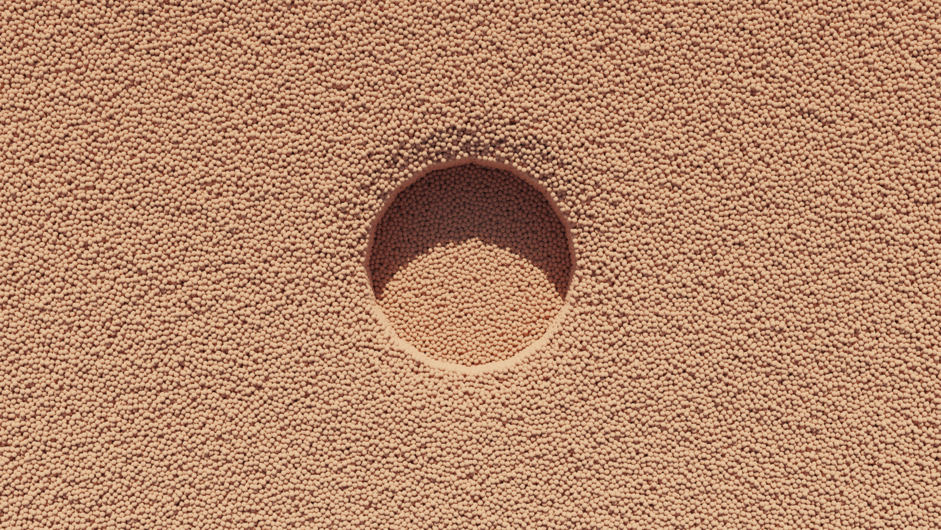}
		\caption{Top view with 0.2 m radius plate.}
	\end{subfigure}
	\begin{subfigure}{0.49\textwidth}
		\centering
		\includegraphics[width=3in]{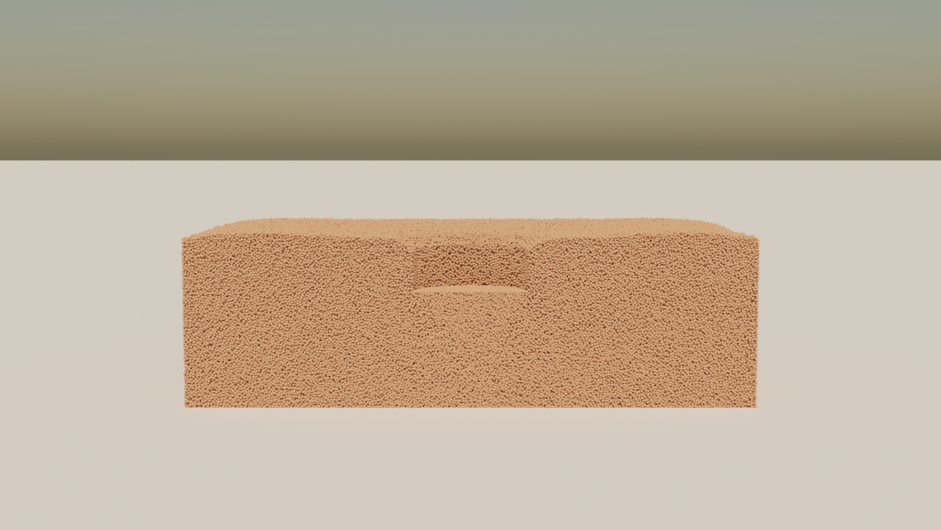}
		\caption{Side view with 0.2 m radius plate.} 
	\end{subfigure}
	\begin{subfigure}{0.49\textwidth}
		\centering
		\includegraphics[width=3in]{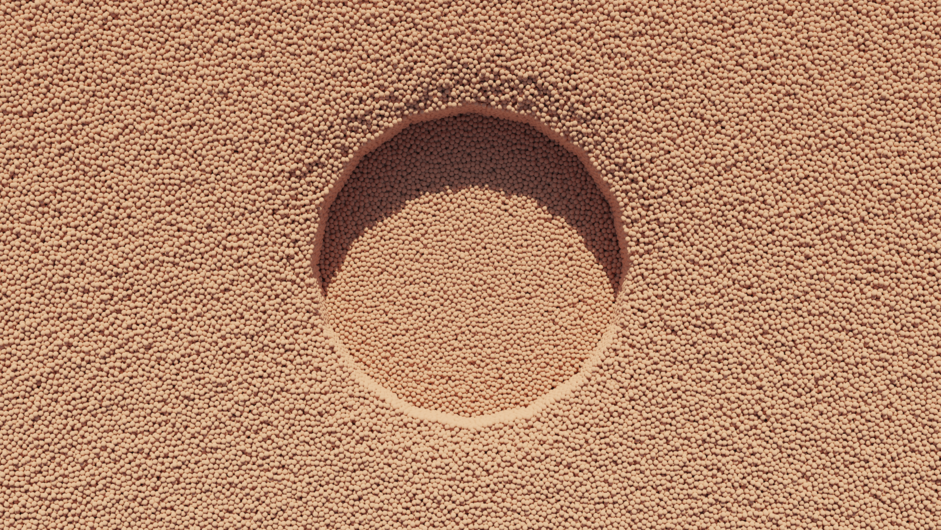}
		\caption{Top view with 0.3 m radius plate.}
	\end{subfigure}
	\begin{subfigure}{0.49\textwidth}
		\centering
		\includegraphics[width=3in]{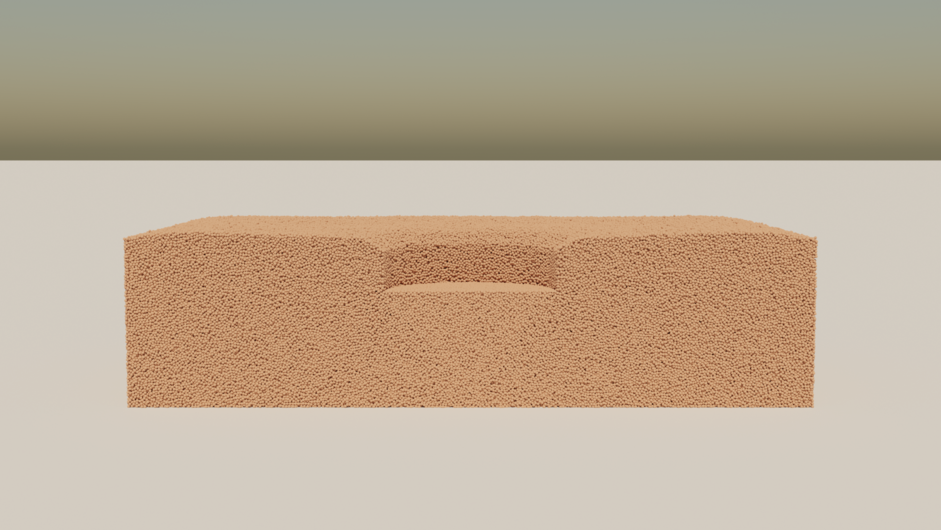}
		\caption{Side view with 0.3 m radius plate.} 
	\end{subfigure}
	\caption{Particle distribution of the sinkage test with different plates.} 
	\label{fig:plate_sinkage_local}
\end{figure}

The plate sinkage virtual test aims to calibrate the parameters $K_c$, $K_\phi$, and $n$ associated with the normal stiffness of the soil in the SCM model. To that end, plates with two different sizes were each pressed down into the soil with three different constant vertical velocities, as shown in Fig.~\ref{fig:sinkage}. Thus, six different complete simulations were performed to produce ground truth data. The radius of the DEM particles was set to 0.005 $\si{m}$, while the density was set to 2650 $\si{kg/m^3}$. The particle-particle and the particle-plate static friction and rolling friction coefficients were set to 0.9. The particle-particle and the particle-plate normal damping coefficients were set to \num{10000} $\si{s^{-1}}$ and tangential damping coefficients were set to \num{2000} $\si{s^{-1}}$. The particle-particle and the particle-plate normal stiffness coefficients and tangential stiffness coefficients were set to \num{10000} $\si{N/m}$.The contact cohesion force between each pair of particles was set to 0.5 times the weight of one individual particle. In the simulation with a smaller size plate (0.2 $\si{m}$ plate radius), the soil bin was of size 2 m $\times$ 2 m $\times$ 0.6 m and contained approximately \num{2760000} particles, while in the simulation with a larger plate (0.3 $\si{m}$ plate radius), the soil bin was 2.4 m $\times$ 2.4 m $\times$ 0.6 m and contained approximately \num{3980000} particles. The three vertical pressing velocities were set to 0.01 m/s, 0.005 m/s, and 0.0025 m/s. To ensure that the final sinkage depth for experiments was identical, the duration of the simulations were set to 20 s, 40 s, and 80 s, respectively. One snapshot of the virtual experiment, run with the smaller plate, is provided in Fig.~\ref{fig:plate_sinkage}. The figure shows both an overall view of the test rig and a zoom-in view of DEM particles around the plate. Figure~\ref{fig:plate_sinkage_local} shows the top and side views of the particle distribution around the plate (the plates were not rendered for better visualization of the particles).

\begin{figure}[h]
	\centering
	\begin{subfigure}{0.49\textwidth}
		\centering
		\includegraphics[width=3.2in]{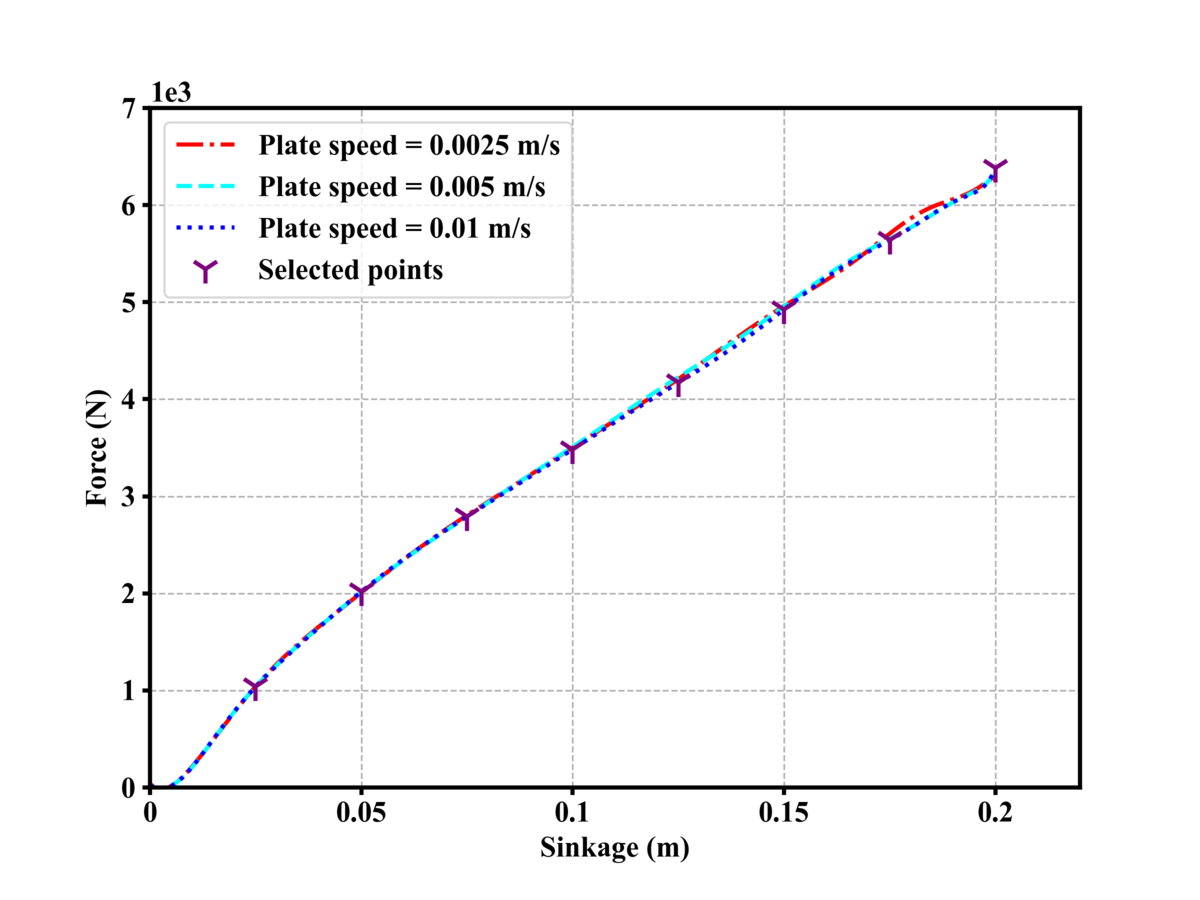}
		\caption{Plate with radius of 0.2 m.}
	\end{subfigure}
	\begin{subfigure}{0.49\textwidth}
		\centering
		\includegraphics[width=3.2in]{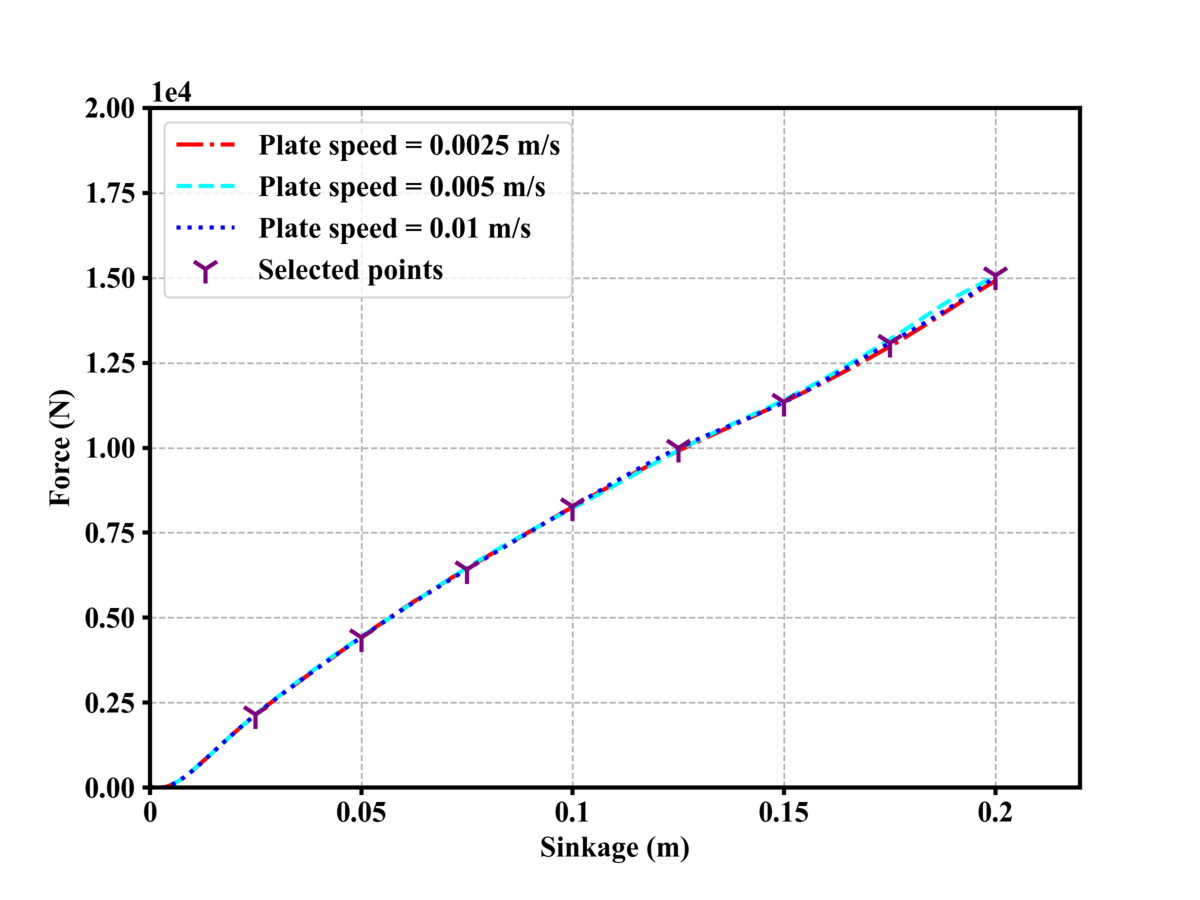}
		\caption{Plate with radius of 0.3 m.} 
	\end{subfigure}
	\caption{Force applied to the plate to maintain the vertical constant speed versus the sinkage depth of the plate with three different pressing velocities. Eight selected points (purple markers) are shown in each curve.} 
	\label{fig:f_sinkage}
\end{figure}

Figures~\ref{fig:f_sinkage} (a) and (b) provide the relationship between the pressing force and the sinkage of the plate. It is noted that: ($i$) to reach the same sinkage depth, the plate with a larger radius needs to be pressed with a larger normal force -- about 2.3 times larger than the force required by the smaller plate; ($ii$) for a given plate size, the pressing force is only related to the plate sinkage, which is consistent with the formula in Eq.~(\ref{equ:bekker}). To get the ground truth data from the high-fidelity DEM simulation of the bevameter tests, we collected pressing force values at eight different sinkage depths, see Fig.~\ref{fig:f_sinkage}. The value of the pressing force is reported in Table~\ref{tab:sinkage_F}, and it was computed by averaging the force values obtained for each of the three different sinkage velocities. Note that given the low sinking rates, the forces required to sink the plate are essentially rate independent.

\begin{table}[h]
	\caption{Values of the selected points (Force vs. Sinkage).} 
	\label{tab:sinkage_F}
	\setlength{\tabcolsep}{8pt}
	\renewcommand{\arraystretch}{1.2}
	\begin{center}
		\begin{tabular}{| c | c | c | c | c | c | c | c | c |} 
			\hline
			Sinkage (m) & 0.025 & 0.05 & 0.075 & 0.1 & 0.125 & 0.15 & 0.175 & 0.2 \\ 
			\hline
			Force (N) - Small plate & 1038 & 2018 & 2792 & 3484 & 4170 & 4924 & 5640 & 6381 \\
			\hline
			Force (N) - Large plate & 2147 & 4422 & 6420 & 8269 & 9989 & 11346 & 13084 & 15071 \\
			\hline
		\end{tabular}
	\end{center}
\end{table}

\begin{figure}[htp]
	\centering
	\begin{subfigure}{0.32\textwidth}
		\centering
		\includegraphics[width=2.3in]{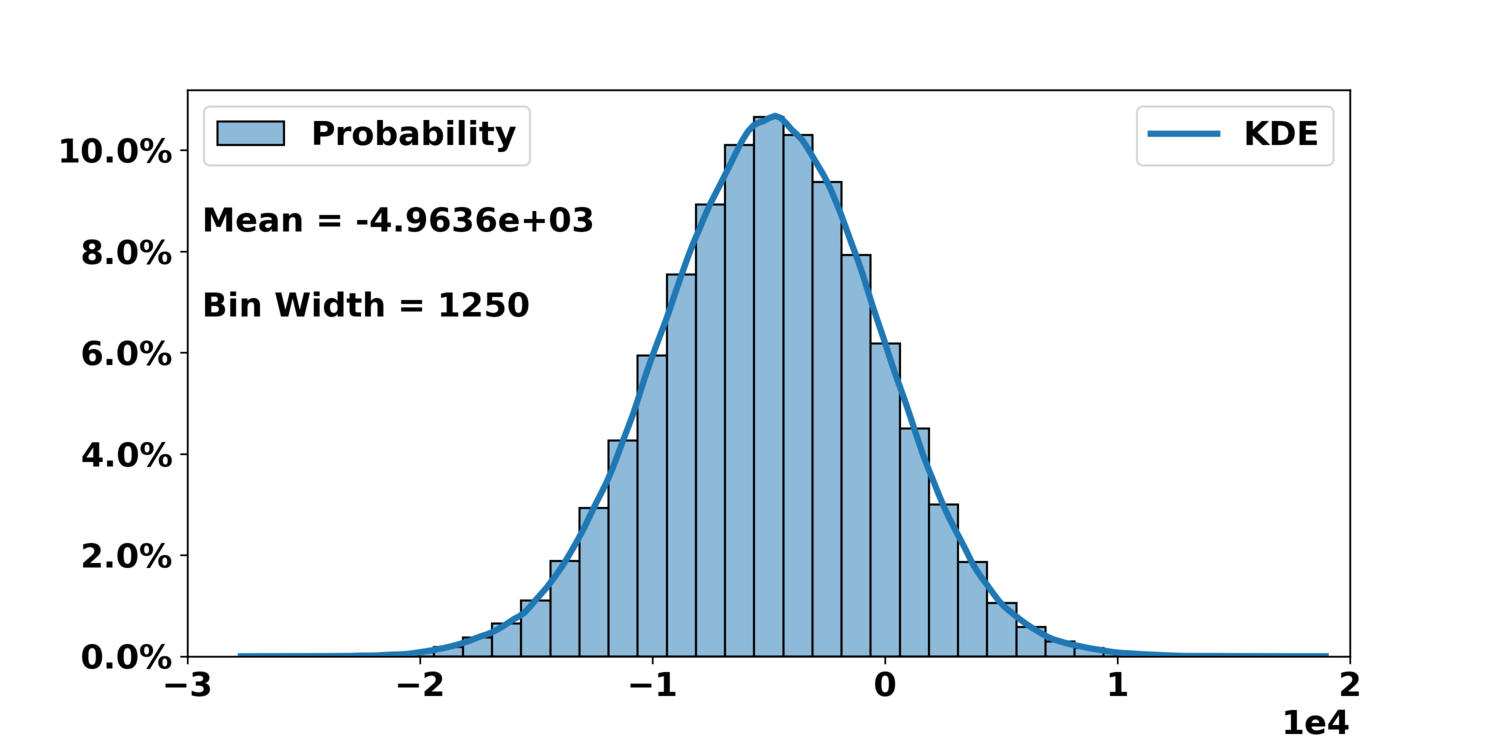}
		\caption{Mean.}
	\end{subfigure}
	\begin{subfigure}{0.32\textwidth}
		\centering
		\includegraphics[width=2.3in]{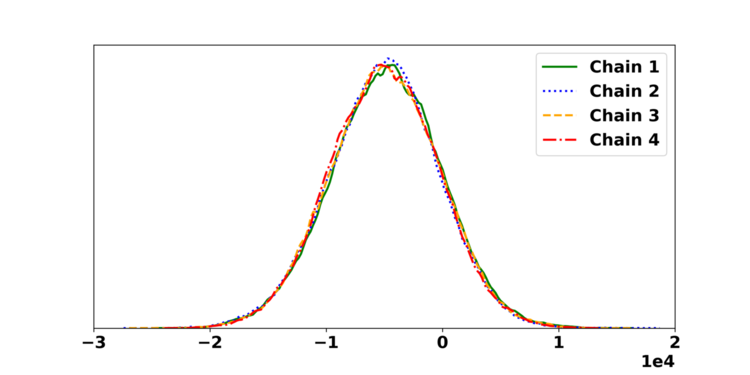}
		\caption{Four chains.} 
	\end{subfigure}
	\begin{subfigure}{0.32\textwidth}
		\centering
		\includegraphics[width=2.3in]{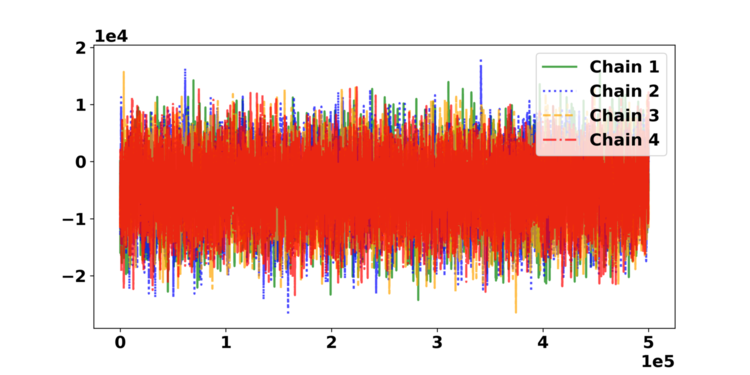}
		\caption{Trace.}
	\end{subfigure}
	\caption{Averaged posterior probability distribution and its kernel density estimation (KDE) of four chains for the parameter $K_c$ with different number of samples.} 
	\label{fig:trace_posterior_kc}
\end{figure}

\begin{figure}[htp]
	\centering
	\begin{subfigure}{0.32\textwidth}
		\centering
		\includegraphics[width=2.3in]{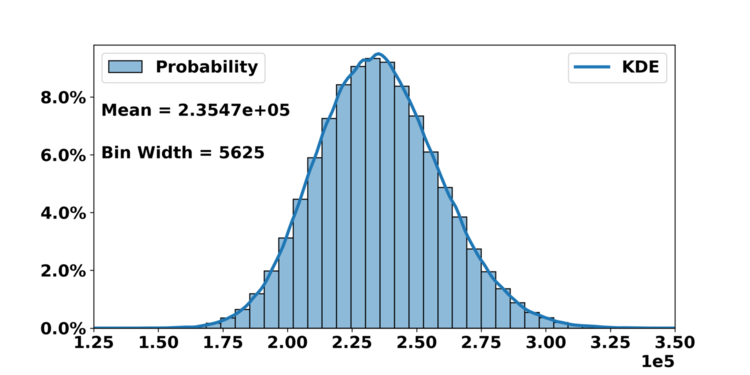}
		\caption{Mean.}
	\end{subfigure}
	\begin{subfigure}{0.32\textwidth}
		\centering
		\includegraphics[width=2.3in]{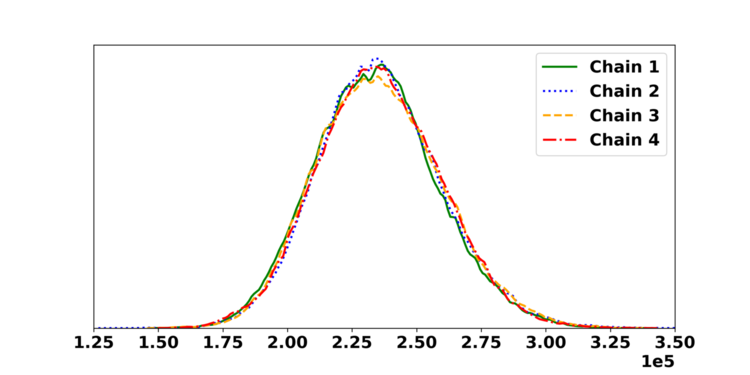}
		\caption{Four chains.} 
	\end{subfigure}
	\begin{subfigure}{0.32\textwidth}
		\centering
		\includegraphics[width=2.3in]{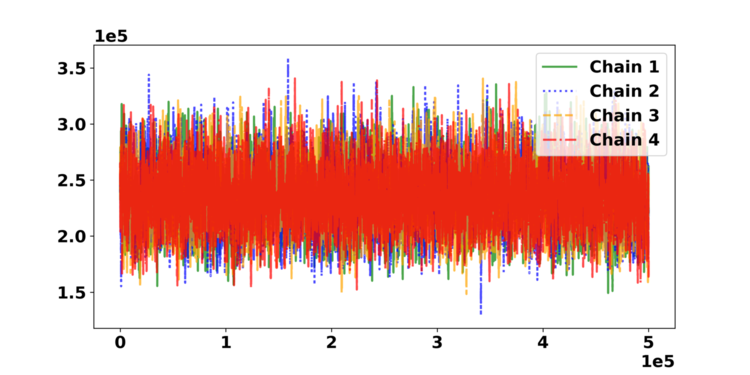}
		\caption{Trace.}
	\end{subfigure}
	\caption{Averaged posterior probability distribution and its kernel density estimation (KDE) of four chains for the parameter $K_{\phi}$ with different number of samples.} 
	\label{fig:trace_posterior_kphi}
\end{figure}

\begin{figure}[htp]
	\centering
	\begin{subfigure}{0.32\textwidth}
		\centering
		\includegraphics[width=2.3in]{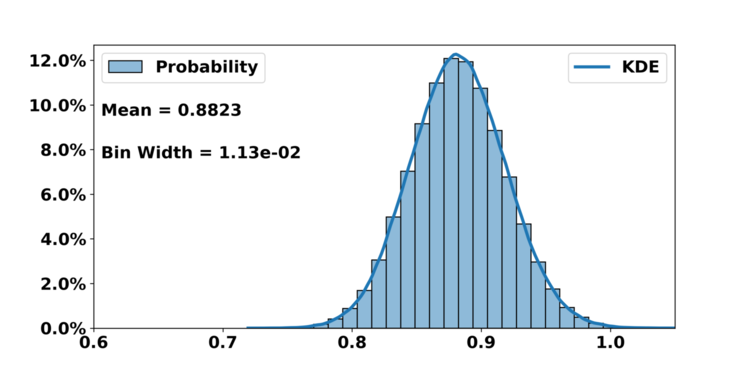}
		\caption{Mean.}
	\end{subfigure}
	\begin{subfigure}{0.32\textwidth}
		\centering
		\includegraphics[width=2.3in]{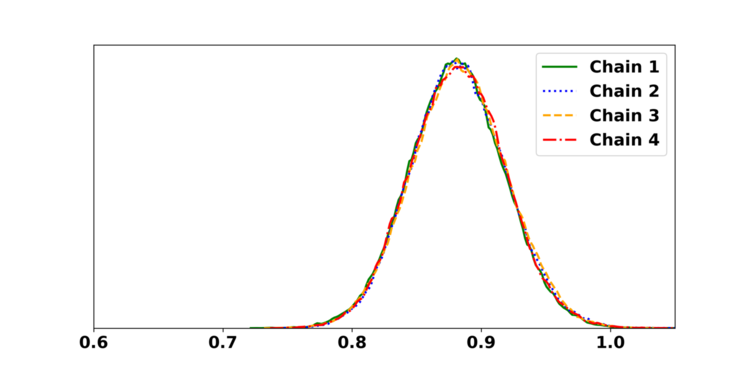}
		\caption{Four chains.} 
	\end{subfigure}
	\begin{subfigure}{0.32\textwidth}
		\centering
		\includegraphics[width=2.3in]{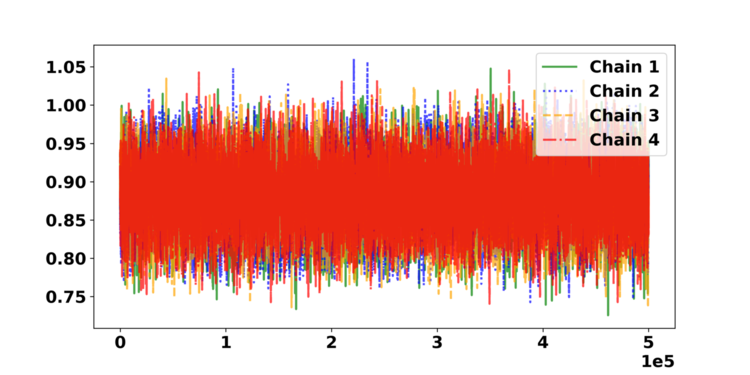}
		\caption{Trace.}
	\end{subfigure}
	\caption{Averaged posterior probability distribution and its kernel density estimation (KDE) of four chains for the parameter $n$ with different number of samples.}
	\label{fig:trace_posterior_n}
\end{figure}

\begin{figure}[htp]
	\centering
	\begin{subfigure}{0.49\textwidth}
		\centering
		\includegraphics[width=3in]{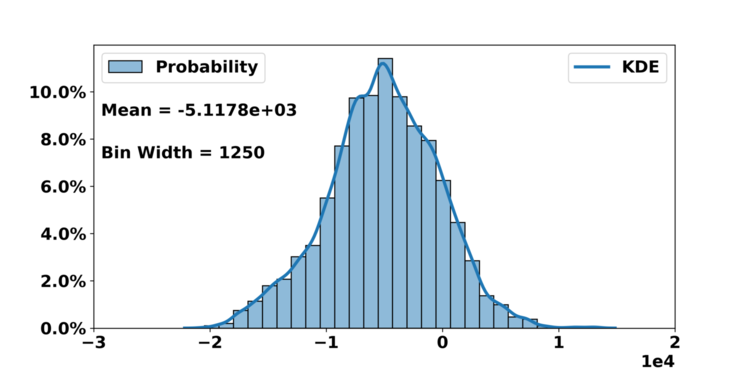}
		\caption{\num{10000} samples.}
	\end{subfigure}
	\begin{subfigure}{0.49\textwidth}
		\centering
		\includegraphics[width=3in]{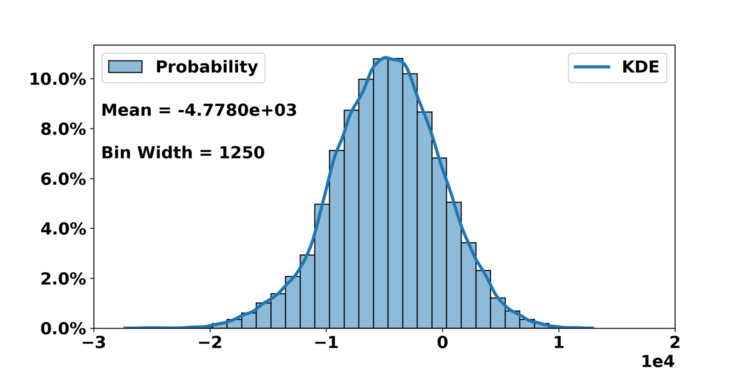}
		\caption{\num{50000} samples.} 
	\end{subfigure}
	\begin{subfigure}{0.49\textwidth}
		\centering
		\includegraphics[width=3in]{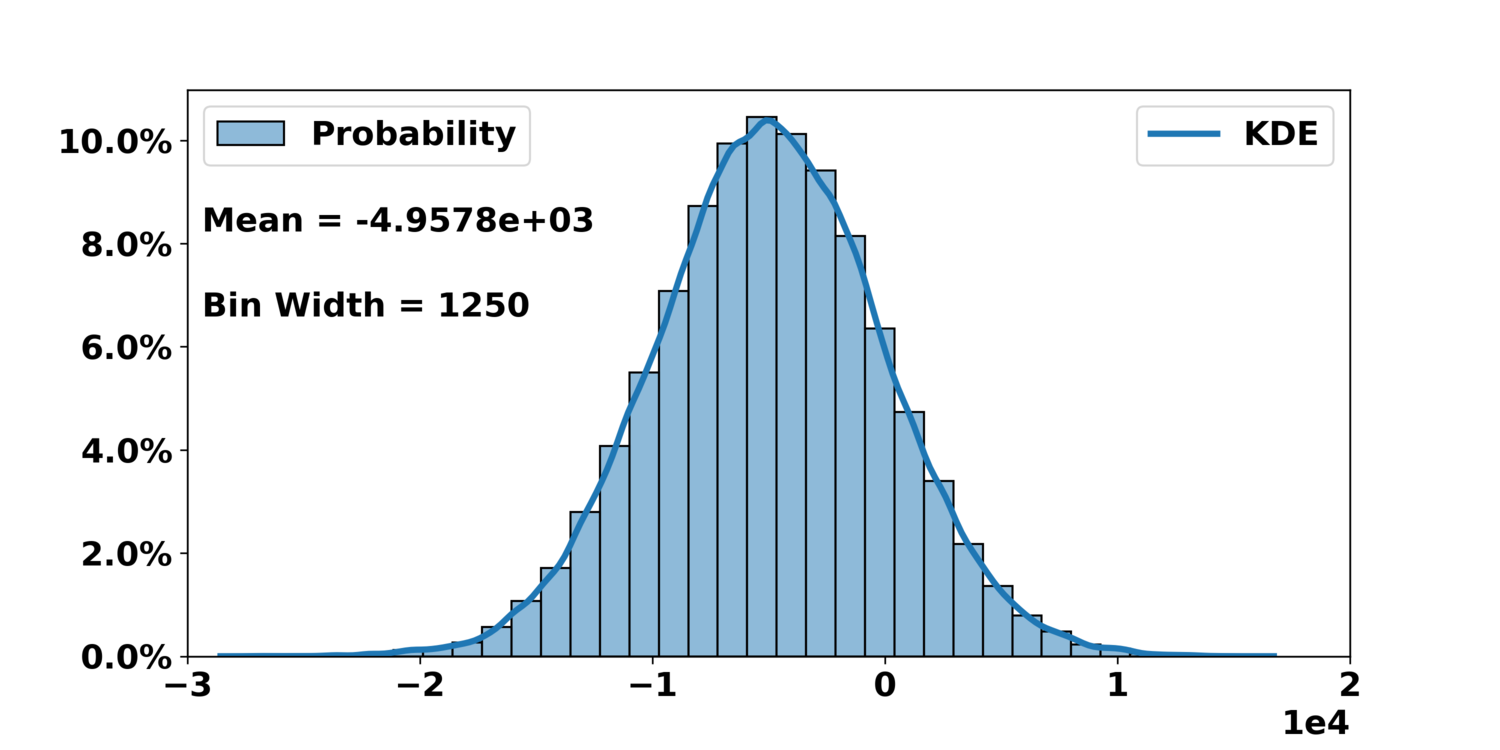}
		\caption{\num{100000} samples.}
	\end{subfigure}
	\begin{subfigure}{0.49\textwidth}
		\centering
		\includegraphics[width=3in]{Post_Mean_500000_samples_K_c.png}
		\caption{\num{500000} samples.} 
	\end{subfigure}
	\caption{Averaged posterior probability distribution and its kernel density estimation (KDE) for the parameter $K_c$ across four chains with different sample sizes.} 
	\label{fig:posterior_kc}
\end{figure}

\begin{figure}[htp]
	\centering
	\begin{subfigure}[t]{0.49\textwidth}
		\centering
		\includegraphics[width=1.5in]{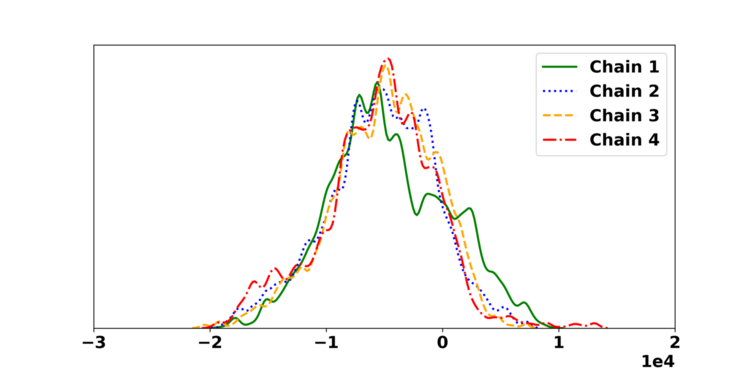}
		\includegraphics[width=1.5in]{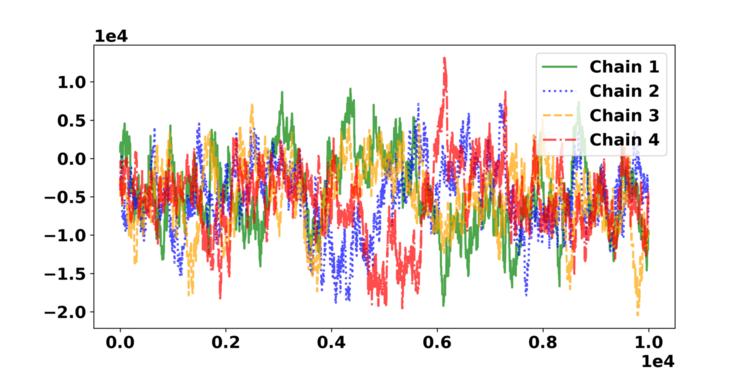}
		\caption{\num{10000} samples.}
	\end{subfigure}
	\begin{subfigure}[t]{0.49\textwidth}
		\centering
		\includegraphics[width=1.5in]{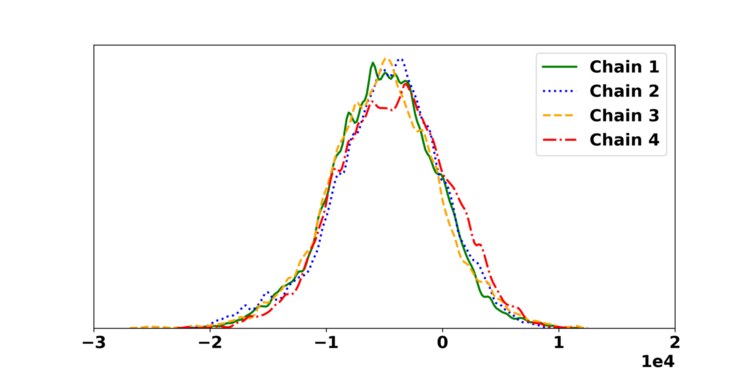}
		\includegraphics[width=1.5in]{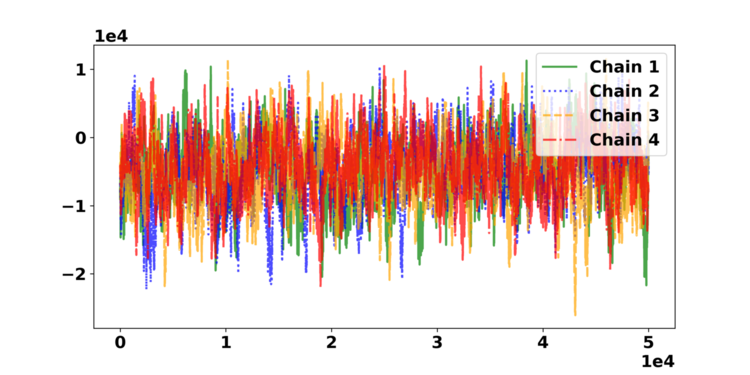}
		\caption{\num{50000} samples.}
	\end{subfigure}
	\begin{subfigure}[t]{0.49\textwidth}
		\centering
		\includegraphics[width=1.5in]{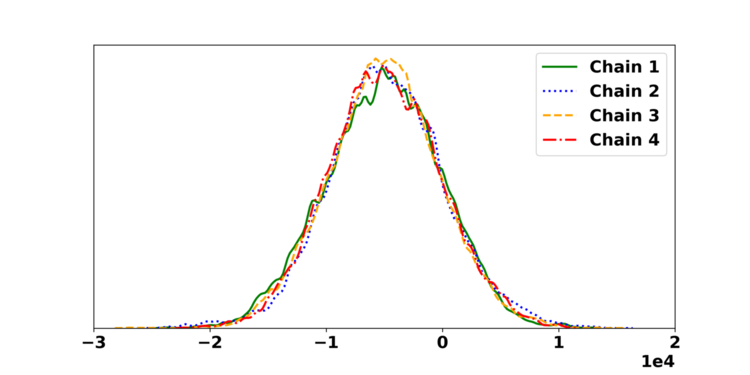}
		\includegraphics[width=1.5in]{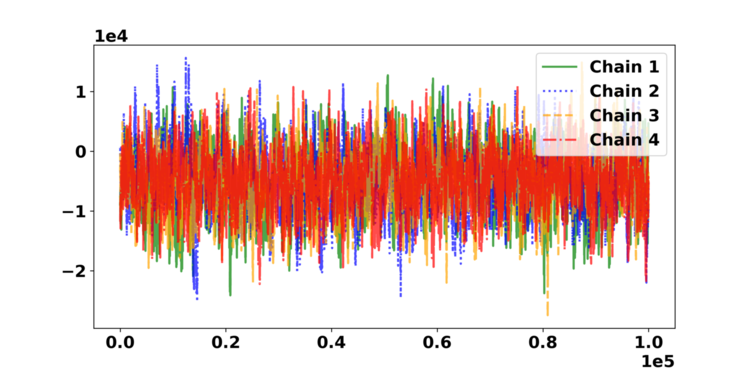}
		\caption{\num{100000} samples.}
	\end{subfigure}
	\begin{subfigure}[t]{0.49\textwidth}
		\centering
		\includegraphics[width=1.5in]{Post_500000_samples_K_c.png}
		\includegraphics[width=1.5in]{Trace_500000_samples_K_c.png}
		\caption{\num{500000} samples}
	\end{subfigure}
	\caption{Posterior probability distribution and trace plot for parameter $K_c$ in four different Metropolis-Hastings chains.}
	\label{fig:trace_posterior_kc_4chain}
\end{figure}

\begin{figure}[htp]
	\centering
	\begin{subfigure}{0.49\textwidth}
		\centering
		\includegraphics[width=3in]{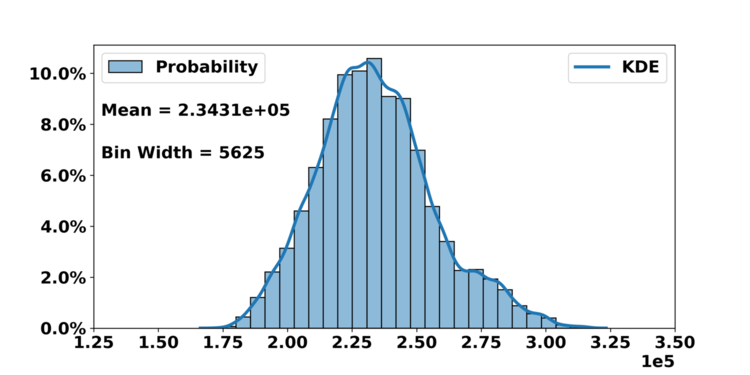}
		\caption{\num{10000} samples.}
	\end{subfigure}
	\begin{subfigure}{0.49\textwidth}
		\centering
		\includegraphics[width=3in]{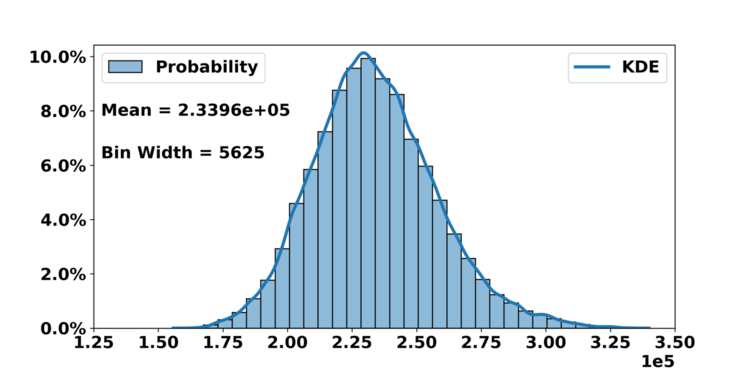}
		\caption{\num{50000} samples.} 
	\end{subfigure}
	\begin{subfigure}{0.49\textwidth}
		\centering
		\includegraphics[width=3in]{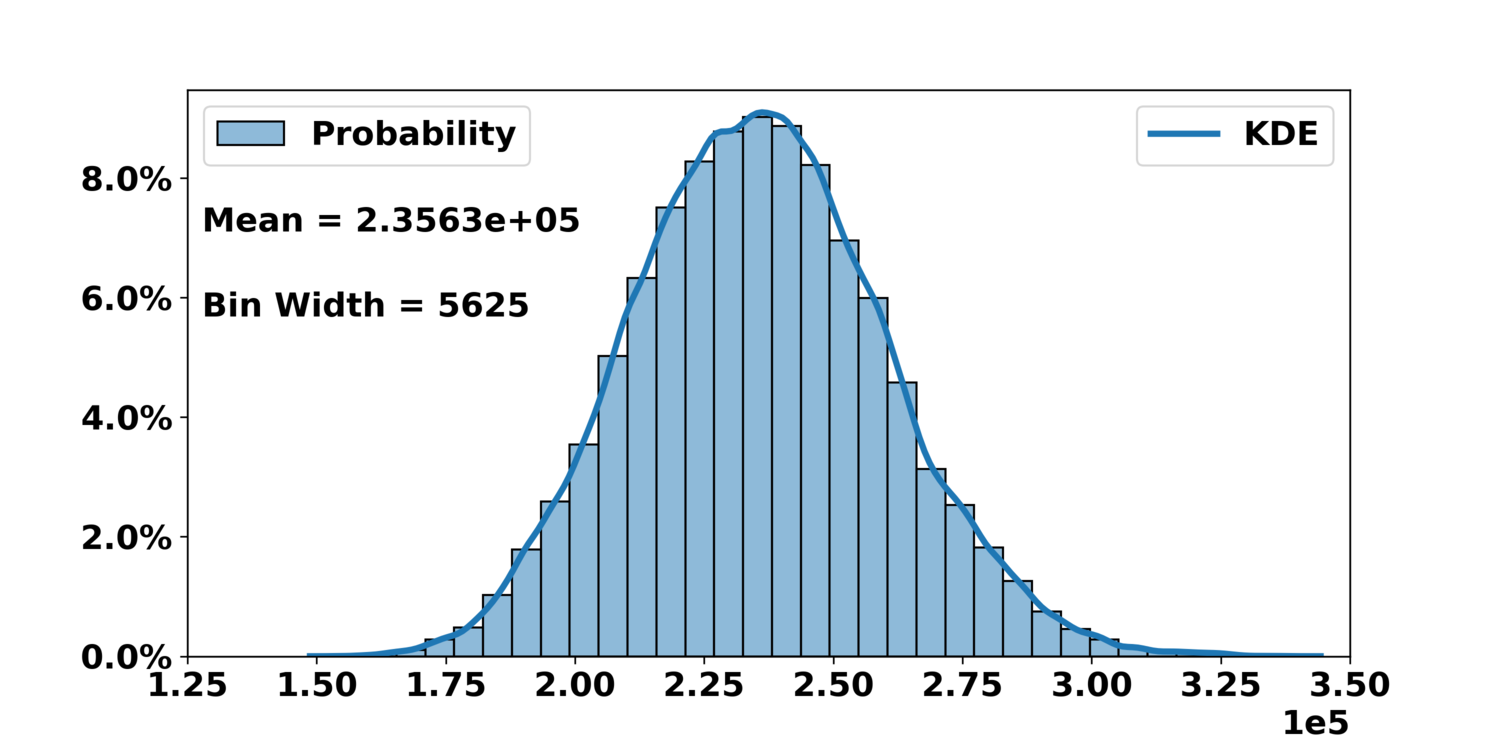}
		\caption{\num{100000} samples.}
	\end{subfigure}
	\begin{subfigure}{0.49\textwidth}
		\centering
		\includegraphics[width=3in]{Post_Mean_500000_samples_K_phi.png}
		\caption{\num{500000} samples.} 
	\end{subfigure}
	\caption{Averaged posterior probability distribution and its kernel density estimation (KDE) for the parameter $K_{\phi}$ across four chains with different sample sizes.} 
	\label{fig:posterior_kphi}
\end{figure}

To calibrate $K_c$, $K_{\phi}$, and $n$ in Eq. (\ref{equ:bekker}), we used the ground truth data in Table~\ref{tab:sinkage_F} as the input of the Bayesian inference framework. As in \cite{hu2024calibration}, the likelihood function was defined as:
\begin{equation}
\mathbb{L} = -\frac{1}{2\sigma^2}\sum_{k=1}^{N} \frac{\left [  F(k) - F^{truth}(k) \right ]^2}{N} \; ,
\end{equation}
where $F^{truth}(k)$ is the ground truth  force of Table \ref{tab:sinkage_F}; $F(k)$ is the SCM force evaluated at a given sinkage via  Eq.~(\ref{equ:bekker}) for a provided sample of the parameters $K_c$, $K_{\phi}$, $n$; $N=16$ is the number of ground truth data points collected; and $\sigma^2$ is a constant variance we expect to see between the ground truth and the evaluation, which was set to 0.01 in this study. To obtain a converged, smooth distribution of each parameter, we ran the sampling loop \num{500000} times. To guarantee convergence, we ran the sampling loop in four different chains as shown in Figs.~\ref{fig:trace_posterior_kc}(b)(c), \ref{fig:trace_posterior_kphi}(b)(c), and \ref{fig:trace_posterior_n}(b)(c). The posterior distributions produced by the four chains converged to the same result. Figures~\ref{fig:trace_posterior_kc}(a), \ref{fig:trace_posterior_kphi}(a), and \ref{fig:trace_posterior_n}(a) show the averaged posterior distribution of the three parameters, whose mean values are -4957, 235605, and 0.883, respectively. The Bayesian inference was run for several numbers of iterations, varied from \num{10000} to \num{500000}. The results are shown in Figs. \ref{fig:posterior_kc}, \ref{fig:trace_posterior_kc_4chain},  \ref{fig:posterior_kphi},  \ref{fig:trace_posterior_kphi_4chain}, \ref{fig:posterior_n}, and \ref{fig:trace_posterior_n_4chain} of the Appendix. This convergence analysis justified the choice of \num{500000} iterations adopted herein.

\begin{figure}[h]
	\centering
	\begin{subfigure}{0.49\textwidth}
		\centering
		\includegraphics[width=3.2in]{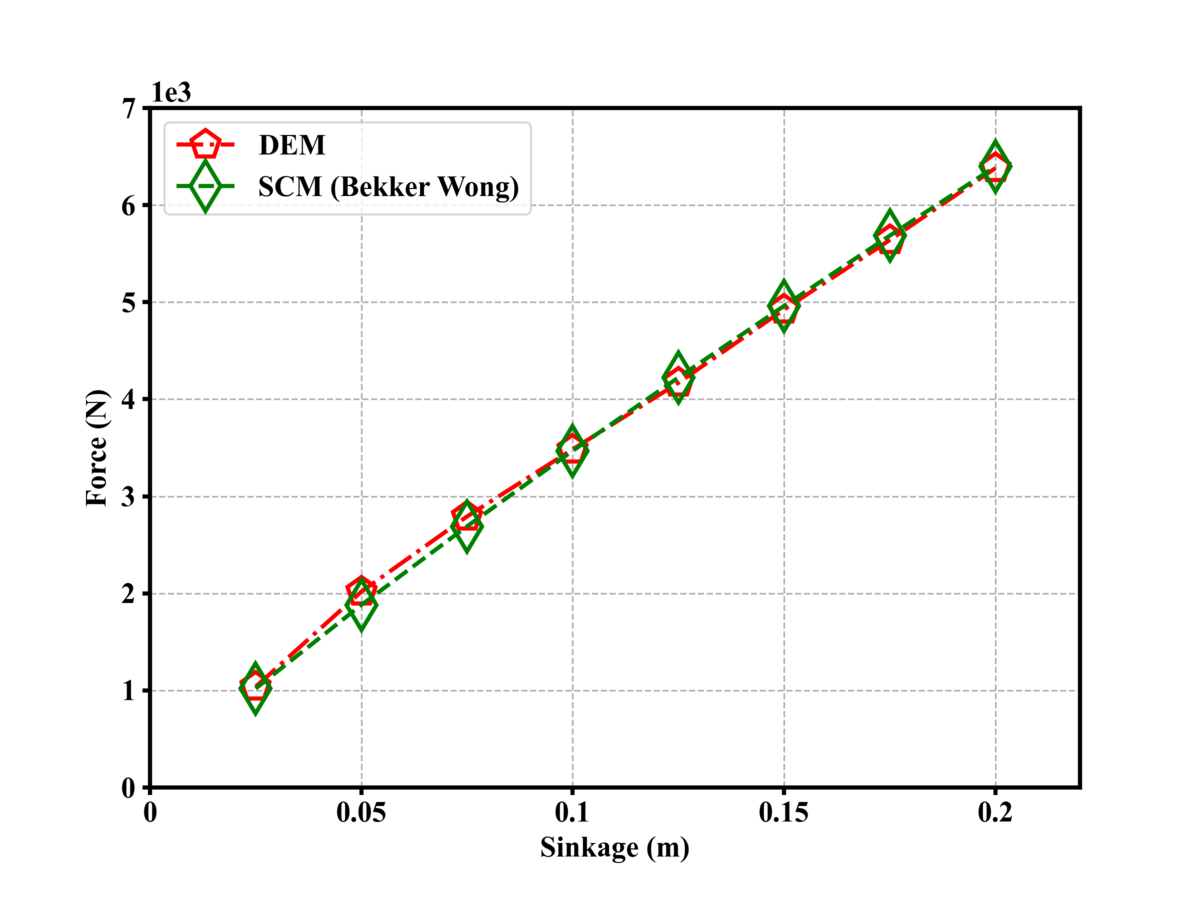}
		\caption{Small plate ($r$ = 0.2 m).}
	\end{subfigure}
	\begin{subfigure}{0.49\textwidth}
		\centering
		\includegraphics[width=3.2in]{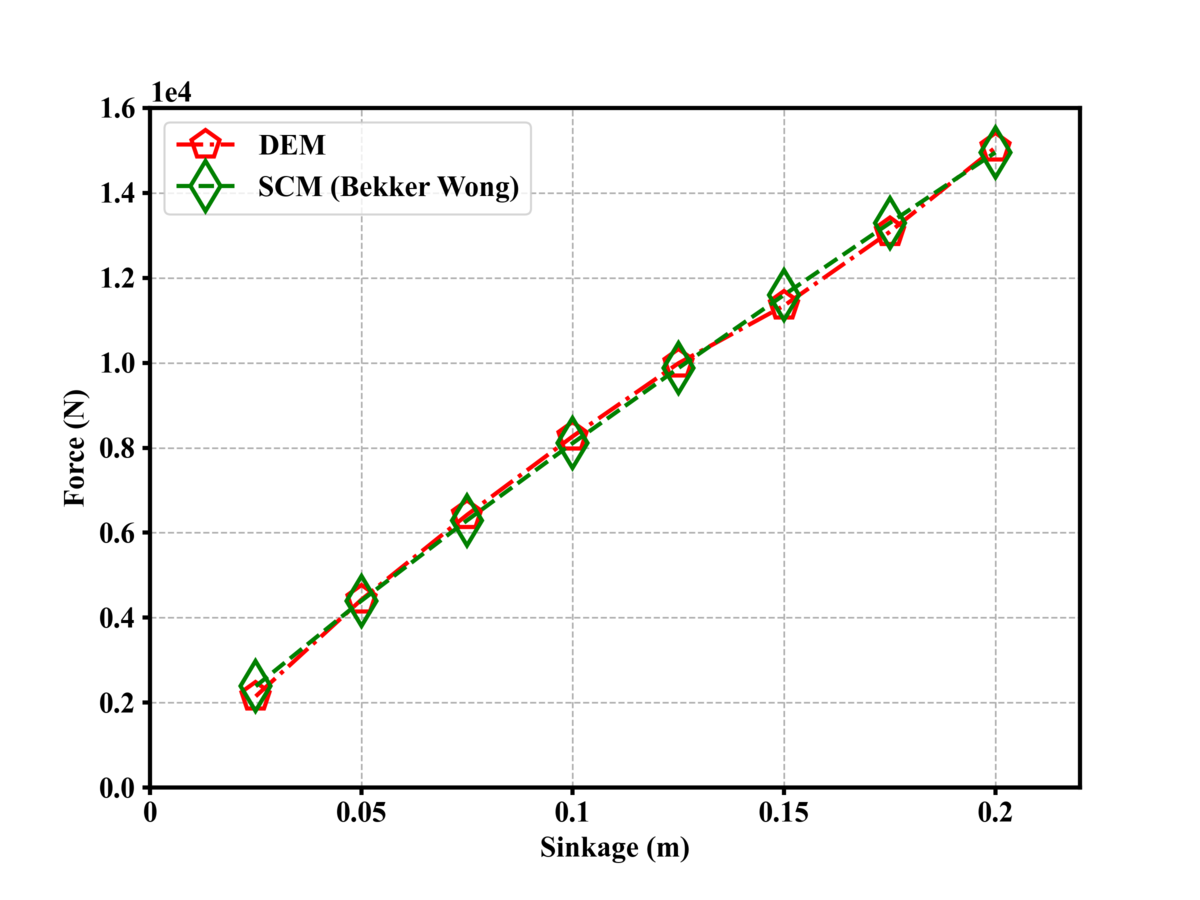}
		\caption{Large plate ($r$ = 0.3 m).} 
	\end{subfigure}
	\caption{A comparison between selected points, DEM vs. SCM. The SCM results are calculated using calibrated values of $K_c$, $K_{\phi}$, and $n$ and the Bekker-Wong formulation of Eq.~(\ref{equ:bekker}).} 
	\label{fig:f_sinkage_dem_scm}
\end{figure}

\begin{figure}[h]
	\centering
	\begin{subfigure}{0.49\textwidth}
		\centering
		\includegraphics[width=3.2in]{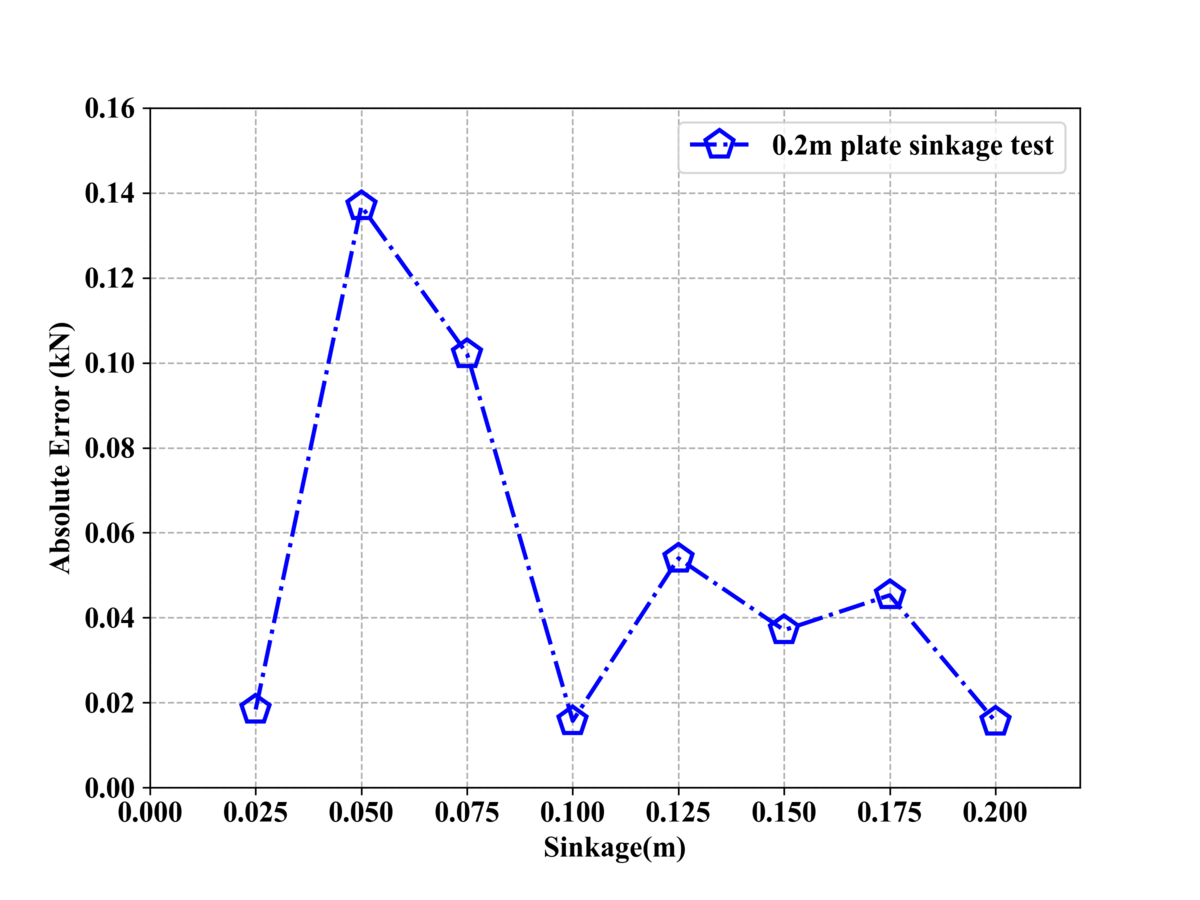}
		\caption{Small plate ($r$ = 0.2 m).}
	\end{subfigure}
	\begin{subfigure}{0.49\textwidth}
		\centering
		\includegraphics[width=3.2in]{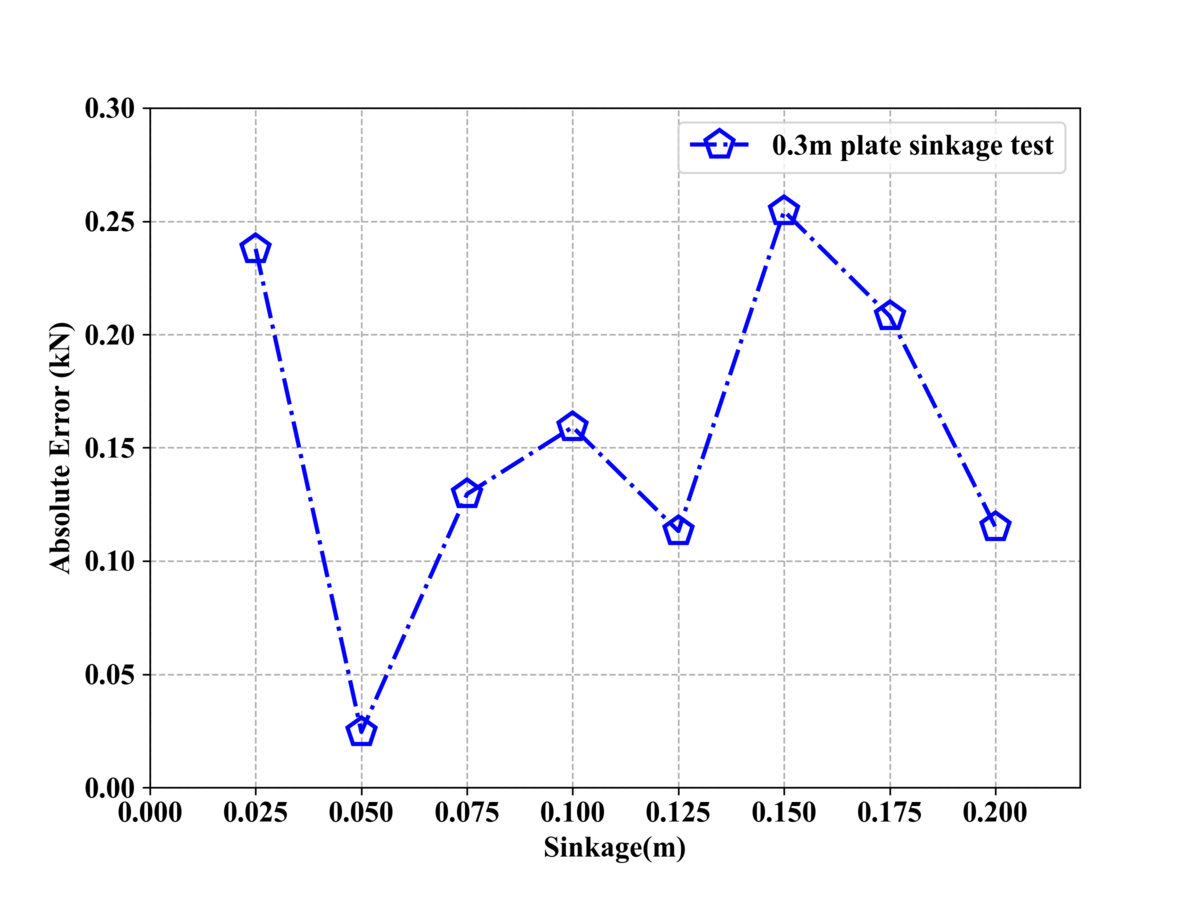}
		\caption{Large plate ($r$ = 0.3 m).} 
	\end{subfigure}
	\caption{Absolute error of the SCM results compared with the ground truth data from DEM simulations.} 
	\label{fig:bekker_error}
\end{figure}

To test the accuracy of the three SCM parameters obtained above, we evaluated the pressing force using Eq.~(\ref{equ:bekker}) with the calibrated values, and subsequently compared the results with the ground truth data in Table~\ref{tab:sinkage_F}. Figure~\ref{fig:f_sinkage_dem_scm} shows the results of the SCM vs. DEM comparison for both the 0.2 m and 0.3 m plates -- the results confirm a good agreement between the sets of SCM and DEM force vs. sinkage results. To further characterize the accuracy of the parameters, the absolute error (defined as $|SCM-DEM|$) at each reference point is shown in Fig.~\ref{fig:bekker_error}. It is noted that the relative error (defined as $|\frac{SCM-DEM}{DEM_{Max}}| \times 100\%$, where $DEM_{Max}$ is the the maximal pressing force shown in the curve) is smaller than 2\% for both plates.

\subsection{Annulus shear virtual test, and the $c$, $\varphi$, and $K_s$ calibration}
\begin{figure}[htp]
	\centering
	\includegraphics[width=6in]{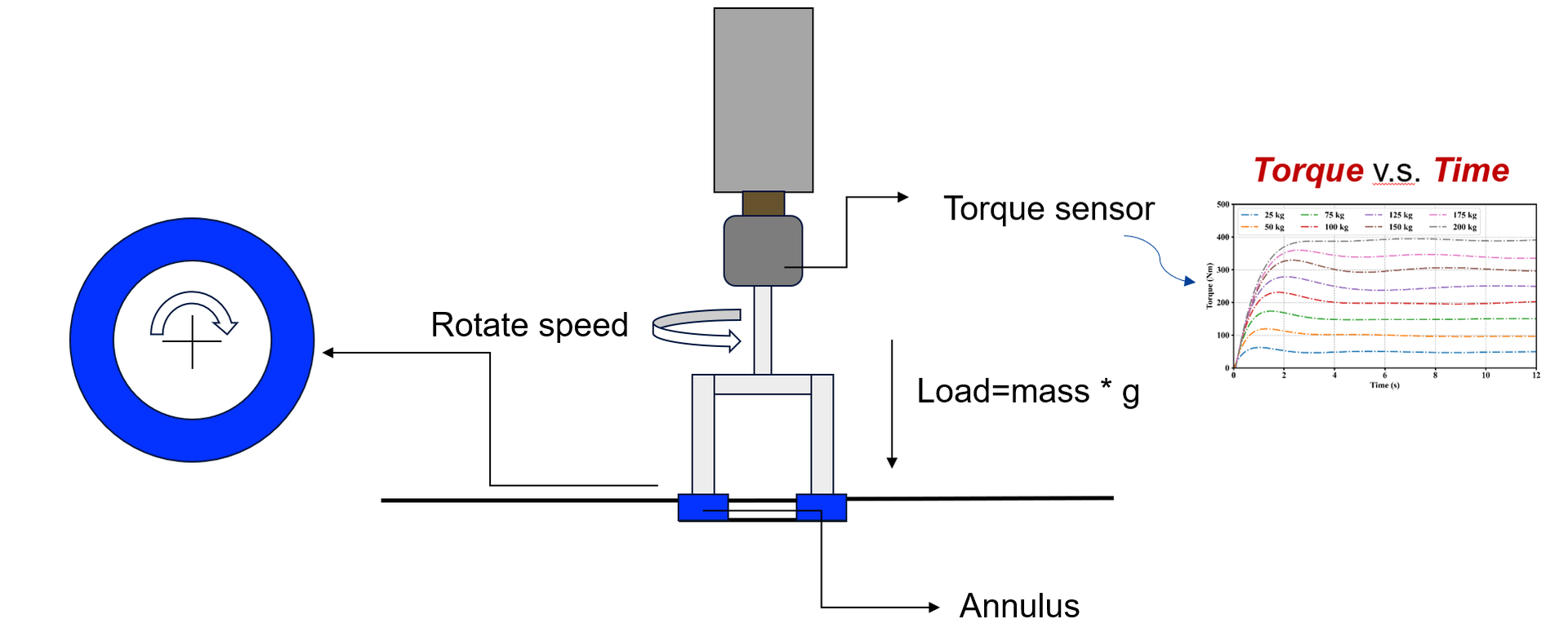}
	\caption{Schematic of the annulus shear test rig.}
	\label{fig:annulus}
\end{figure}

\begin{figure}[htp]
	\centering
	\begin{subfigure}{0.49\textwidth}
		\centering
		\includegraphics[width=3in]{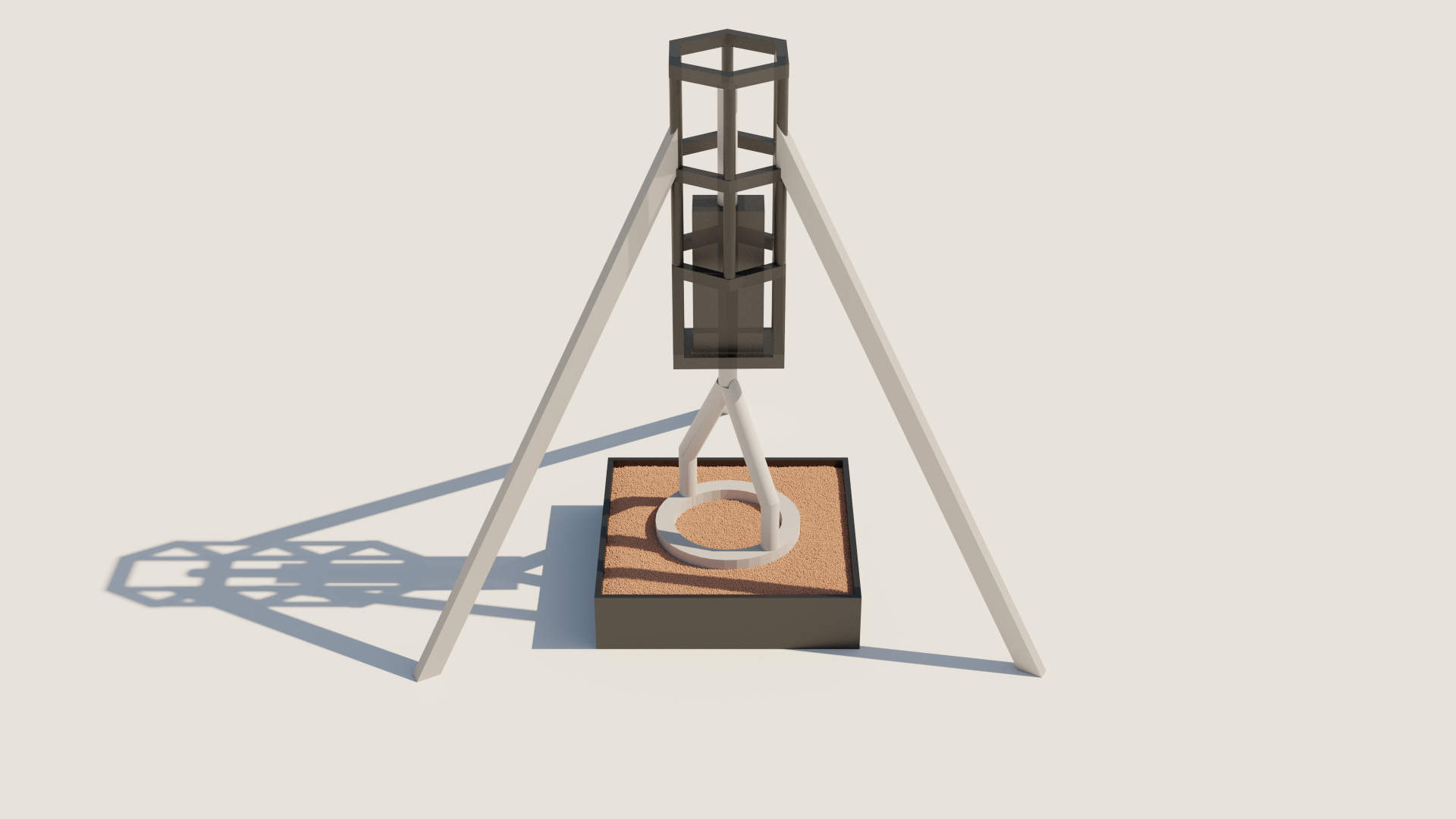}
		\caption{Overall view.}
	\end{subfigure}
	\begin{subfigure}{0.49\textwidth}
		\centering
		\includegraphics[width=3in]{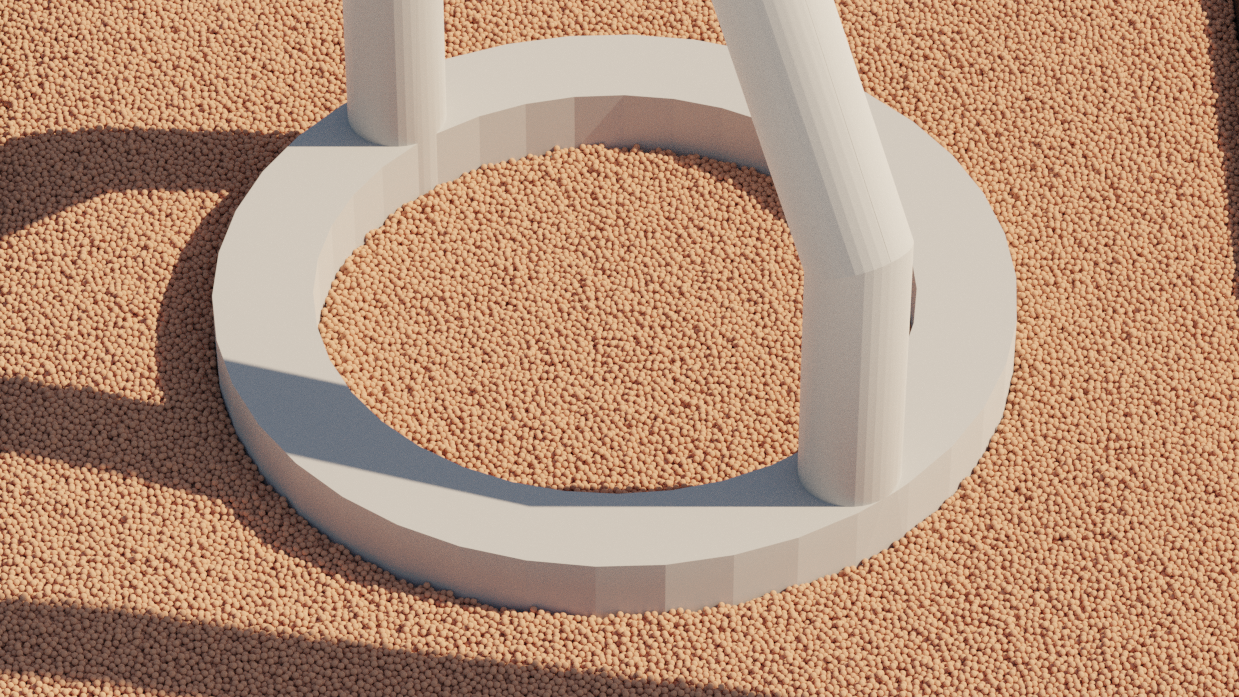}
		\caption{Zoomed-in perspective.}
	\end{subfigure}
	\caption{Annulus with 0.6 m external radius and 0.45 m inner radius. Animation of the annulus shear test simulation is provided in the supplementary materials.} 
	\label{fig:annulus_rotation}
\end{figure}

The shear virtual test is carried out to collect the data used to calibrate the cohesion coefficient $c$, friction coefficient $\varphi$, and the Janosi coefficient $K_s$ in the expression of $\tau$ and $\tau_{\text{max}}$, see Eq.~(\ref{equ:janosi}). The schematic of the annulus shear test rig is shown in Fig.~\ref{fig:annulus}. The external radius of the annulus was 0.6 m, the inner one was 0.45 m, while its thickness was 0.15 m. The annulus, which experiences a certain load, is in contact with the DEM soil while a motor turns the annulus (via an axle) at a constant angular velocity of 1 degree per second. The annulus loads are as induced by several masses that ranged from 25 kg to 200 kg (mass was kept constant for each test). We record the time history of the motor torque required to keep the annulus rotating at constant angular velocity. The duration of all simulations was set to 12 s, which was long enough to reach steady state. The DEM soil had exactly the same properties it did in the simulations for the plate sinkage test. The soil bin used had the same size as the plate sinkage test with a smaller plate: 2 m $\times$ 2 m $\times$ 0.6 m. For a mass of 100 kg, a frame of the virtual test is shown in Fig.~\ref{fig:annulus_rotation}, which provides an overall view of the test rig and a zoom-in around the DEM particles neighboring the annulus. 

\begin{figure}[htp]
	\centering
	\includegraphics[width=4in]{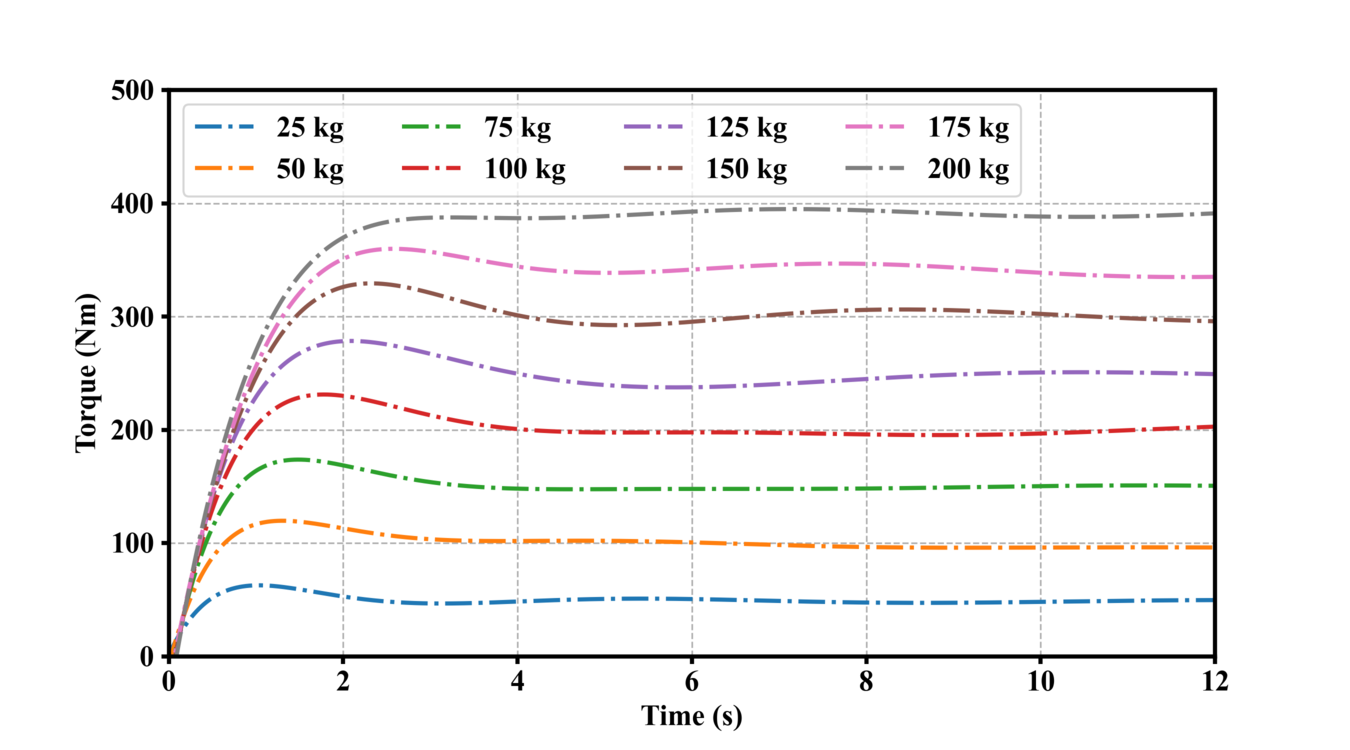}
	\caption{Time histories of the total torque applied on the annulus under different loads.}
	\label{fig:Torque_time}
\end{figure}

Figure~\ref{fig:Torque_time} shows the time histories of the total torque needed to rotate the annulus -- different masses yield different torque histories. It is noted that: ($i$) the higher the load is, the larger the torque needed to rotate the annulus; and ($ii$) at the start of the simulations, the torque rapidly increased from zero to a peak value, where it then essentially plateaued. To collect ground truth data from the high-fidelity DEM simulation of the shear test, we registered the required rotating torque at $t = $ 1 s, 2 s, and 3 s for eight different mass values.  The torque needed at steady state was the one at $t =$ 5 s. The results are shown in Tables~\ref{tab:load_T} and \ref{tab:load_T_1s2s3s}.

\begin{table}[htp]
	\caption{Rotating torque at the steady state for cases with different loads (varied from 25 kg to 200 kg).} 
	\label{tab:load_T}
	\setlength{\tabcolsep}{10pt}
	\renewcommand{\arraystretch}{1.2}
	\begin{center}
		\begin{tabular}{| c | c | c | c | c | c | c | c | c |} 
			\hline
			Load (kg) & 25 & 50 & 75 & 100 & 125 & 150 & 175 & 200 \\ 
			\hline
			Steady Torque (N$\cdot$m) & 49 & 100 & 151 & 203 & 250 & 303 & 342 & 387\\
			\hline
		\end{tabular}
	\end{center}
\end{table}

\begin{table}[htp]
	\caption{Rotating torque at t = 1s, 2s, and 3s for cases with different loads (varied from 25 kg to 200 kg).} 
	\label{tab:load_T_1s2s3s}
	\setlength{\tabcolsep}{10pt}
	\renewcommand{\arraystretch}{1.2}
	\begin{center}
		\begin{tabular}{| c | c | c | c | c | c | c | c | c |} 
			\hline
			Load (kg) & 25 & 50 & 75 & 100 & 125 & 150 & 175 & 200 \\ 
			\hline
			Torque at 1~\si{s} (N$\cdot$m)  & 62 & 117 & 164 & 203 & 229 & 246 & 256 & 270 \\
			\hline
			Torque at 2~\si{s} (N$\cdot$m) & 53 & 113 & 169 & 230 & 278 & 326 & 351 & 370 \\
			\hline
			Torque at 3~\si{s} (N$\cdot$m) & 47 & 103 & 154 & 213 & 267 & 321 & 357 & 387 \\
			\hline
		\end{tabular}
	\end{center}
\end{table}

\begin{figure}[htp]
	\centering
	\begin{subfigure}{0.32\textwidth}
		\centering
		\includegraphics[width=2.3in]{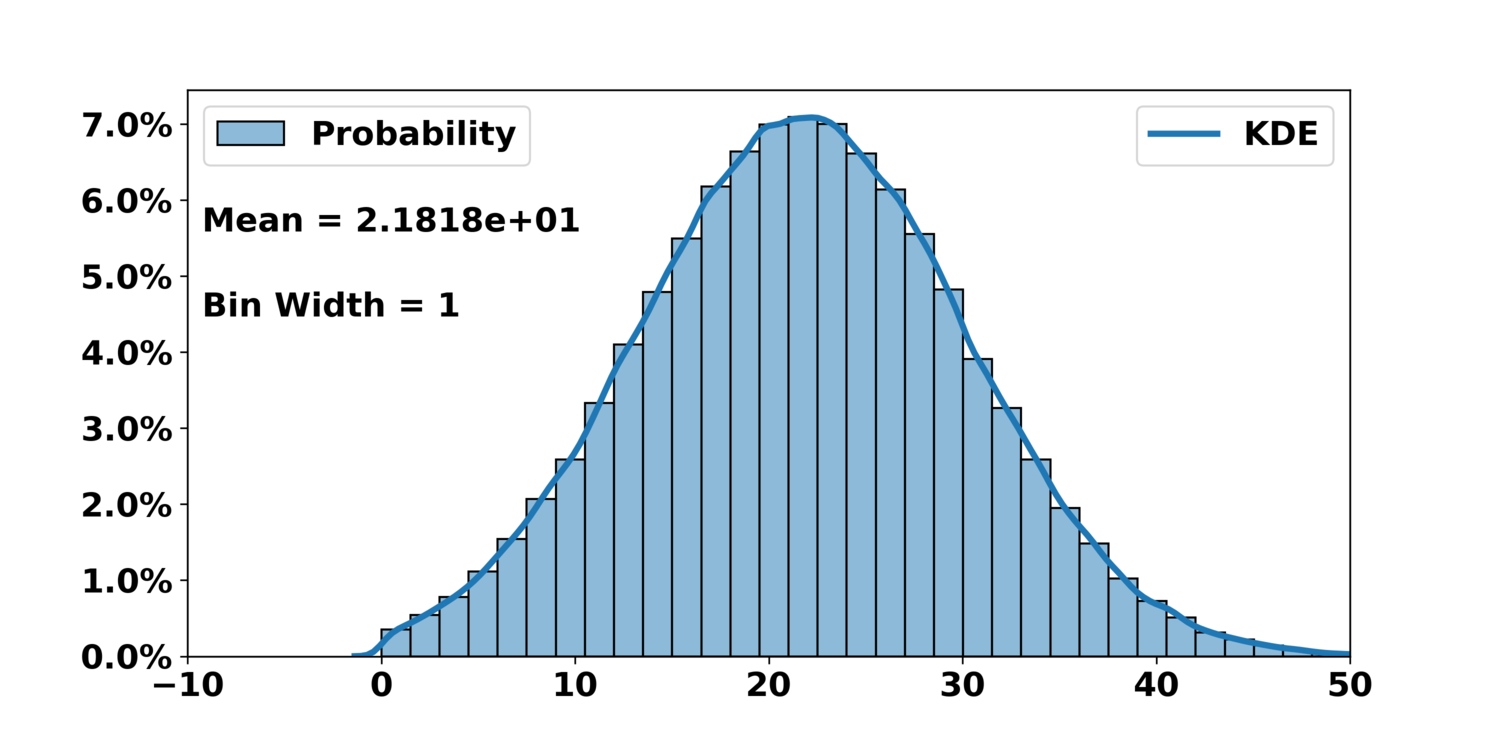}
		\caption{Mean.}
	\end{subfigure}
	\begin{subfigure}{0.32\textwidth}
		\centering
		\includegraphics[width=2.3in]{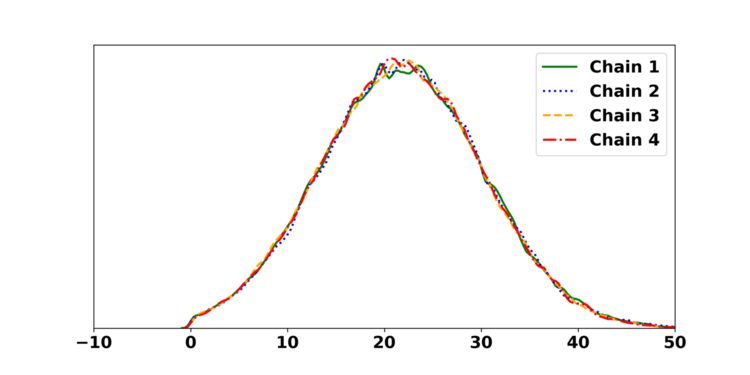}
		\caption{Four chains.} 
	\end{subfigure}
	\begin{subfigure}{0.32\textwidth}
		\centering
		\includegraphics[width=2.3in]{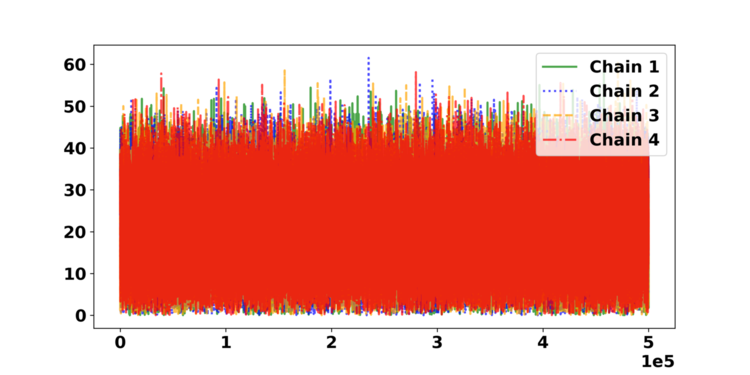}
		\caption{Trace.}
	\end{subfigure}
	\caption{Averaged posterior probability distribution and its kernel density estimation (KDE) of four chains for the parameter $c$ with different number of samples.} 
	\label{fig:trace_posterior_cohesion}
\end{figure}

\begin{figure}[htp]
	\centering
	\begin{subfigure}{0.32\textwidth}
		\centering
		\includegraphics[width=2.3in]{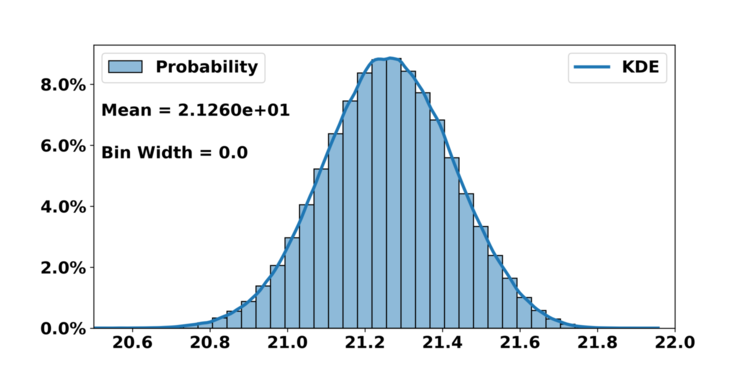}
		\caption{Mean.}
	\end{subfigure}
	\begin{subfigure}{0.32\textwidth}
		\centering
		\includegraphics[width=2.3in]{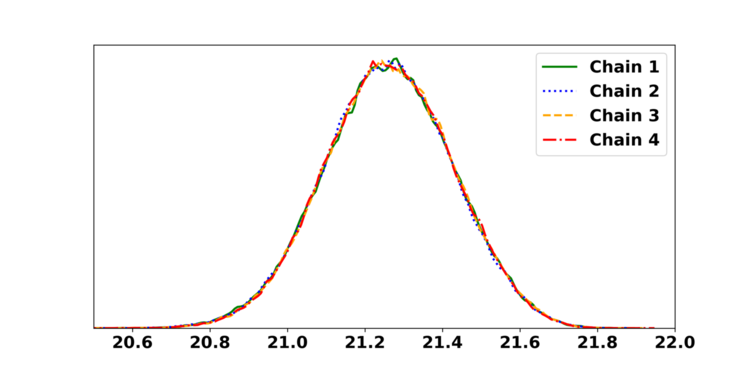}
		\caption{Four chains.} 
	\end{subfigure}
	\begin{subfigure}{0.32\textwidth}
		\centering
		\includegraphics[width=2.3in]{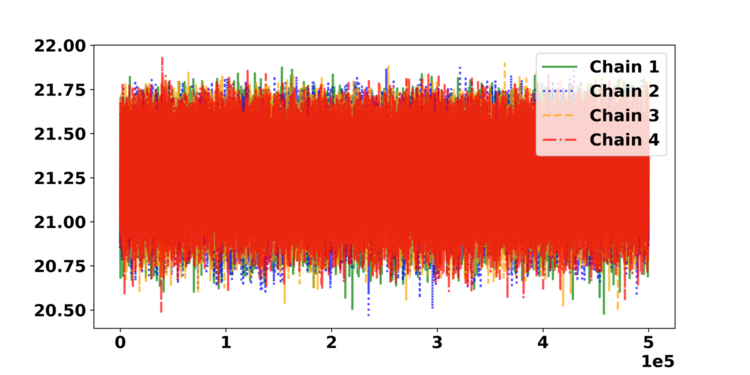}
		\caption{Trace.}
	\end{subfigure}
	\caption{Averaged posterior probability distribution and its kernel density estimation (KDE) of four chains for the parameter ${\varphi}$ with different number of samples.} 
	\label{fig:trace_posterior_phi}
\end{figure}

\begin{figure}[htp]
	\centering
	\begin{subfigure}{0.32\textwidth}
		\centering
		\includegraphics[width=2.3in]{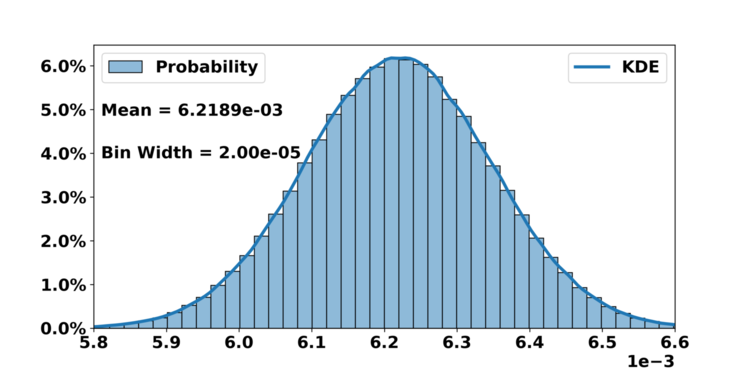}
		\caption{Mean.}
	\end{subfigure}
	\begin{subfigure}{0.32\textwidth}
		\centering
		\includegraphics[width=2.3in]{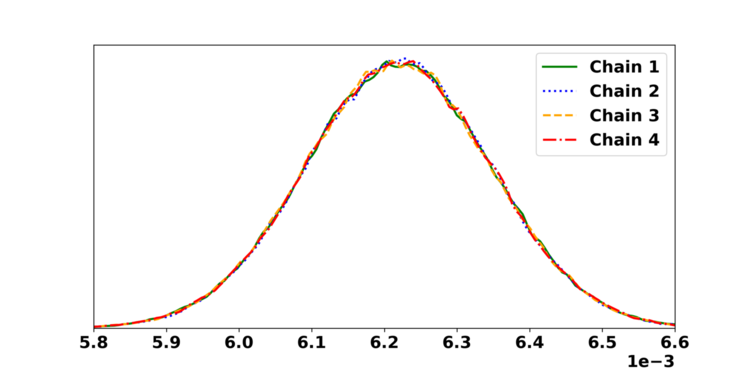}
		\caption{Four chains.} 
	\end{subfigure}
	\begin{subfigure}{0.32\textwidth}
		\centering
		\includegraphics[width=2.3in]{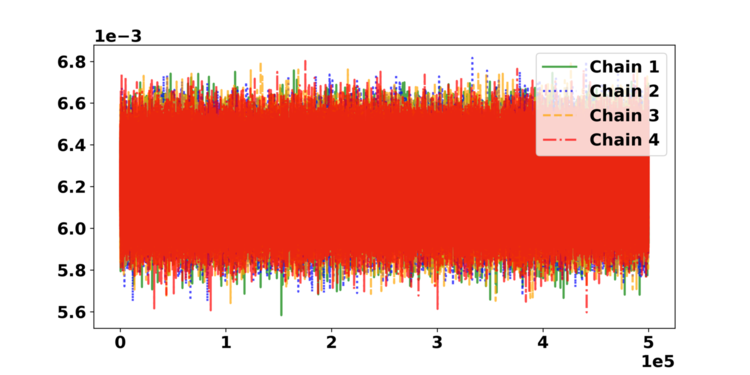}
		\caption{Trace.}
	\end{subfigure}
	\caption{Averaged posterior probability distribution and its kernel density estimation (KDE) of four chains for the parameter $K_s$ with different number of samples.} 
	\label{fig:trace_posterior_Ks}
\end{figure}

\begin{figure}[htp]
	\centering
	\begin{subfigure}{0.49\textwidth}
		\centering
		\includegraphics[width=3in]{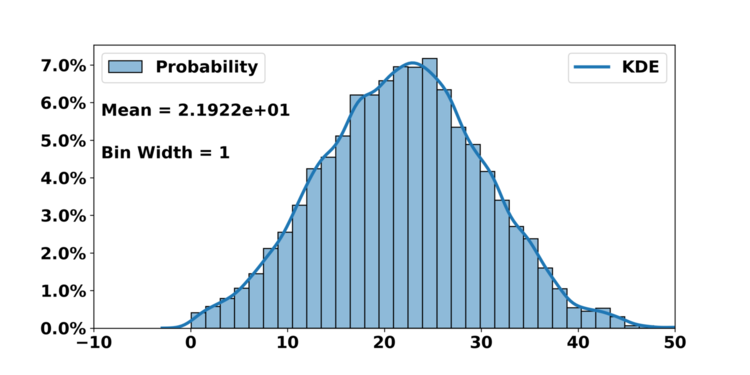}
		\caption{\num{10000} samples.}
	\end{subfigure}
	\begin{subfigure}{0.49\textwidth}
		\centering
		\includegraphics[width=3in]{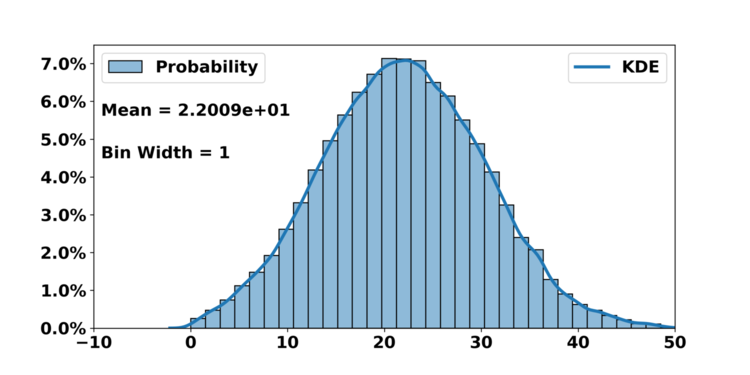}
		\caption{\num{50000} samples.} 
	\end{subfigure}
	\begin{subfigure}{0.49\textwidth}
		\centering
		\includegraphics[width=3in]{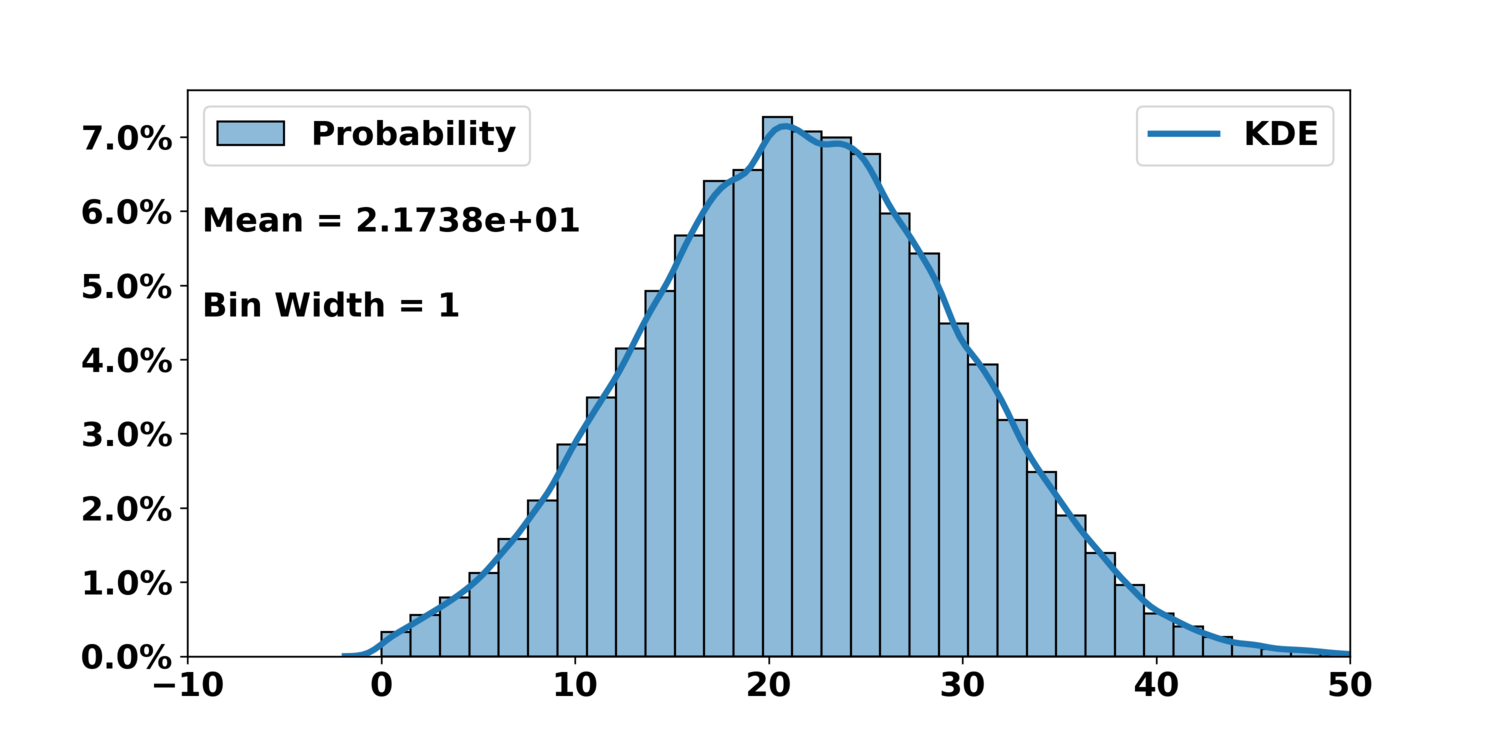}
		\caption{\num{100000} samples.}
	\end{subfigure}
	\begin{subfigure}{0.49\textwidth}
		\centering
		\includegraphics[width=3in]{Post_Mean_500000_samples_cohesion.png}
		\caption{\num{500000} samples.} 
	\end{subfigure}
	\caption{Averaged posterior probability distribution and its kernel density estimation (KDE) for the parameter $c$ across four chains with different sample sizes.} 
	\label{fig:posterior_c}
\end{figure}

\begin{figure}[htp]
	\centering
	\begin{subfigure}[t]{0.49\textwidth}
		\centering
		\includegraphics[width=1.5in]{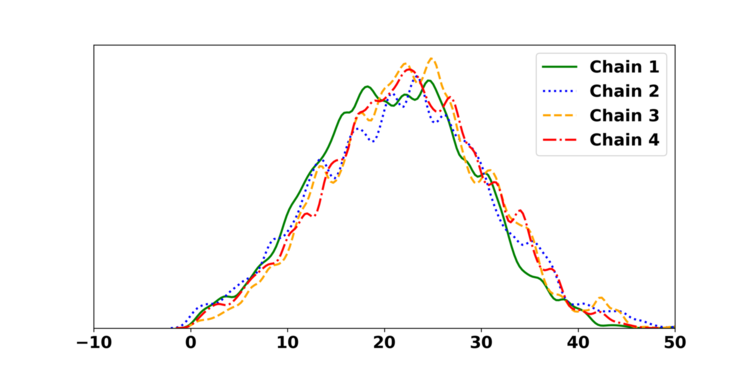}
		\includegraphics[width=1.5in]{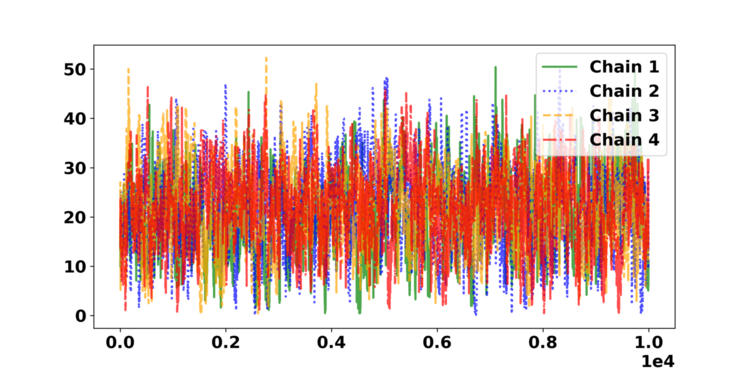}
		\caption{\num{10000} samples.}
	\end{subfigure}
	\begin{subfigure}[t]{0.49\textwidth}
		\centering
		\includegraphics[width=1.5in]{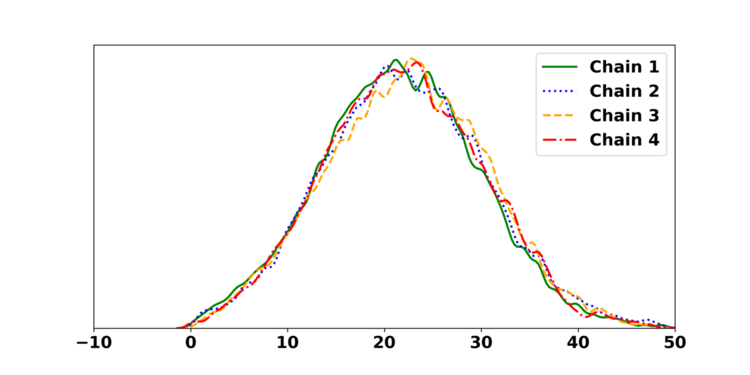}
		\includegraphics[width=1.5in]{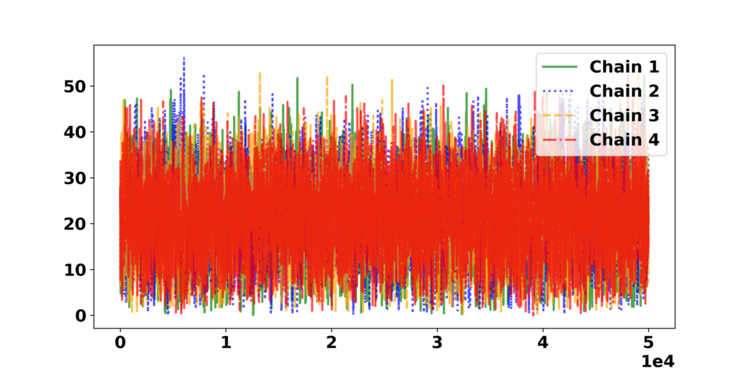}
		\caption{\num{50000} samples.}
	\end{subfigure}
	\begin{subfigure}[t]{0.49\textwidth}
		\centering
		\includegraphics[width=1.5in]{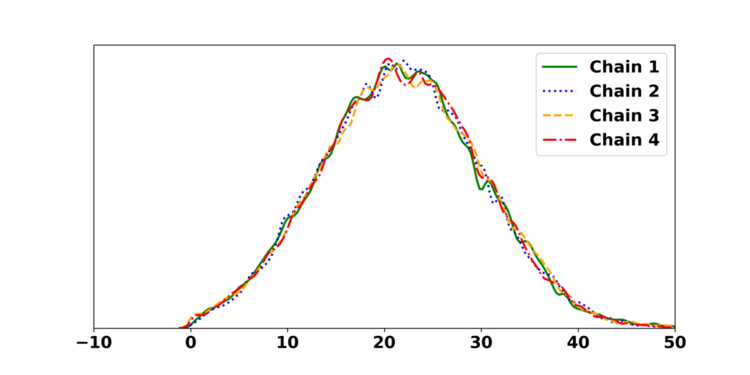}
		\includegraphics[width=1.5in]{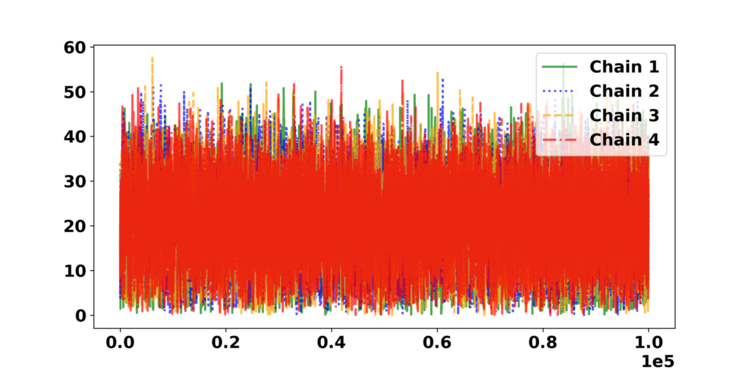}
		\caption{\num{100000} samples.}
	\end{subfigure}
	\begin{subfigure}[t]{0.49\textwidth}
		\centering
		\includegraphics[width=1.5in]{Post_500000_samples_cohesion.png}
		\includegraphics[width=1.5in]{Trace_500000_samples_cohesion.png}
		\caption{\num{500000} samples.}
	\end{subfigure}
	\caption{Posterior probability distribution and trace plot for parameter $c$ in four different Metropolis-Hastings chains.}
	\label{fig:trace_posterior_c_4chain}
\end{figure}

\begin{figure}[htp]
	\centering
	\begin{subfigure}{0.49\textwidth}
		\centering
		\includegraphics[width=3in]{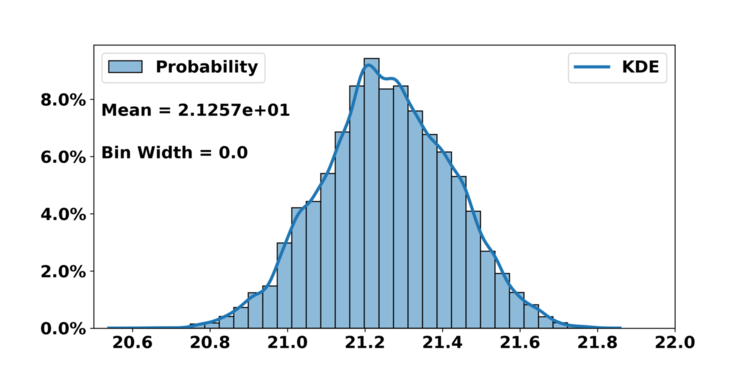}
		\caption{\num{10000} samples.}
	\end{subfigure}
	\begin{subfigure}{0.49\textwidth}
		\centering
		\includegraphics[width=3in]{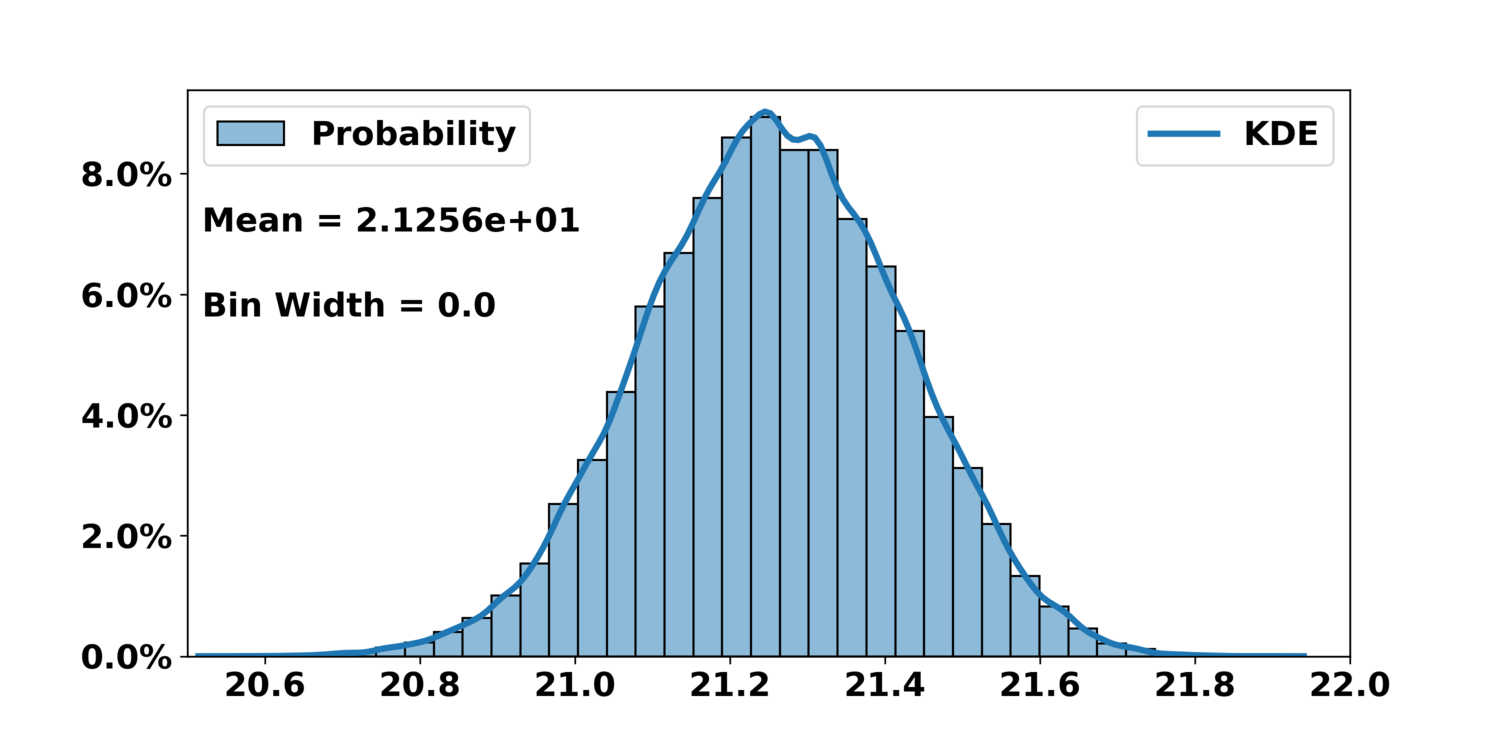}
		\caption{\num{50000} samples.} 
	\end{subfigure}
	\begin{subfigure}{0.49\textwidth}
		\centering
		\includegraphics[width=3in]{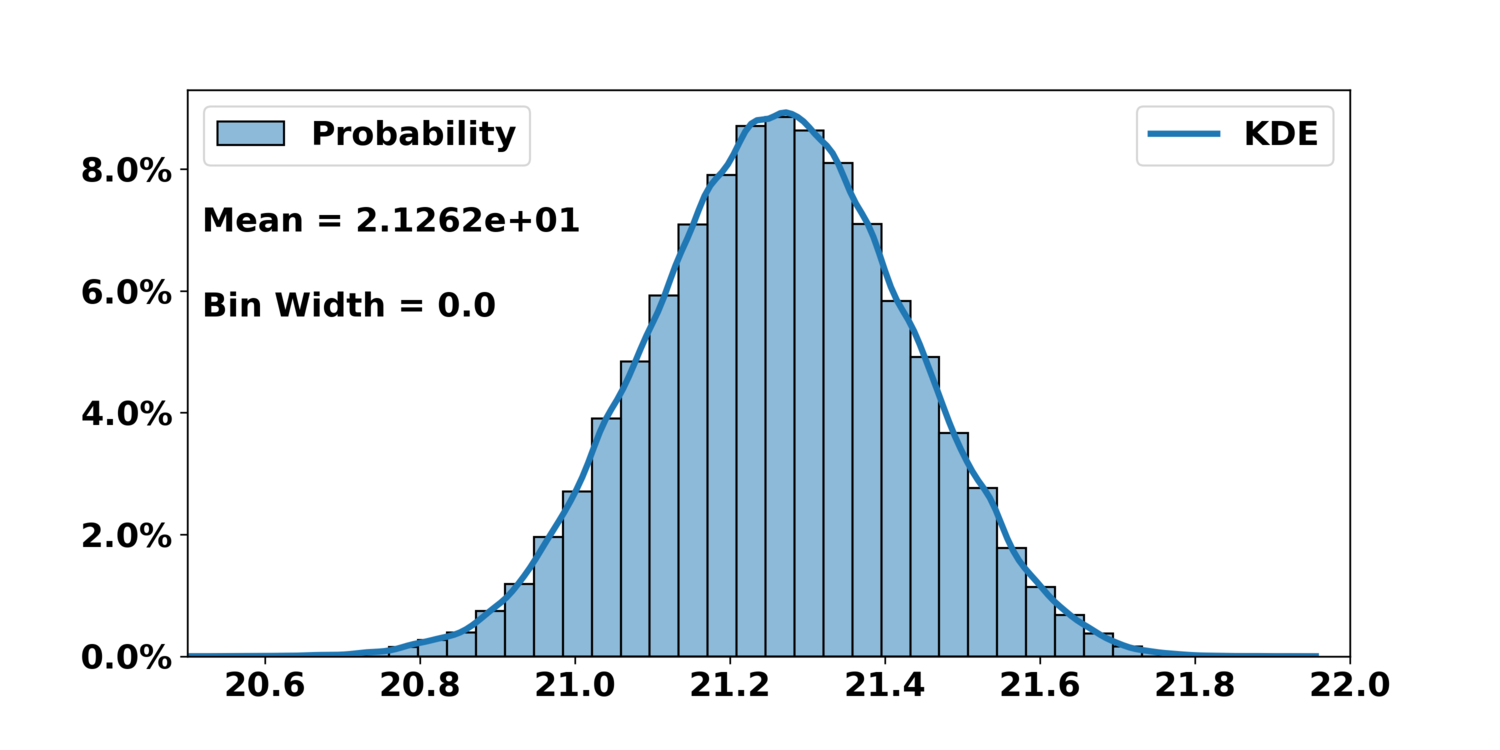}
		\caption{\num{100000} samples.}
	\end{subfigure}
	\begin{subfigure}{0.49\textwidth}
		\centering
		\includegraphics[width=3in]{Post_Mean_500000_samples_phi.png}
		\caption{\num{500000} samples.} 
	\end{subfigure}
	\caption{Averaged posterior probability distribution and its kernel density estimation (KDE) for the parameter $\varphi$ across four chains with different sample sizes.} 
	\label{fig:posterior_phi}
\end{figure}

The $c$, $\varphi$, and $K_s$ parameters were calibrated in a two-step algorithm. First, $c$ and $\varphi$ were calibrated using the ground truth data in Table~\ref{tab:load_T}, which is the steady state value of the torque for each load. Subsequently, $K_s$ was calibrated using the ground truth data in Table~\ref{tab:load_T_1s2s3s}, which is the torque recorded at 1 s, 2 s, and 3 s under different loads.  The Bayesian likelihood function for the calibration was defined as:
\begin{equation}
\mathbb{L} = -\frac{1}{2\sigma^2}\sum_{k=1}^{N} \frac{\left [  T(k) - T^{truth}(k) \right ]^2}{N} \; , 
\end{equation}
where $T^{truth}(k)$ is the ground truth  torque recorded in Tables~\ref{tab:load_T} and \ref{tab:load_T_1s2s3s}; $T(k)$ is the torque evaluated using Eqs.~(\ref{equ:janosi}); $N$ is the total number of  the ground truth data, which is eight in the calibration of $c$ and $\varphi$, and 24 in the calibration of $K_s$; and $\sigma^2$ was set to 0.01. Again, we ran the sampling loop for \num{500000} iterations. Four parallel inference chains were run to assess convergence, see Figs.~\ref{fig:trace_posterior_cohesion}(b)(c), \ref{fig:trace_posterior_phi}(b)(c), and \ref{fig:trace_posterior_Ks}(b)(c). The sampling distribution in all four chains converged to the same result. Figures~\ref{fig:trace_posterior_cohesion}(a), \ref{fig:trace_posterior_phi}(a), and \ref{fig:trace_posterior_Ks}(a) show the averaged posterior distribution of the three parameters, whose mean values are 21.872, 21.259, and 0.0062, respectively. The Bayesian inference was run for several iteration counts, varied from \num{10000} to \num{500000}. The results are shown in Figs. \ref{fig:posterior_c}, \ref{fig:trace_posterior_c_4chain}, \ref{fig:posterior_phi}, \ref{fig:trace_posterior_phi_4chain}, \ref{fig:posterior_ks} and \ref{fig:trace_posterior_ks_4chain} of the Appendix. This convergence analysis justified the choice of \num{500000} iterations adopted herein.

\begin{figure}[h]
	\centering
	\begin{subfigure}{0.32\textwidth}
		\centering
		\includegraphics[width=2.3in]{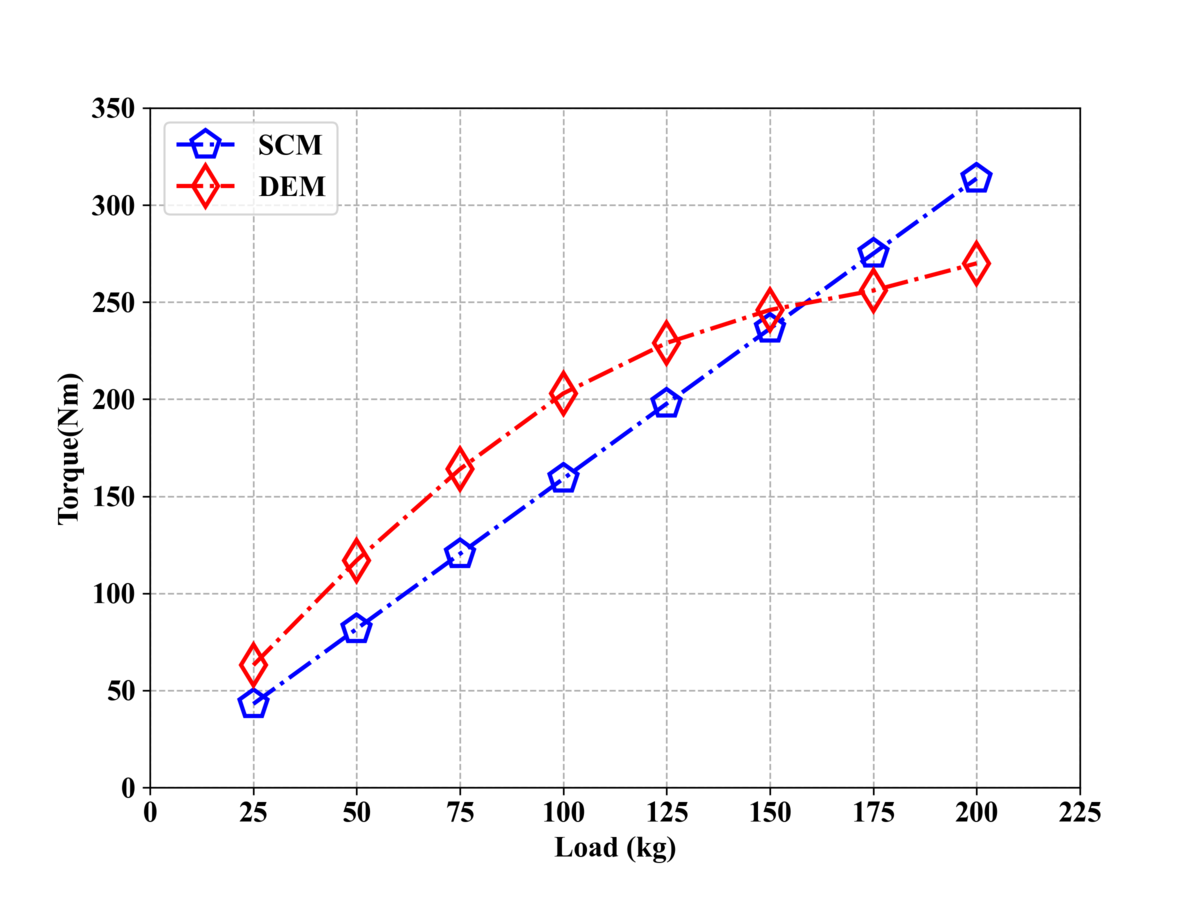}
		\caption{Time = 1s.}
	\end{subfigure}
	\begin{subfigure}{0.32\textwidth}
		\centering
		\includegraphics[width=2.3in]{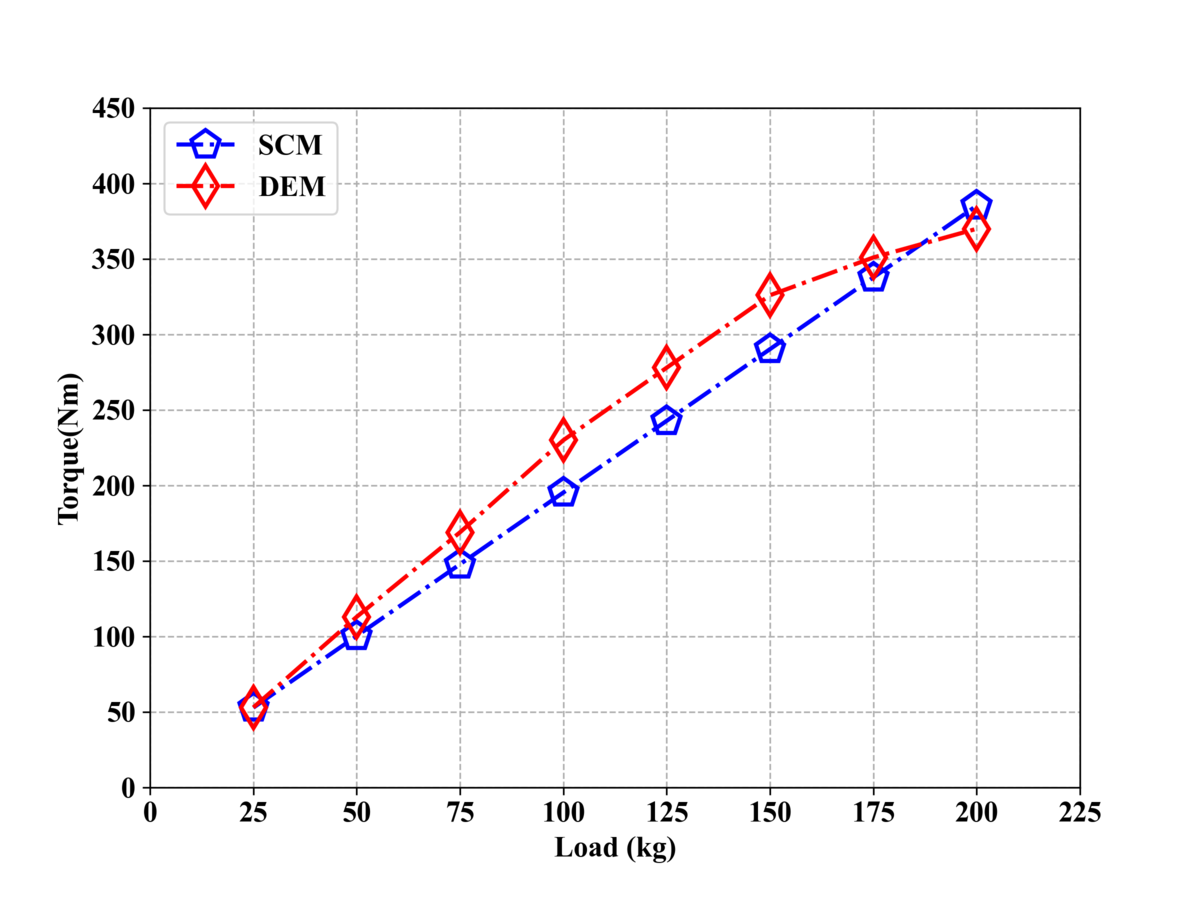}
		\caption{Time = 2s.} 
	\end{subfigure}
	\begin{subfigure}{0.32\textwidth}
		\centering
		\includegraphics[width=2.3in]{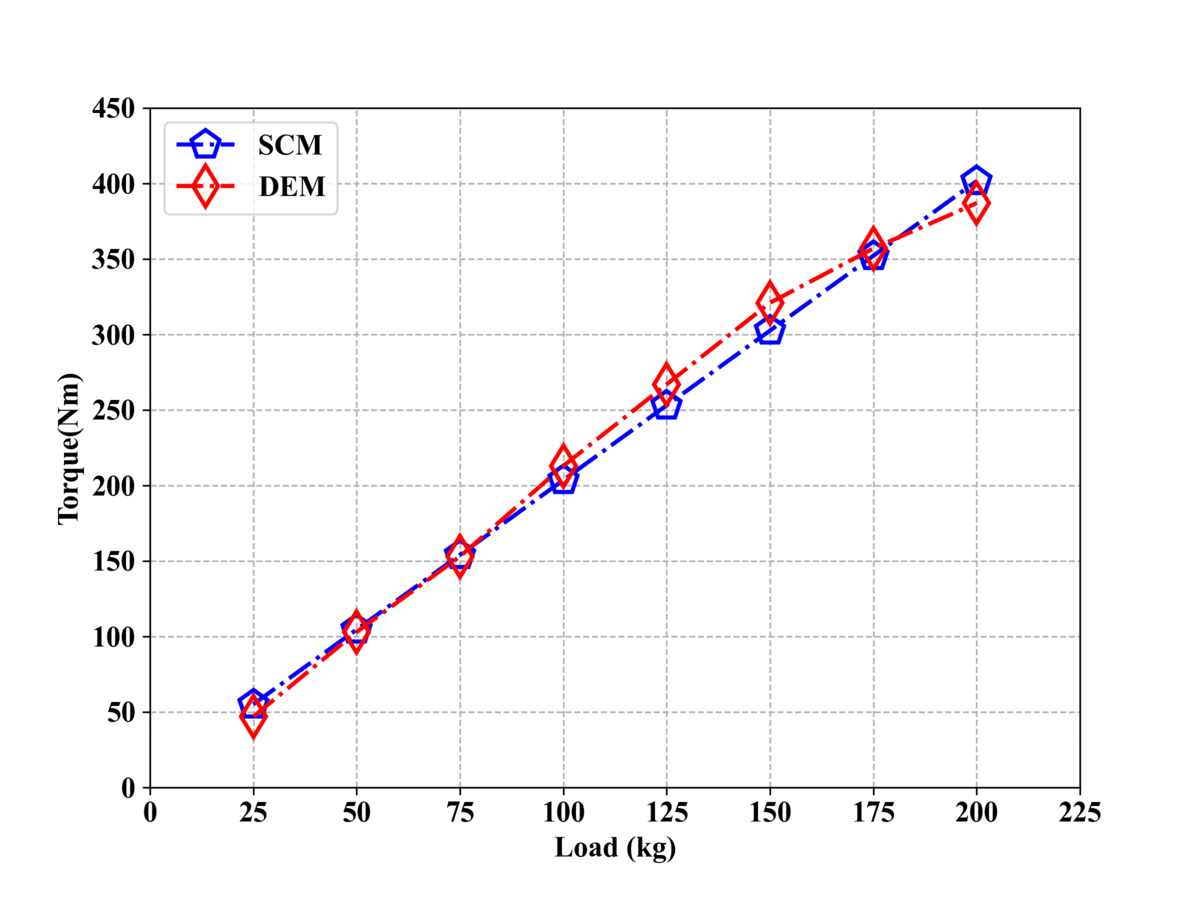}
		\caption{Time = 3s.}
	\end{subfigure}
	\caption{DEM vs. SCM comparison of results at 1 s, 2 s, and 3 s.} 
	\label{fig:shear_123}
\end{figure}

\begin{figure}[h]
	\centering
	\begin{subfigure}{0.32\textwidth}
		\centering
		\includegraphics[width=2.3in]{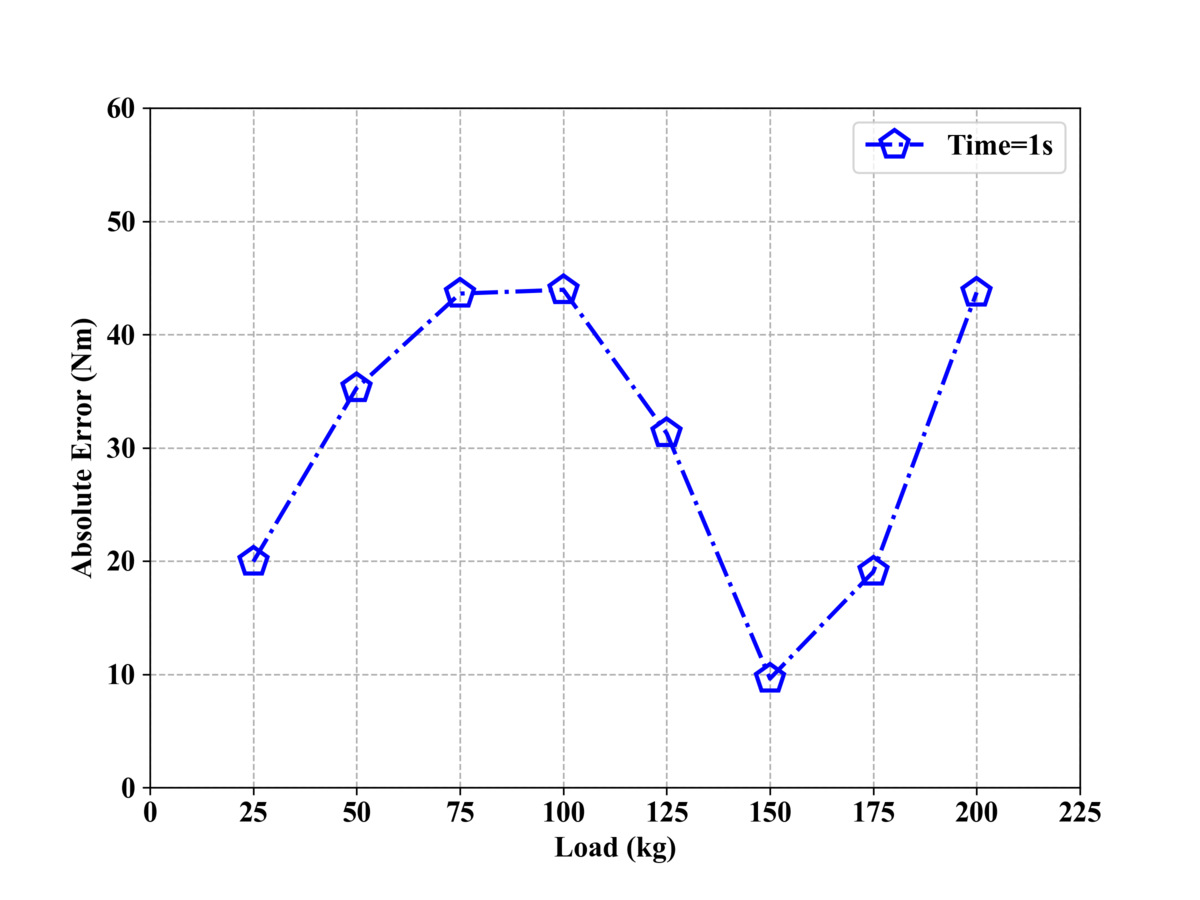}
		\caption{Time = 1s.}
	\end{subfigure}
	\begin{subfigure}{0.32\textwidth}
		\centering
		\includegraphics[width=2.3in]{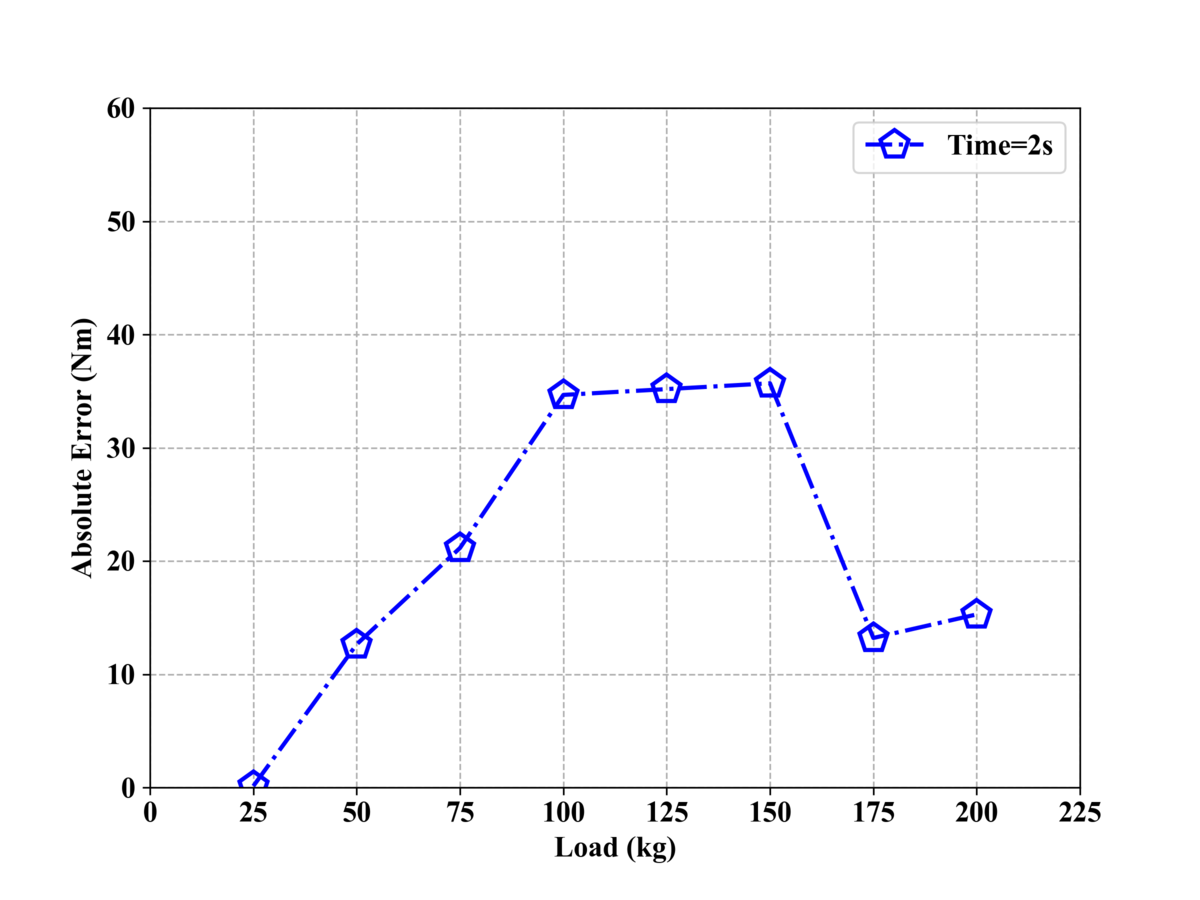}
		\caption{Time = 2s.} 
	\end{subfigure}
	\begin{subfigure}{0.32\textwidth}
		\centering
		\includegraphics[width=2.3in]{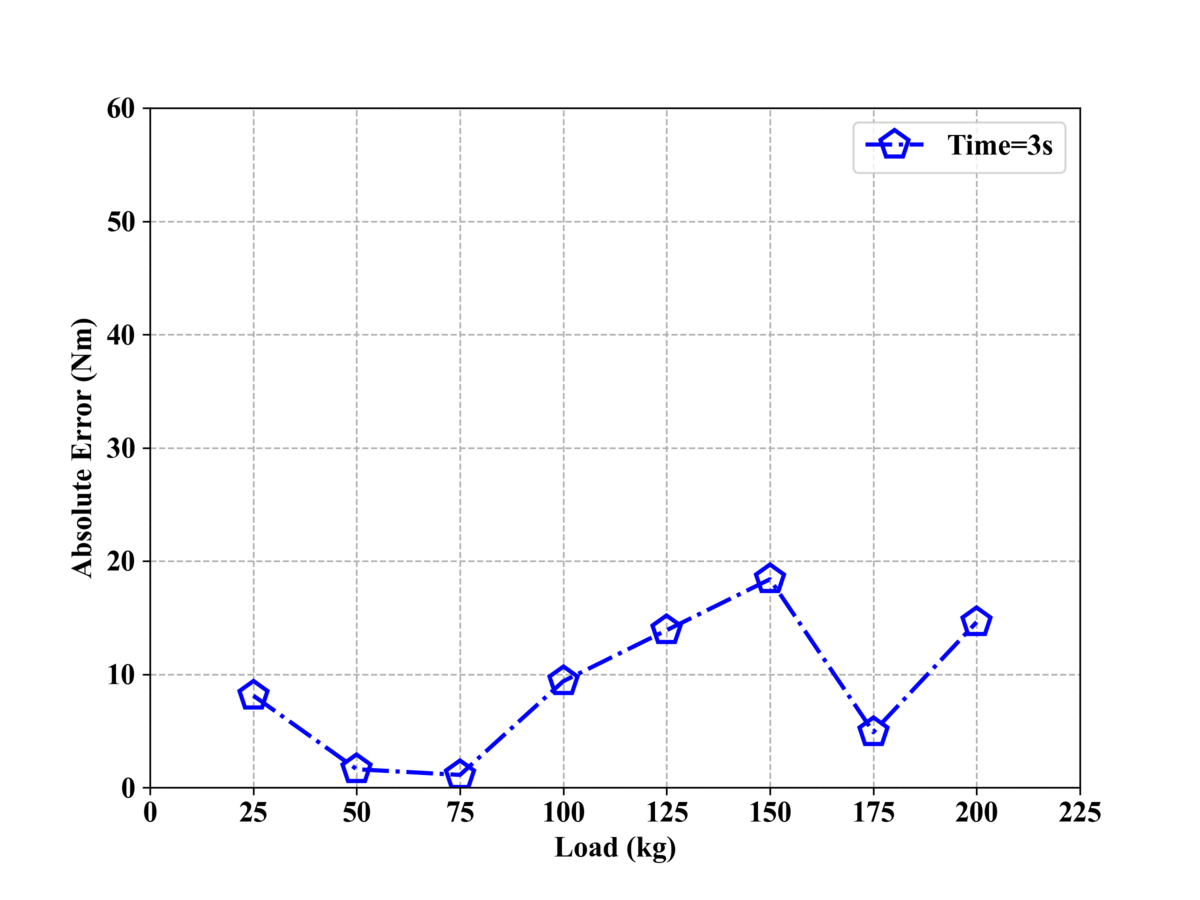}
		\caption{Time = 3s.}
	\end{subfigure}
	\caption{Absolute error of the SCM results compared with the DEM solution at 1 s, 2 s, and 3 s.} 
	\label{fig:shear_error_123}
\end{figure}

\begin{figure}[htp]
	\centering
	\includegraphics[width=3.5in]{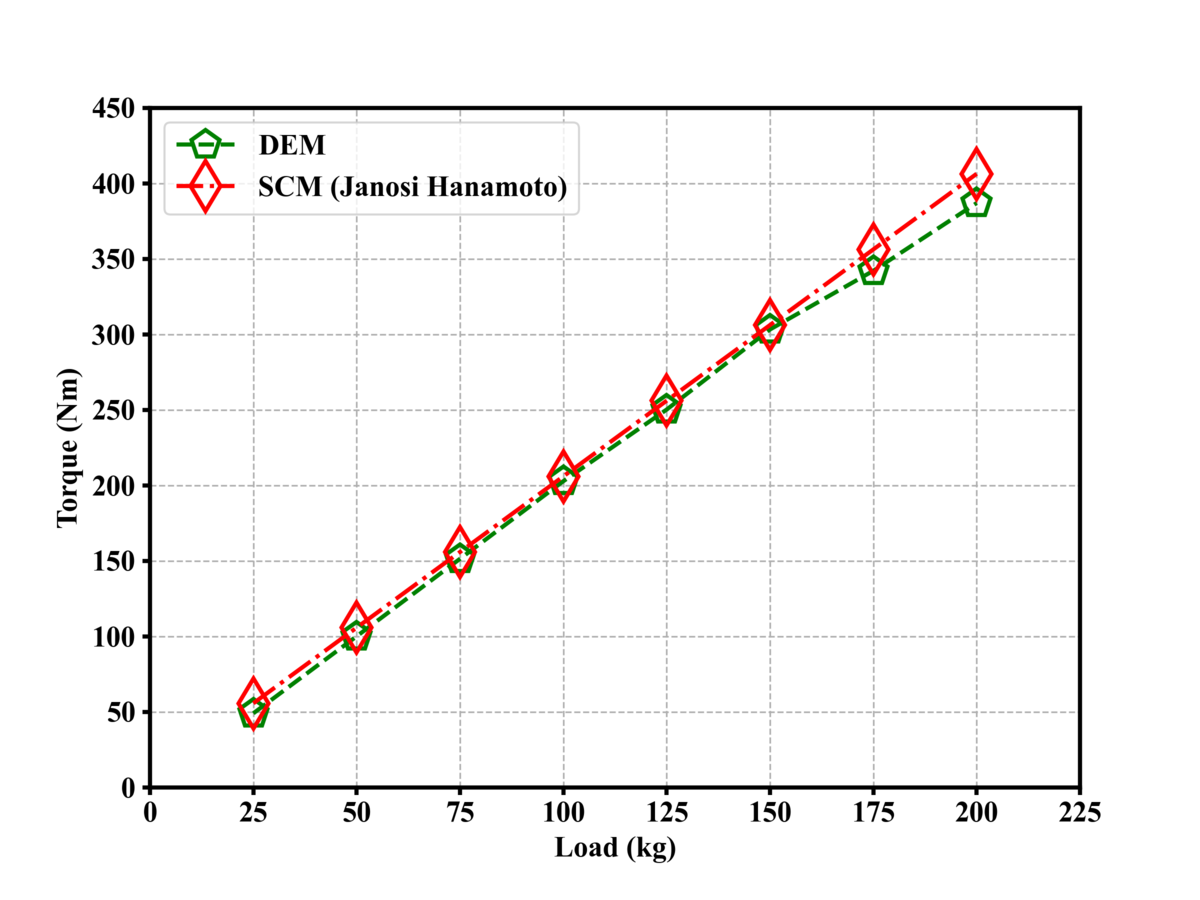}
	\caption{DEM vs. SCM comparison for selected points at the steady state. }
	\label{fig:shear_T}
\end{figure}

\begin{figure}[htp]
	\centering
	\includegraphics[width=3.5in]{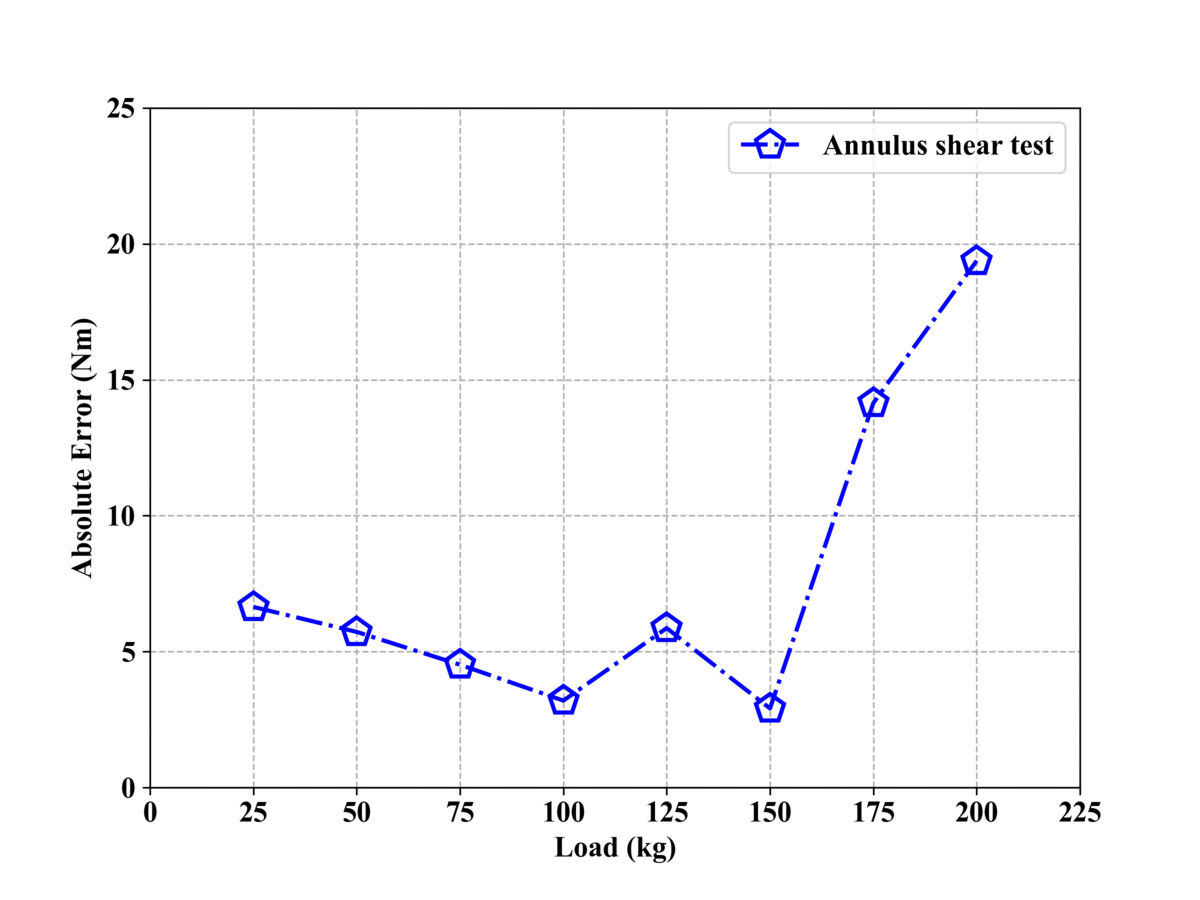}
	\caption{DEM vs. SCM absolute error comparison at the steady state.}
	\label{fig:shear_T_error}
\end{figure}

To assess the accuracy of the SCM model when using calibrated parameters, we evaluated the torque using Eq.~(\ref{equ:janosi}) with the calibrated $c$ and $\varphi$, and compared the results with the ground truth data in Table~\ref{tab:load_T}. We then evaluated the torque using $\tau_{\text{max}}$ of Eq.~(\ref{equ:janosi}) with the calibrated $K_s$ and compared with the ground truth data in Table~\ref{tab:load_T_1s2s3s}. For various masses, Fig.~\ref{fig:shear_123} compares the SCM torque values with the DEM ground truth data at 1 s, 2 s, and 3 s. To further demonstrate the accuracy of the parameters, we show the absolute error (defined as for the plate test) in Fig.~\ref{fig:shear_error_123}. The error is relatively larger at t = 1 s and smaller at t = 3 s. Figure~\ref{fig:shear_T} compares SCM and DEM results in terms of steady state torque. The absolute error at steady state is shown in Fig.~\ref{fig:shear_T_error}; the relative error is less than 5\%.

\begin{figure}[htp]
	\centering
	\includegraphics[width=5in]{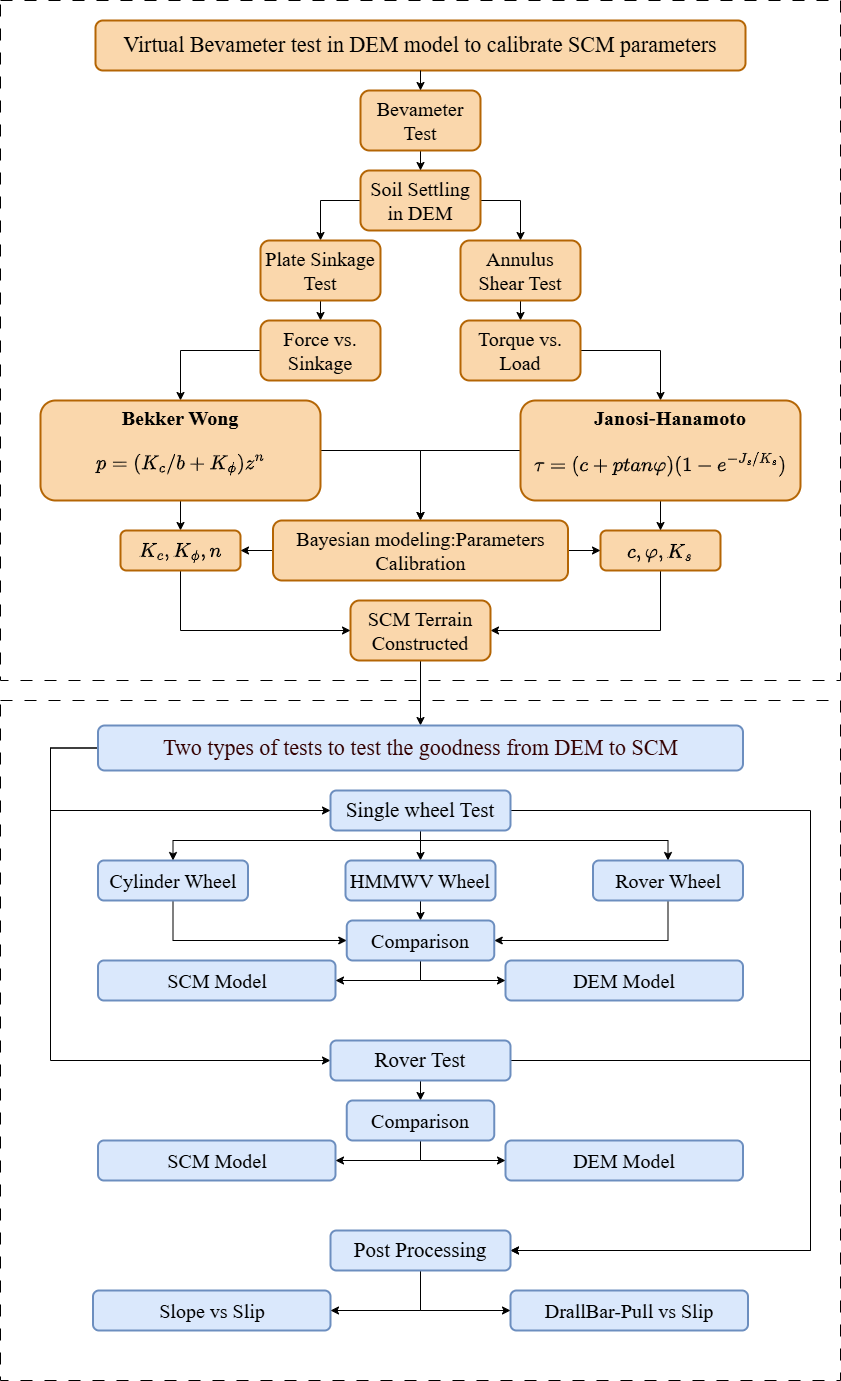}
	\caption{Flowchart of the proposed approach draws on the virtual bevameter test rig modeled in a multibody system framework and the DEM approach. The Bayesian inference framework was employed to calibrated the SCM parameters. Both single wheel and full rover simulations were performed to test the accuracy of the parameters.}
	\label{fig:flowchart}
\end{figure}

The final SCM parameter values are provided in Table~\ref{tab:scm_para}. The workflow for the calibration approach is summarized in Fig.~\ref{fig:flowchart}. The calibration process has three stages: ($i$) ground truth data was generated using the virtual bevameter test rig, which was modeled using DEM in a multibody dynamics framework; ($ii$) the SCM parameters were calibrated using Bayesian inference, which worked off the data collected in the virtual bevameter test; and ($iii$) the accuracy of the calibrated SCM parameters was verified using both the Bekker-Wong and Janosi-Hanamoto formulas.

\begin{table}[htp]
	\caption{Calibrated values of the SCM parameters using data generated from DEM simulations.} 
	\label{tab:scm_para}
	\setlength{\tabcolsep}{10pt}
	\renewcommand{\arraystretch}{1.2}
	\begin{center}
		\begin{tabular}{| c | c | c | c | c | c | c |} 
			\hline
			SCM parameter & $K_c$~(\si{N/m^{n+1}}) & $K_{\phi}$~(\si{N/m^{n+2}}) & $n$ & $c$~(\si{Pa}) & $\varphi$~(\si{deg}) & $K_s$~(\si{m}) \\ 
			\hline
			Calibrated value & -4957 & 235605 & 0.883 & 21.872 & 21.259 & 0.0062 \\
			\hline
		\end{tabular}
	\end{center}
\end{table}

\FloatBarrier
\section{Validation of the calibrated SCM model}\label{sec:validation}
In a real-life setting, once an SCM terramechanics model is calibrated through a physical bevameter test, it can be used for simulations on terrains similar to those that provided the ground truth data. In the same vein, this section demonstrates that parameters calibrated through a \textit{virtual} bevameter test enable SCM terramechanics to accurately capture the response of a vehicle operating on the terrain used to generate the ground truth calibration data. Specifically, we show that once calibrated, SCM terramechanics produces results that are qualitatively identical to the DEM terramechanics results for single wheel and full vehicle tests. The experiments considered are the draw-bar pull (DBP) and slip-slope tests. For these two experiments, the SCM and DEM simulations used the same multibody systems, in the same framework; everything was identical, except that the terrain was handled by SCM and then DEM. 

\subsection{Single wheel virtual experiments}\label{subsec:singlewheel}
\begin{figure}[htp]
	\centering
	\includegraphics[width=6in]{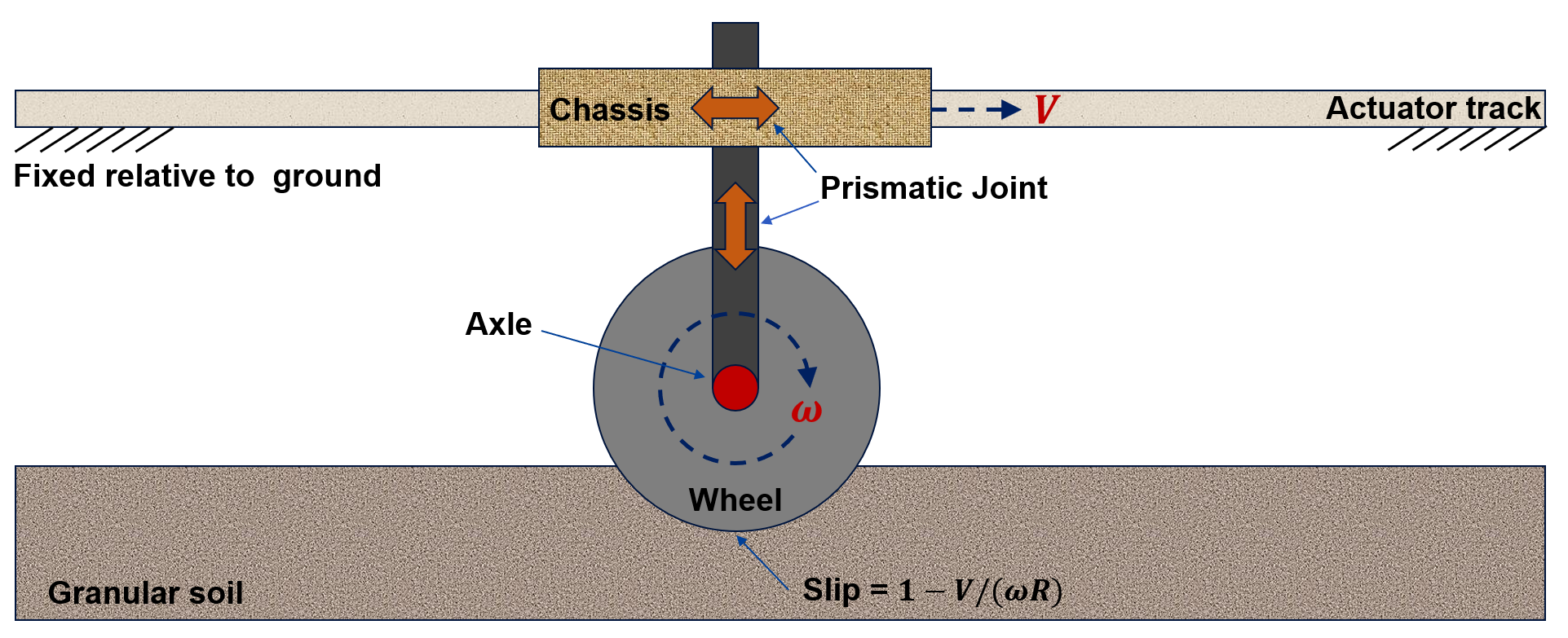}
	\caption{Schematic of the single wheel test rig; both angular and translational velocities can be controlled in the test rig.}
	\label{fig:single_wheel_setup}
\end{figure}

\begin{figure}[htp]
	\centering
	\begin{subfigure}{0.32\textwidth}
		\centering
		\includegraphics[width=1.6in]{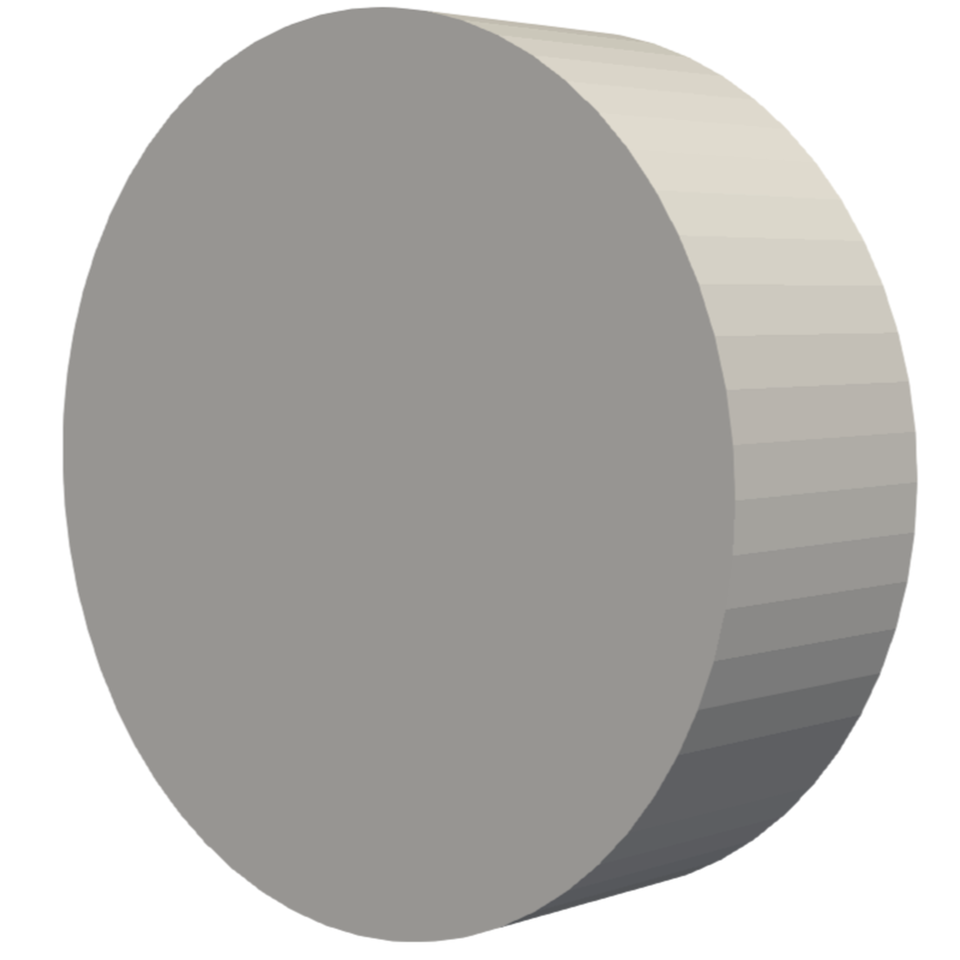}
		\caption{Cylindrical wheel.}
	\end{subfigure}
	\begin{subfigure}{0.32\textwidth}
		\centering
		\includegraphics[width=1.6in]{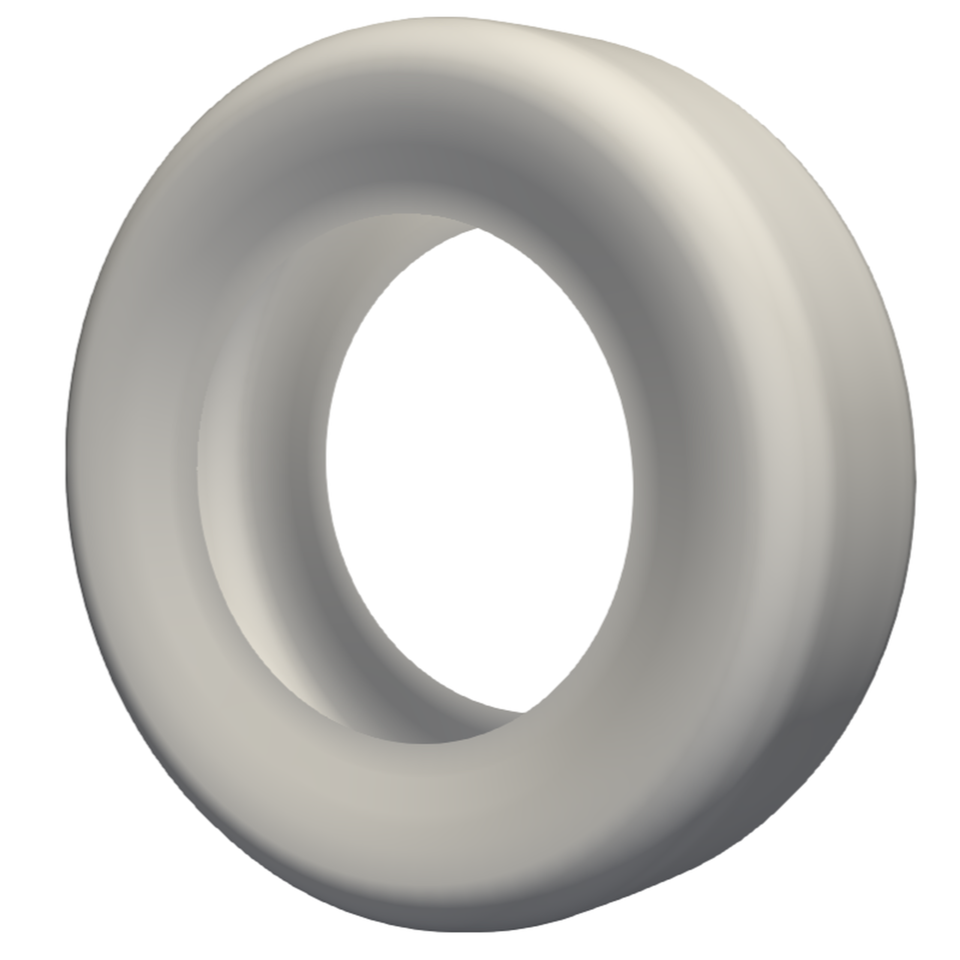}
		\caption{HMMWV wheel.} 
	\end{subfigure}
	\begin{subfigure}{0.32\textwidth}
		\centering
		\includegraphics[width=1.6in]{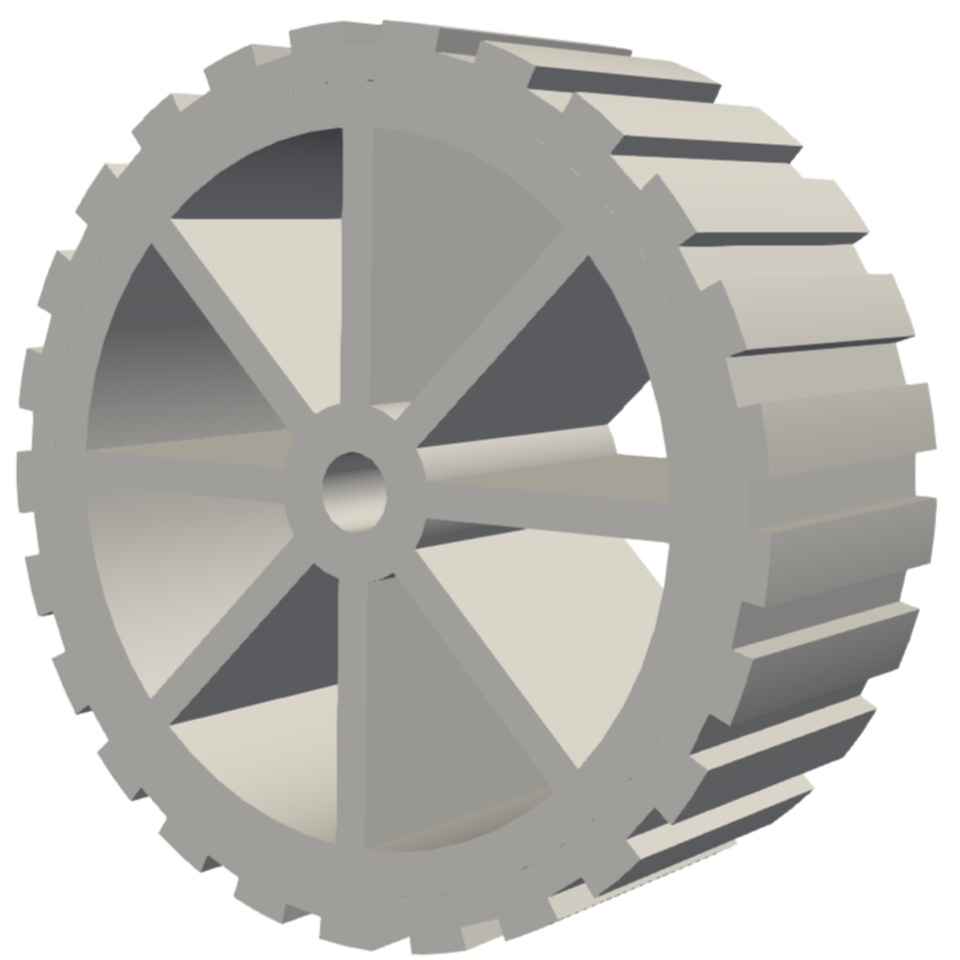}
		\caption{Rover wheel.} 
	\end{subfigure}
	\caption{Three wheels with different geometries. } 
	\label{fig:Three_wheels}
\end{figure}

In the single wheel test, the wheel's \textit{constant} translational $v$ and angular $\omega$ velocities are controlled to obtained a prescribed slip, where the wheel's slip is defined as $s = 1 - \frac{v}{\omega r}$, where $r$ is the wheel's radius. The single-wheel test rig is shown schematically in Fig.~\ref{fig:single_wheel_setup}. A fixed actuator track supports the free translational motion of a wheel assembly at a speed $v$; at the same time, the wheel has its angular velocity controlled. Owing to a primsatic joint, the wheel can slide up and down relative to the actuator track. In this set of experiments, the wheel's slip varied from 0.0 to 0.8 in nine different tests for each type of wheel (three different wheels were tested, see Fig.~\ref{fig:Three_wheels}). To ensure that the wheel moved a predefined distance in all experiments, the translational velocity of wheel was fixed to $v = 1.0~\si{m/s}$, while the angular velocity was enforced to assume the value $\omega = \frac{v}{r(1-s)}$. The DBP force was measured as the force necessary to apply at the center of the wheel to enable it to maintain the predefined slip. To evaluate the performance of the wheel in sloped deformable terrain, the equivalent traction slope was defined as $slope = \text{arctan}(DBP/Load)$, where the load is the weight produced by the mass of the whole system under gravitational pull. For each slip value, the simulation lasted for 15 s, enough to achieve steady state conditions. In the simulation, the soil bin was of size 16 m $\times$ 1 m $\times$ 0.25 m and it contained approximately \num{4600000} particles.

The first wheel used in this set of experiments was a cylindrical wheel. The second wheel came from a HMMWV vehicle and it had a complex geometry described using a triangle mesh; note that the geometry of the wheel did not change, the HMMWV wheel was assumed rigid. The radius of the first two wheels was 0.47 m and their width was 0.3 m. The third wheel was a rover wheel with grousers; its geometry was also described using a triangle mesh. The radius of the rover wheel was 0.25 m and its width was 0.2 m. The mass of the three wheel assemblies was 20 kg. 

\begin{figure}[htp]
	\centering
	\includegraphics[width=6in]{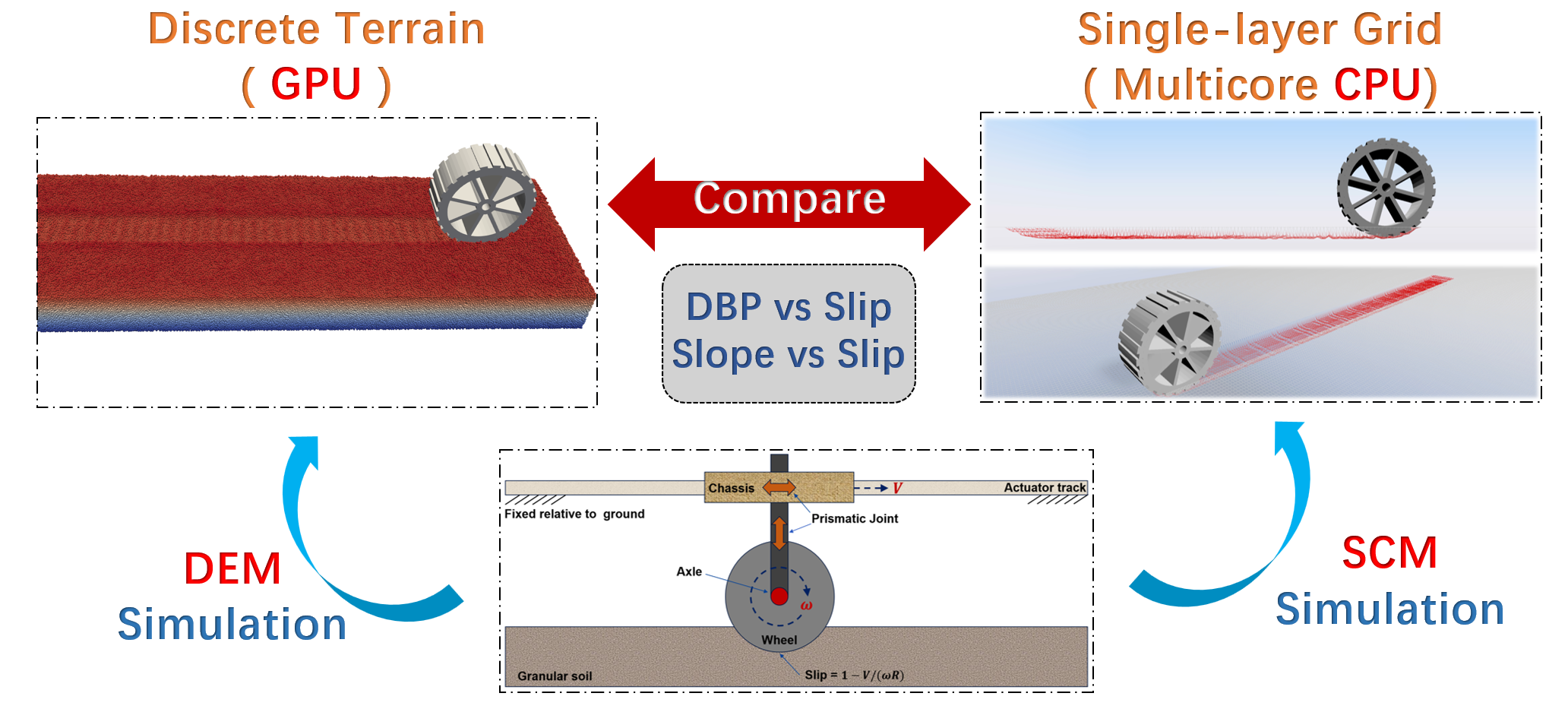}
	\caption{Single wheel test in DEM and SCM.}
	\label{fig:SCM_DEM_Singlewheel}
\end{figure}

\begin{figure}[htp]
	\centering
	\begin{subfigure}{0.49\textwidth}
		\centering
		\includegraphics[width=3.2in]{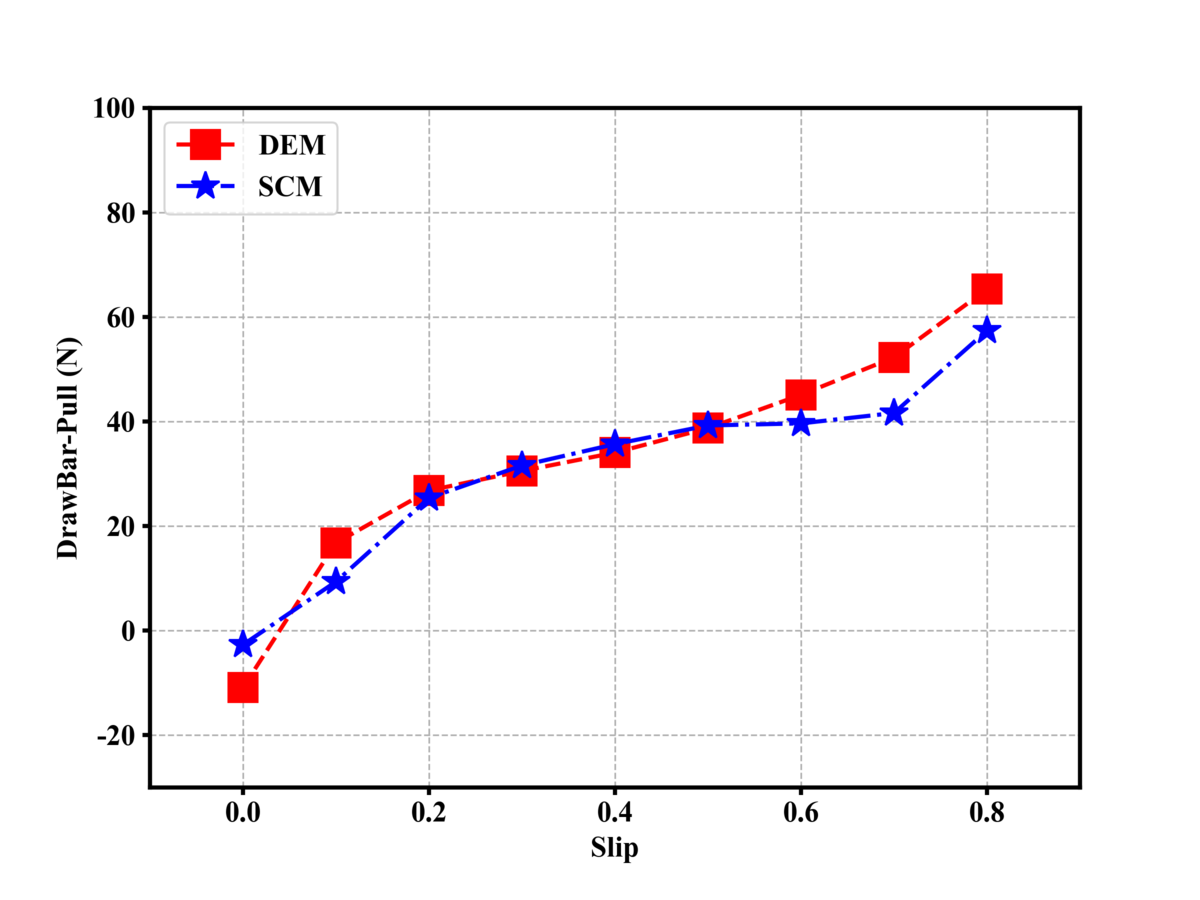}
		\caption{DBP force.}
	\end{subfigure}
	\begin{subfigure}{0.49\textwidth}
		\centering
		\includegraphics[width=3.2in]{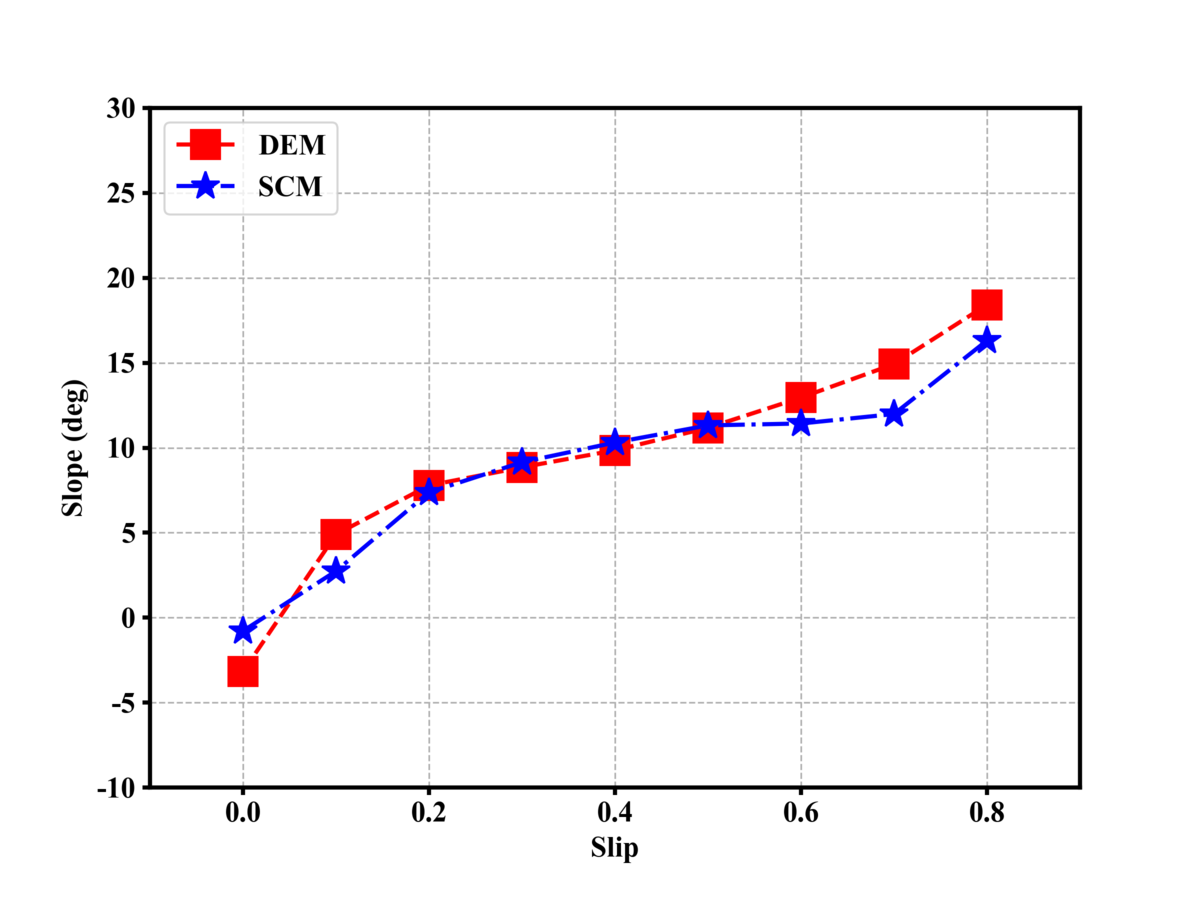}
		\caption{Terrain slope.} 
	\end{subfigure}
	\caption{Single wheel test (rigid cylinder wheel): DBP force and terrain slope vs. slip ratio curves for a normal load associated with 20~$\si{kg}$ under Earth gravitational pull. The SCM simulations were performed with parameters from Table~\ref{tab:scm_para}.} 
	\label{fig:dem_vs_scm_cylinder_wheel}
\end{figure}

\begin{figure}[htp]
	\centering
	\begin{subfigure}{0.49\textwidth}
		\centering
		\includegraphics[width=3.2in]{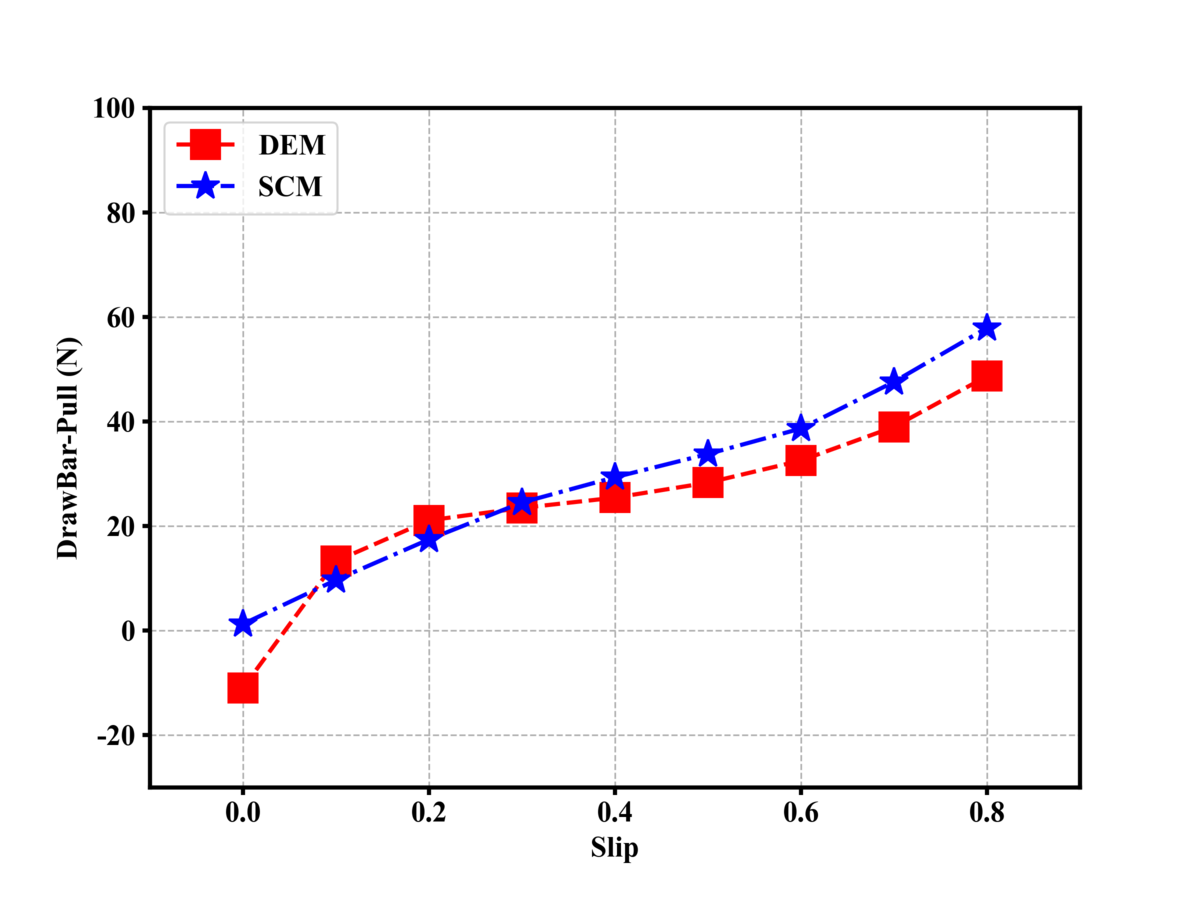}
		\caption{DBP force.}
	\end{subfigure}
	\begin{subfigure}{0.49\textwidth}
		\centering
		\includegraphics[width=3.2in]{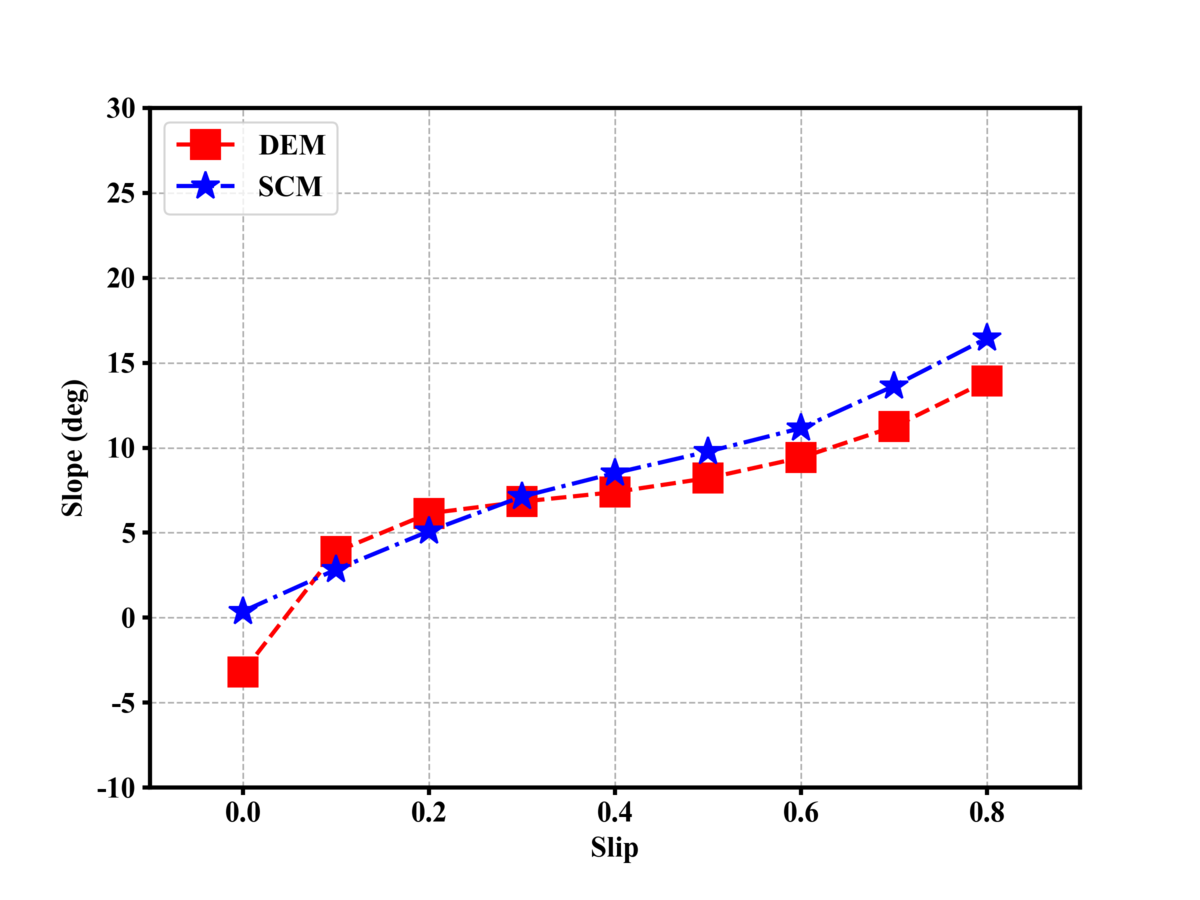}
		\caption{Terrain slope.} 
	\end{subfigure}
	\caption{Single wheel test (rigid HMMWV wheel): DBP force and terrain slope vs. slip ratio curves for a normal load associated with 20~$\si{kg}$ under Earth gravitational pull. The SCM simulations were performed with parameters from Table~\ref{tab:scm_para}.} 
	\label{fig:dem_vs_scm_hmmwv_wheel}
\end{figure}

\begin{figure}[htp]
	\centering
	\begin{subfigure}{0.49\textwidth}
		\centering
		\includegraphics[width=3.2in]{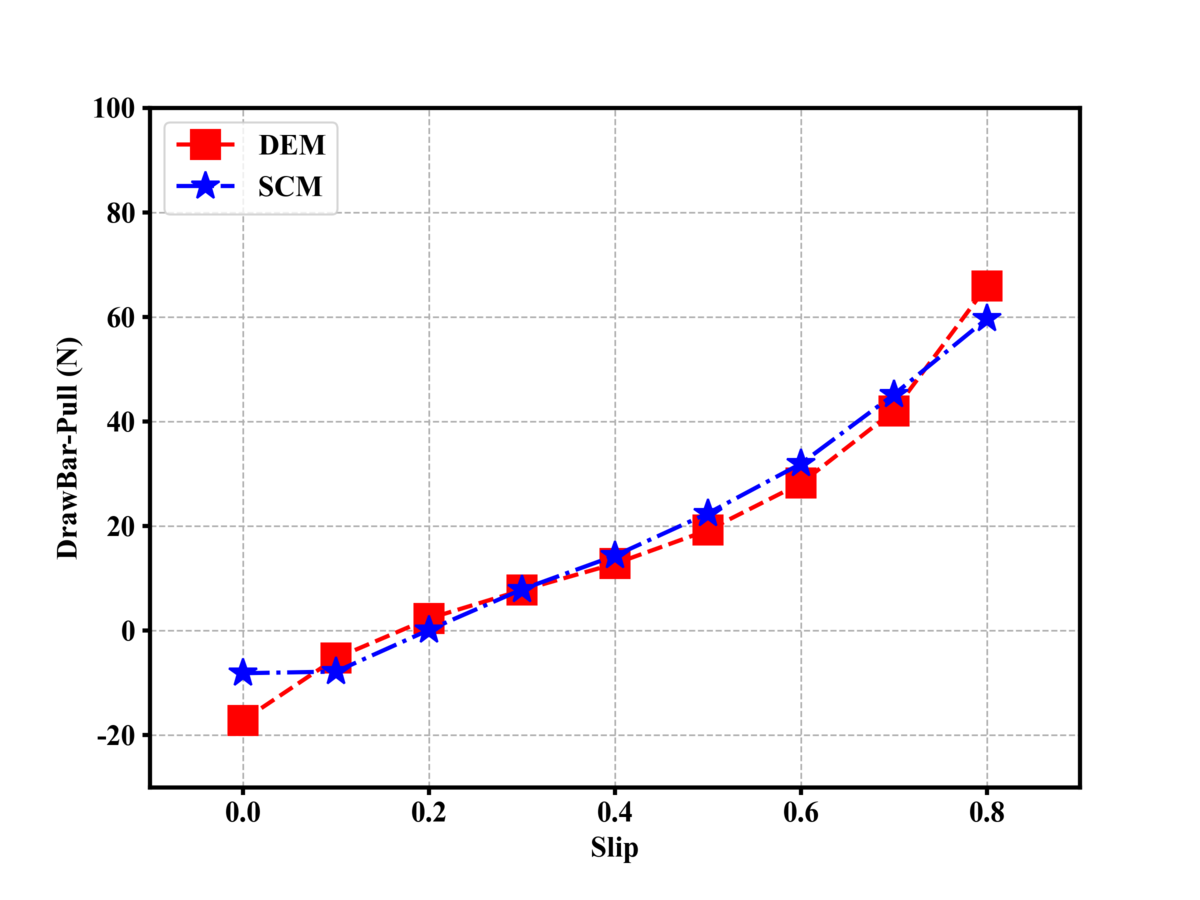}
		\caption{DBP force.}
	\end{subfigure}
	\begin{subfigure}{0.49\textwidth}
		\centering
		\includegraphics[width=3.2in]{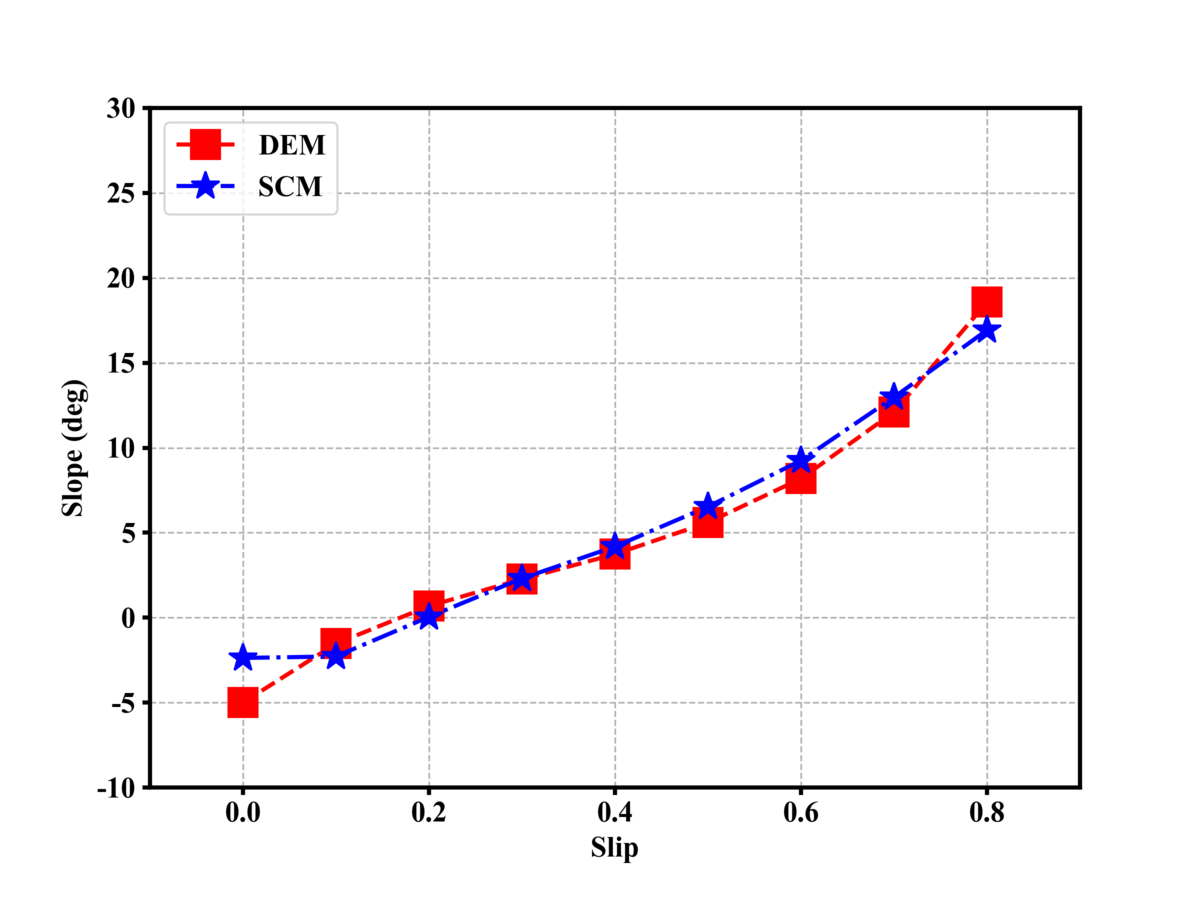}
		\caption{Terrain slope.} 
	\end{subfigure}
	\caption{Single wheel test (rigid rover wheel): DBP force and terrain slope vs. slip ratio curves for a normal load associated with 20~$\si{kg}$ under Earth gravitational pull. The SCM simulations were performed with parameters from Table~\ref{tab:scm_para}.} 
	\label{fig:dem_vs_scm_rover_wheel}
\end{figure}

\begin{figure}[htp]
	\centering
	\begin{subfigure}{0.3\textwidth}
		\centering
		\includegraphics[width=2in]{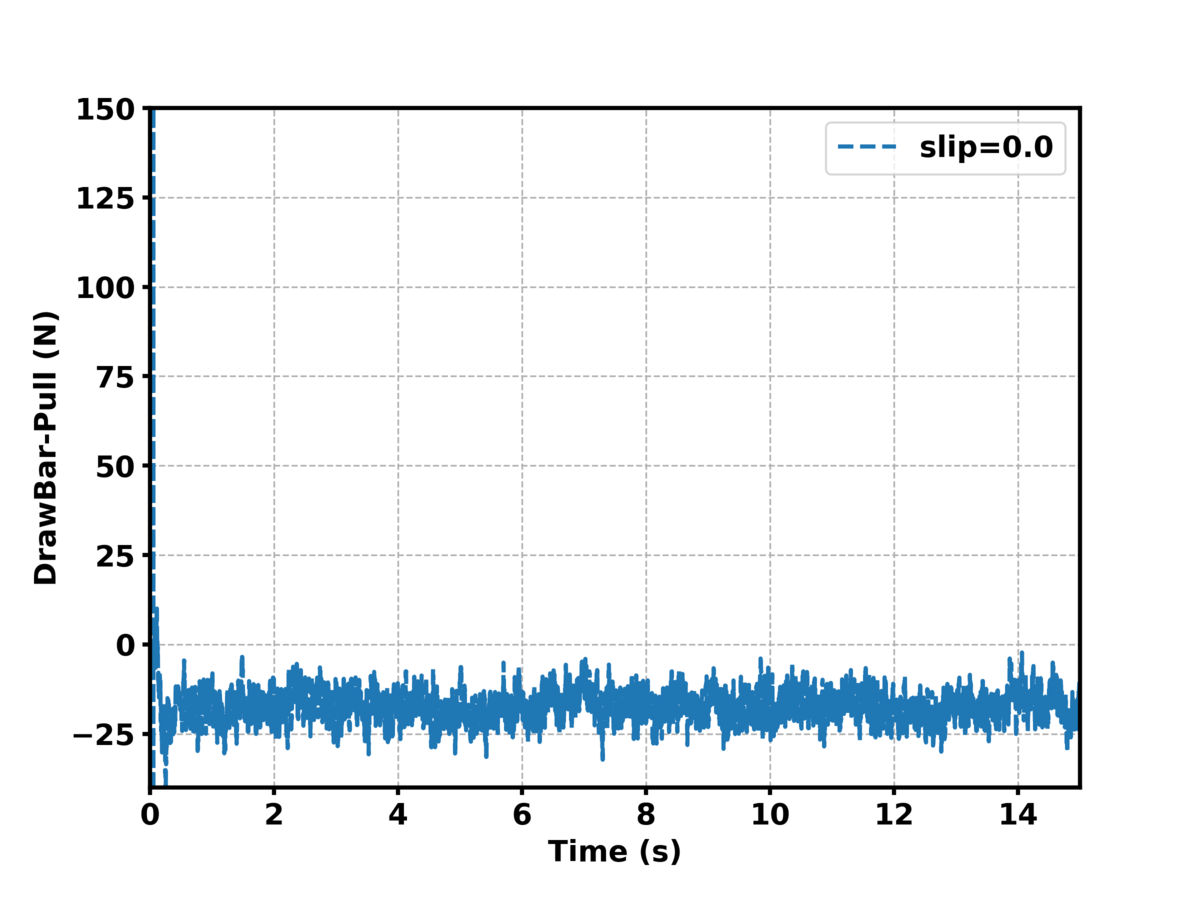}
		\caption{slip=0.0}
	\end{subfigure}
	\begin{subfigure}{0.3\textwidth}
		\centering
		\includegraphics[width=2in]{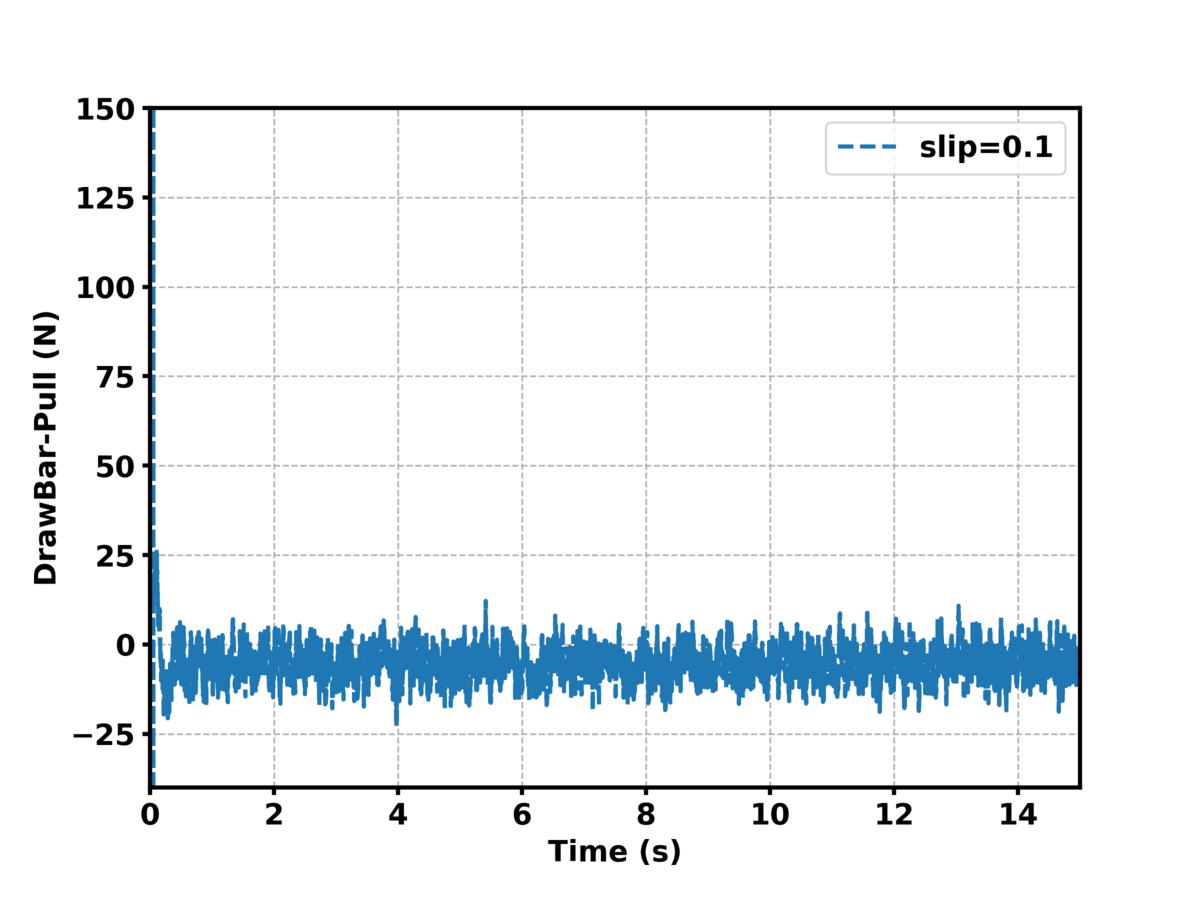}
		\caption{slip=0.1}
	\end{subfigure}
	\begin{subfigure}{0.3\textwidth}
		\centering
		\includegraphics[width=2in]{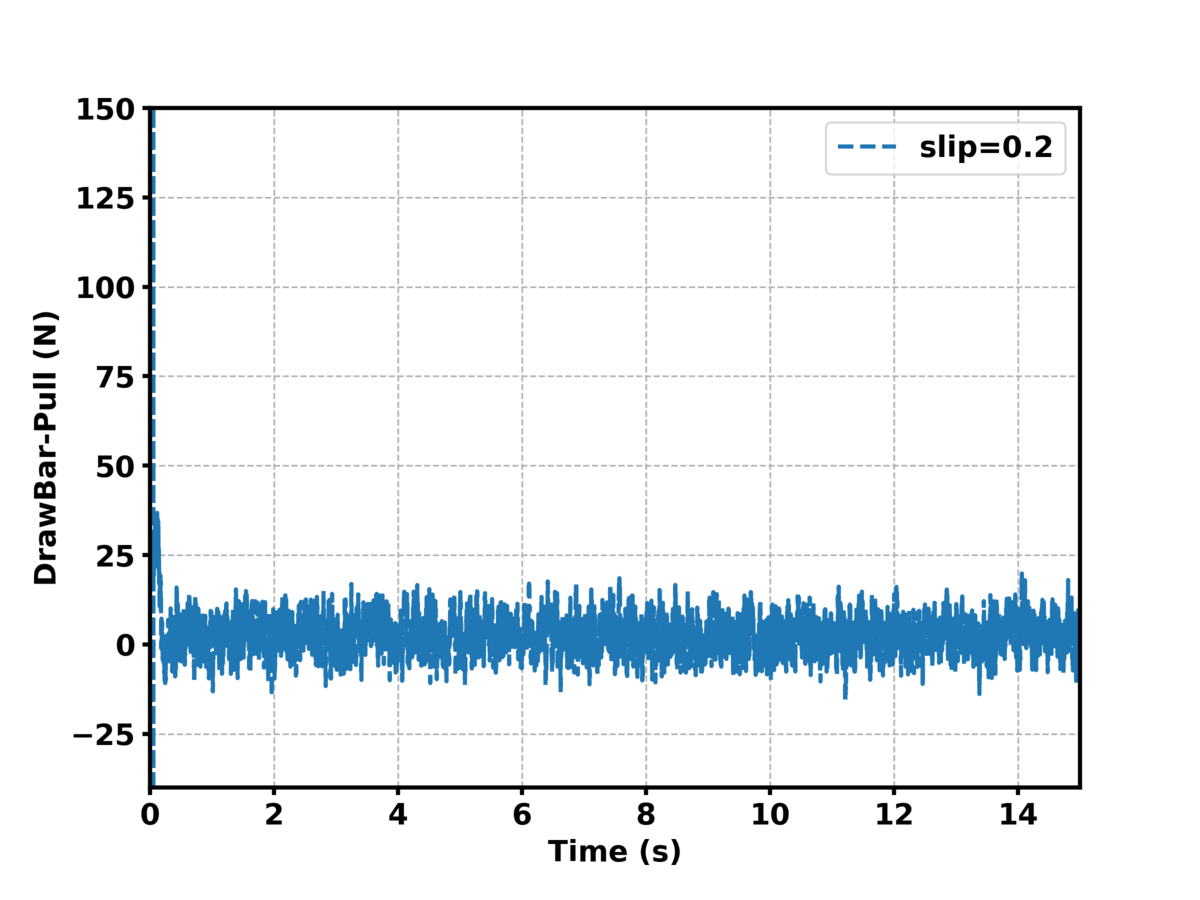}
		\caption{slip=0.2} 
	\end{subfigure}
	
	\qquad
	
	\begin{subfigure}{0.3\textwidth}
		\centering
		\includegraphics[width=2in]{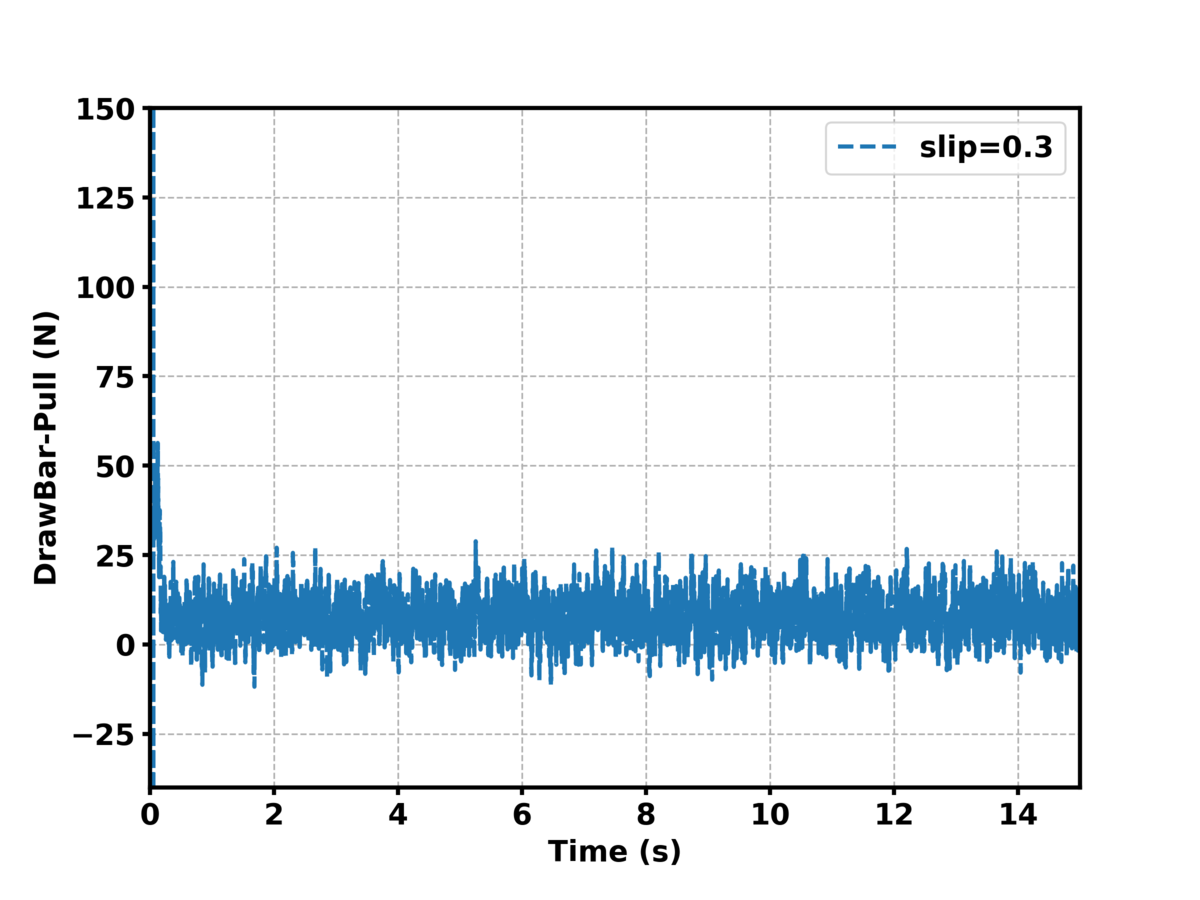}
		\caption{slip=0.3}
	\end{subfigure}
	\begin{subfigure}{0.3\textwidth}
		\centering
		\includegraphics[width=2in]{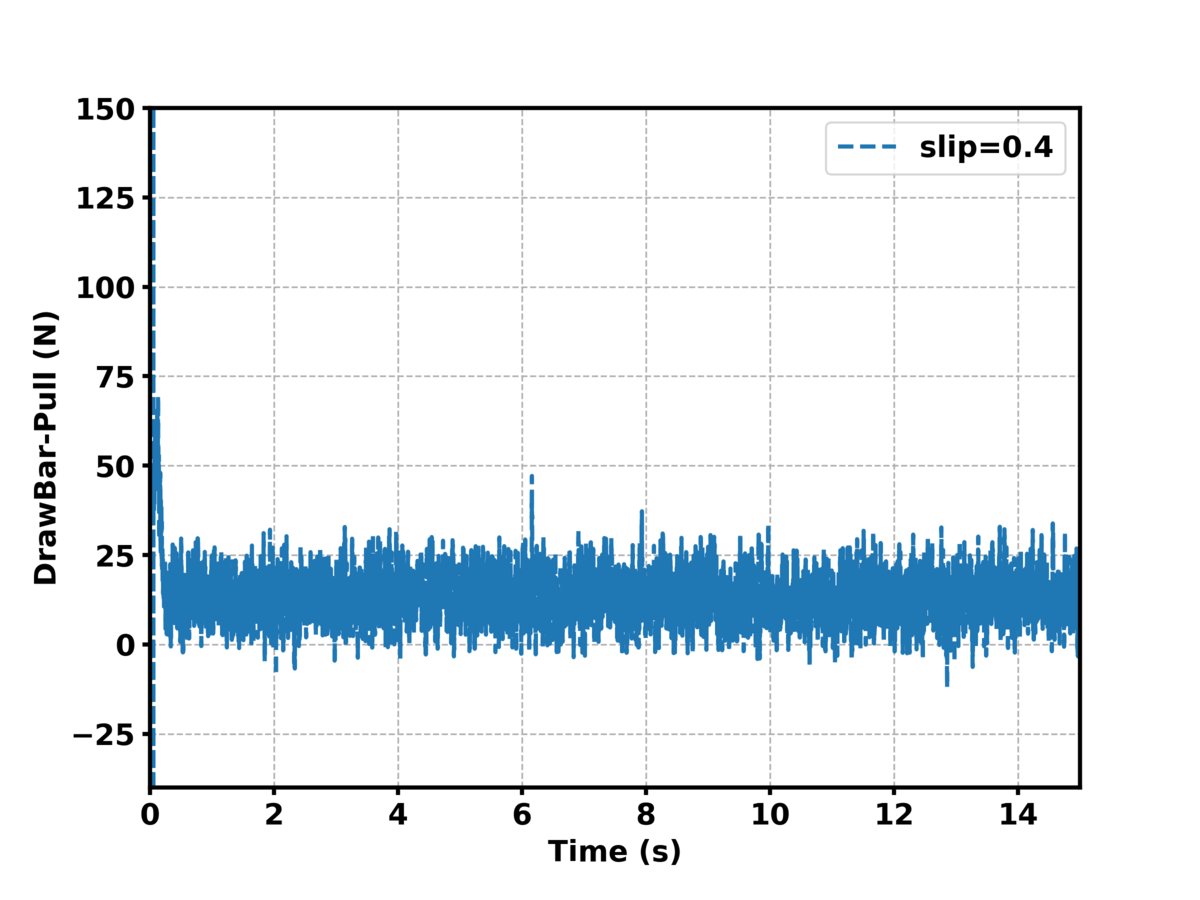}
		\caption{slip=0.4}
	\end{subfigure}
	\begin{subfigure}{0.3\textwidth}
		\centering
		\includegraphics[width=2in]{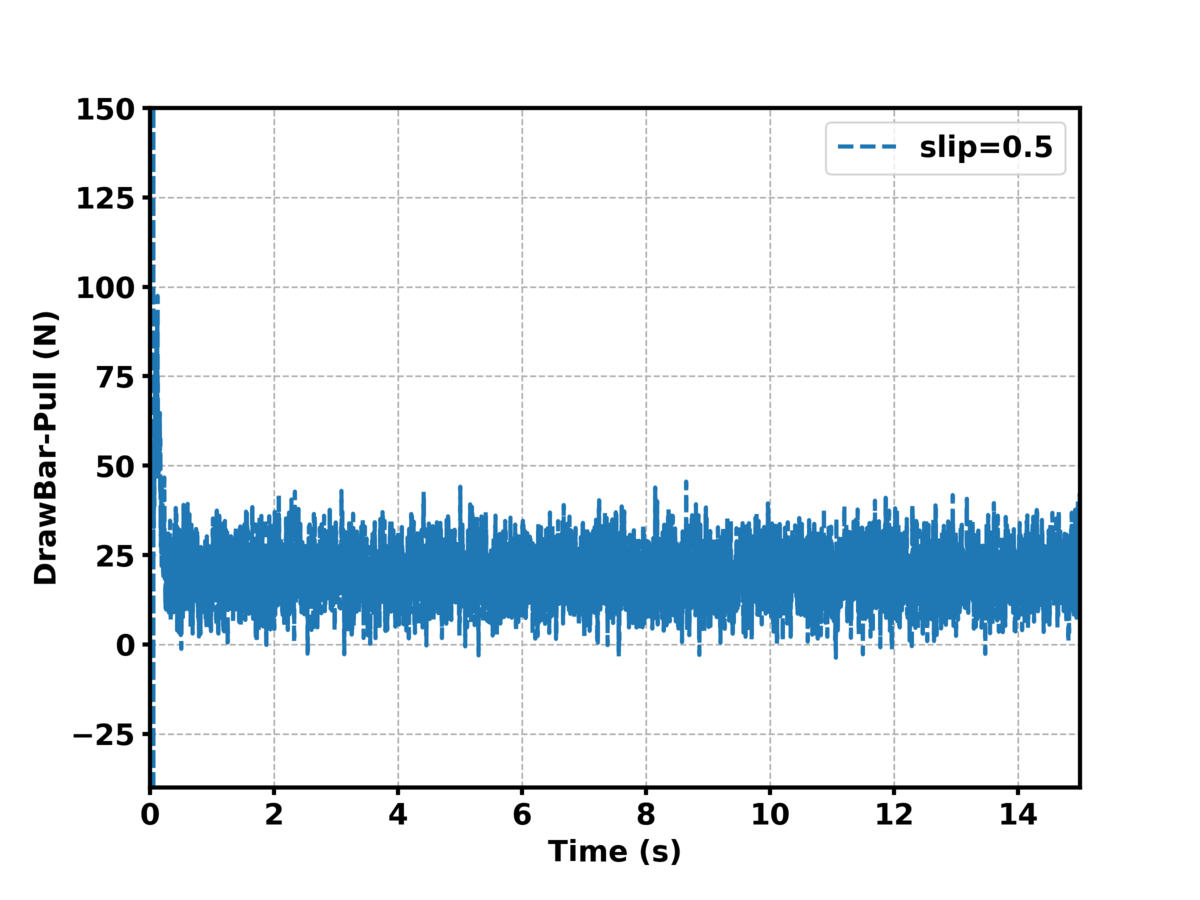}
		\caption{slip=0.5} 
	\end{subfigure}
	
	\qquad
	
	\begin{subfigure}{0.3\textwidth}
		\centering
		\includegraphics[width=2in]{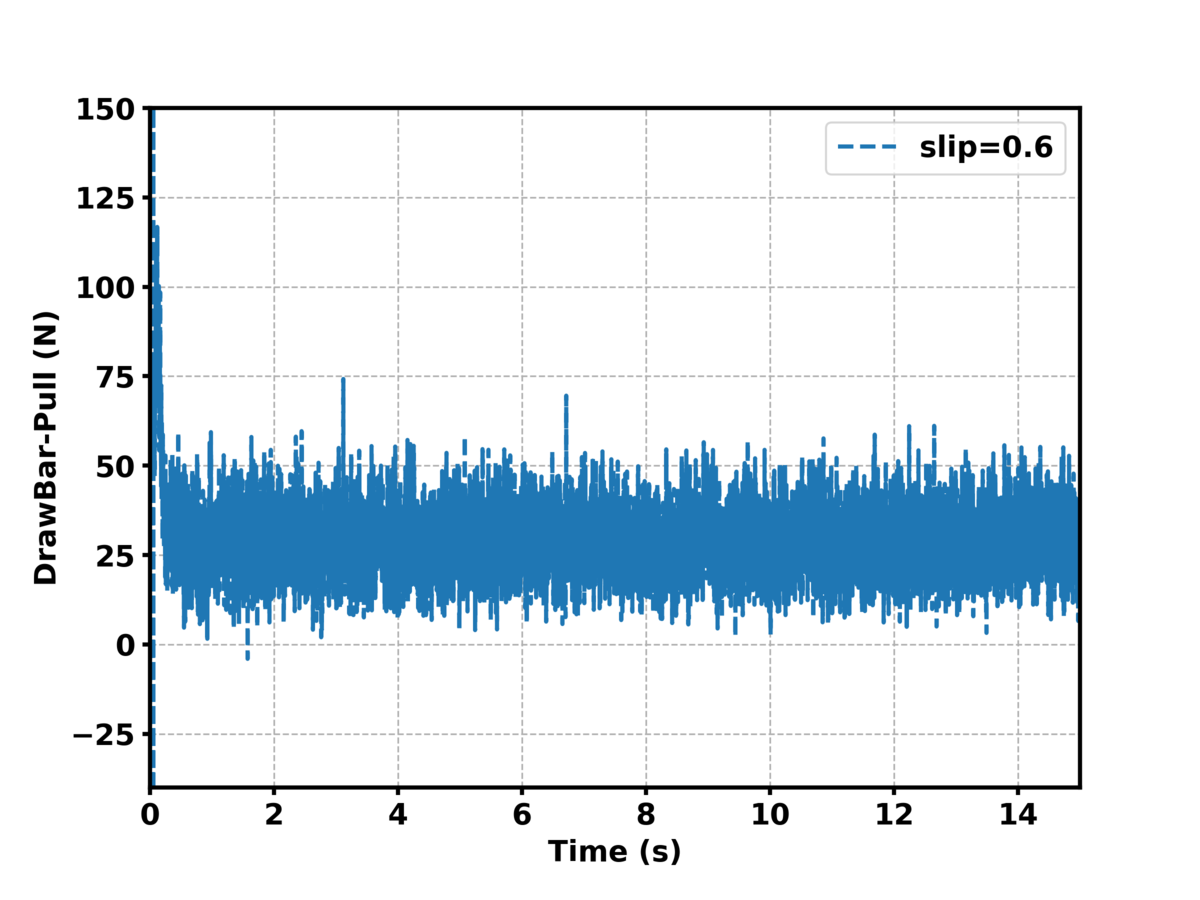}
		\caption{slip=0.6}
	\end{subfigure}
	\begin{subfigure}{0.3\textwidth}
		\centering
		\includegraphics[width=2in]{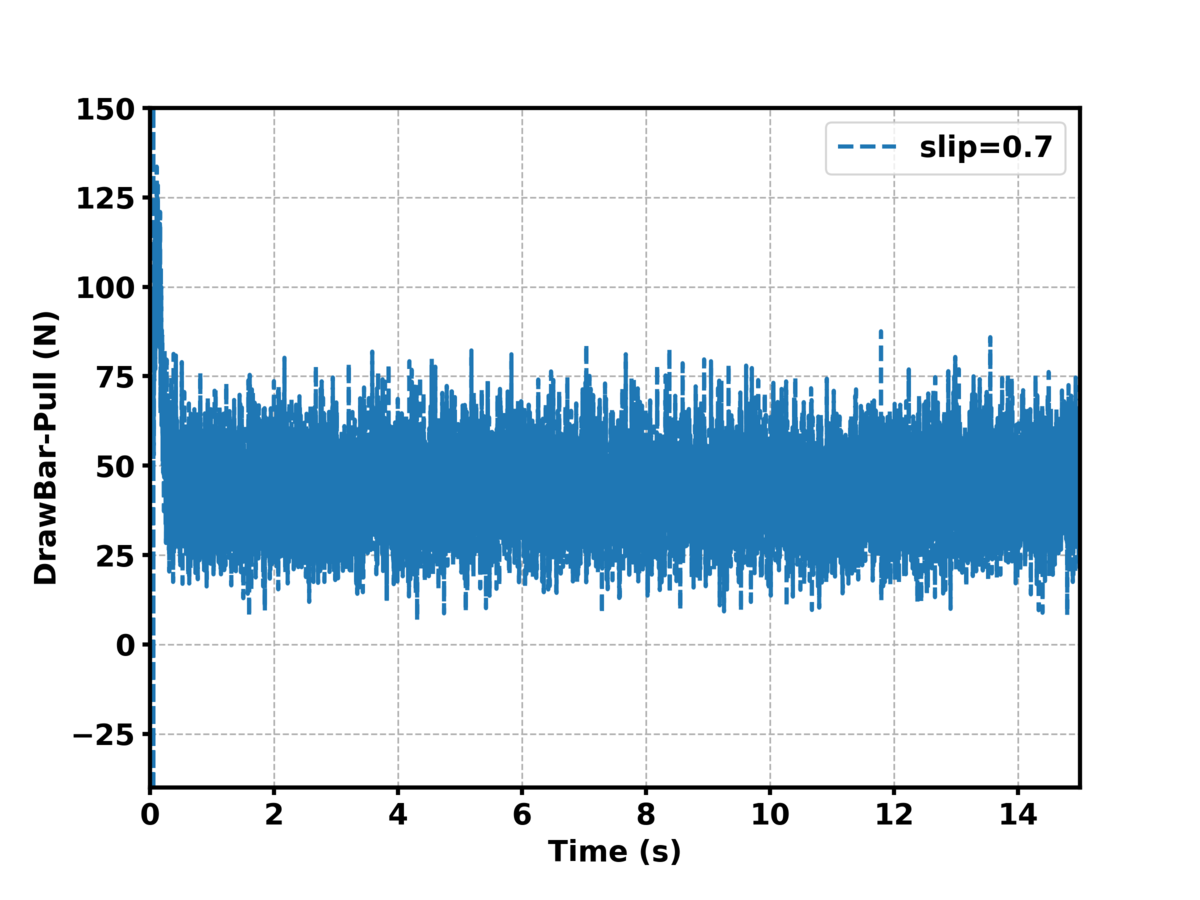}
		\caption{slip=0.7}
	\end{subfigure}
	\begin{subfigure}{0.3\textwidth}
		\centering
		\includegraphics[width=2in]{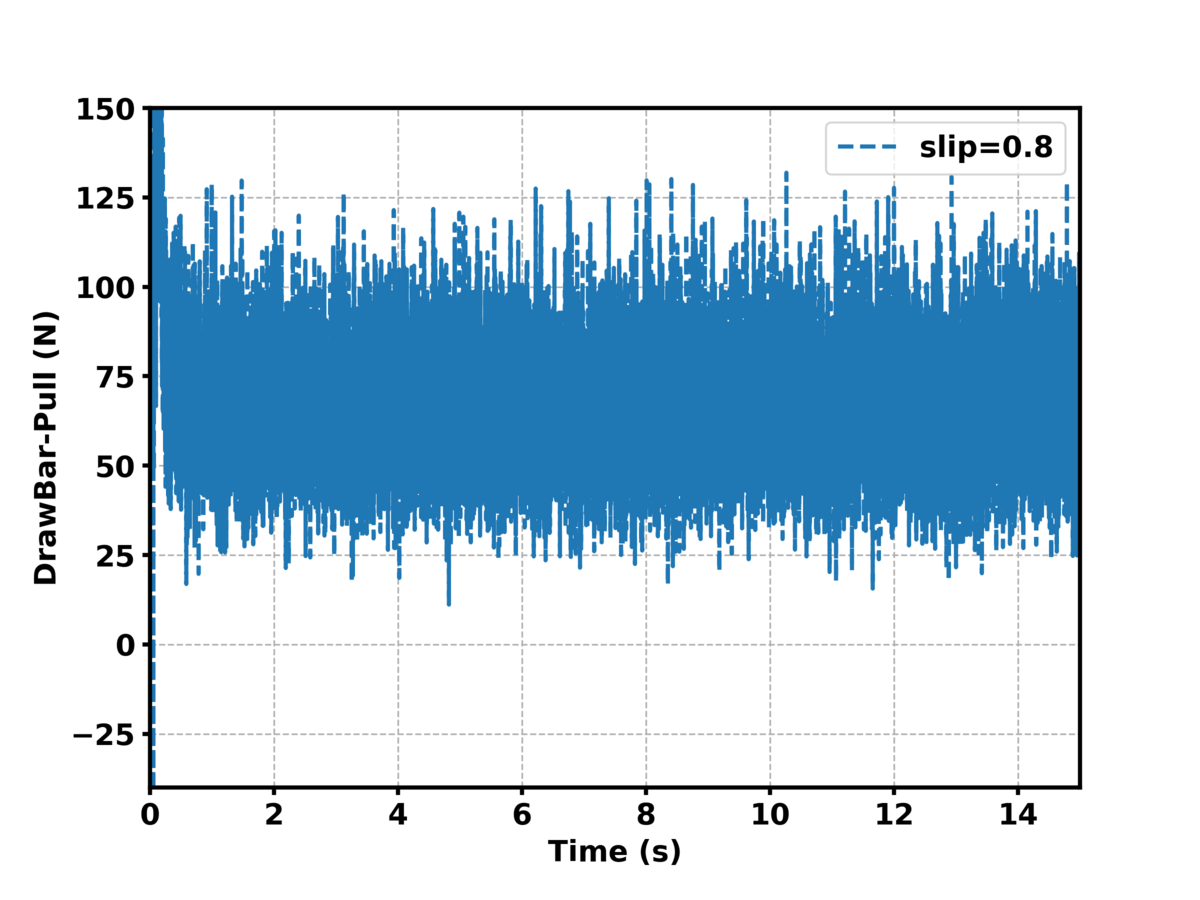}
		\caption{slip=0.8} 
	\end{subfigure}
	
	\caption{Time histories of DBP force with different wheel slip ratio in the DEM test for the rover wheel.} 
	\label{fig:Time_histories_DBP}
\end{figure}

\begin{figure}[htp]
	\centering
	\begin{subfigure}{0.9\textwidth}
		\centering
		\includegraphics[width=6in]{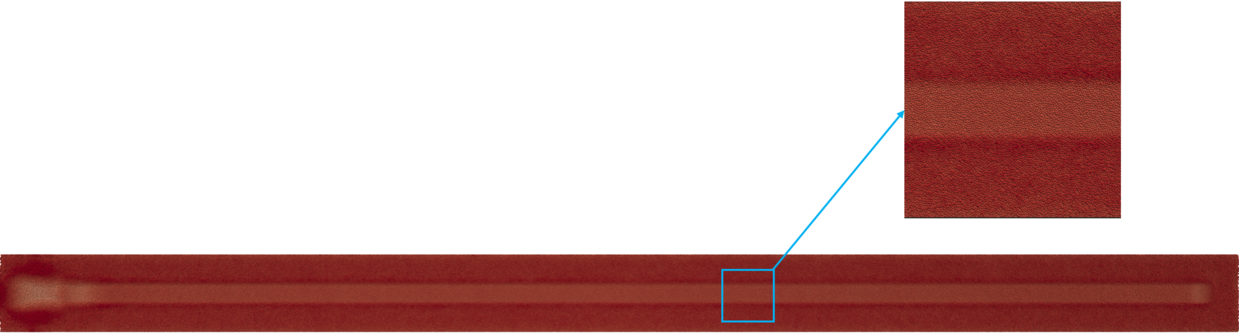}
	\end{subfigure}
	\caption{Track of the HMMWV wheel. Animation of the single wheel test is provided in the supplementary materials.} 
\label{fig:Track_HMMWV_wheel}
\end{figure}

\begin{figure}[htp]
\centering
\begin{subfigure}{0.9\textwidth}
	\centering
	\includegraphics[width=6in]{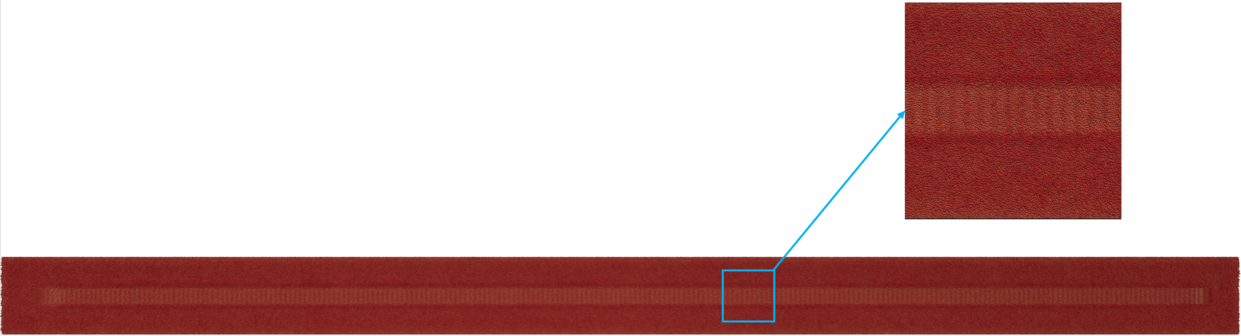}
\end{subfigure}
\caption{Track of the rover wheel. Animation of the single wheel test is provided in the supplementary materials.}
\label{fig:Track_rover_wheel}
\end{figure}

For each of the three wheel assemblies, we first ran simulations using SCM with the calibrated parameters obtained in Section \ref{sec:calibration}. Subsequently we ran high-fidelity DEM terramechanics employing the soil parameters used to produced the ground truth data. Both DBP force and traction slope were measured in the SCM and DEM simulations to compare the two approaches; the methodology is summarized in Fig.~\ref{fig:SCM_DEM_Singlewheel}. The DBP force and traction slope vs. slip ratio curves are shown in Figs.~\ref{fig:dem_vs_scm_cylinder_wheel}, \ref{fig:dem_vs_scm_hmmwv_wheel}, and \ref{fig:dem_vs_scm_rover_wheel}. Both the DBP force and the traction slope for SCM and DEM compare well. As an example of raw simulation results, the single rover wheel on DEM terrain results for the DBP force are shown in Fig.~\ref{fig:Time_histories_DBP}. As expected, the higher the slip ratio, the stronger the force oscillations. To simulate the 15 s single wheel rolling process, SCM simulation took 30 seconds, while the DEM simulation took 4800 seconds. To better reveal the terrain deformation caused by the wheel grousers, terrain profiles are shown in Figs.~\ref{fig:Track_HMMWV_wheel} and \ref{fig:Track_rover_wheel}. The terrain deformation caused by the grousers can be observed in Fig.~\ref{fig:Track_rover_wheel}.

\FloatBarrier

\subsection{Full vehicle virtual experiments}
\label{subsec:fullrover}
\begin{figure}[htp]
	\centering
	\begin{subfigure}{0.49\textwidth}
	\centering
	\includegraphics[width=3.2in]{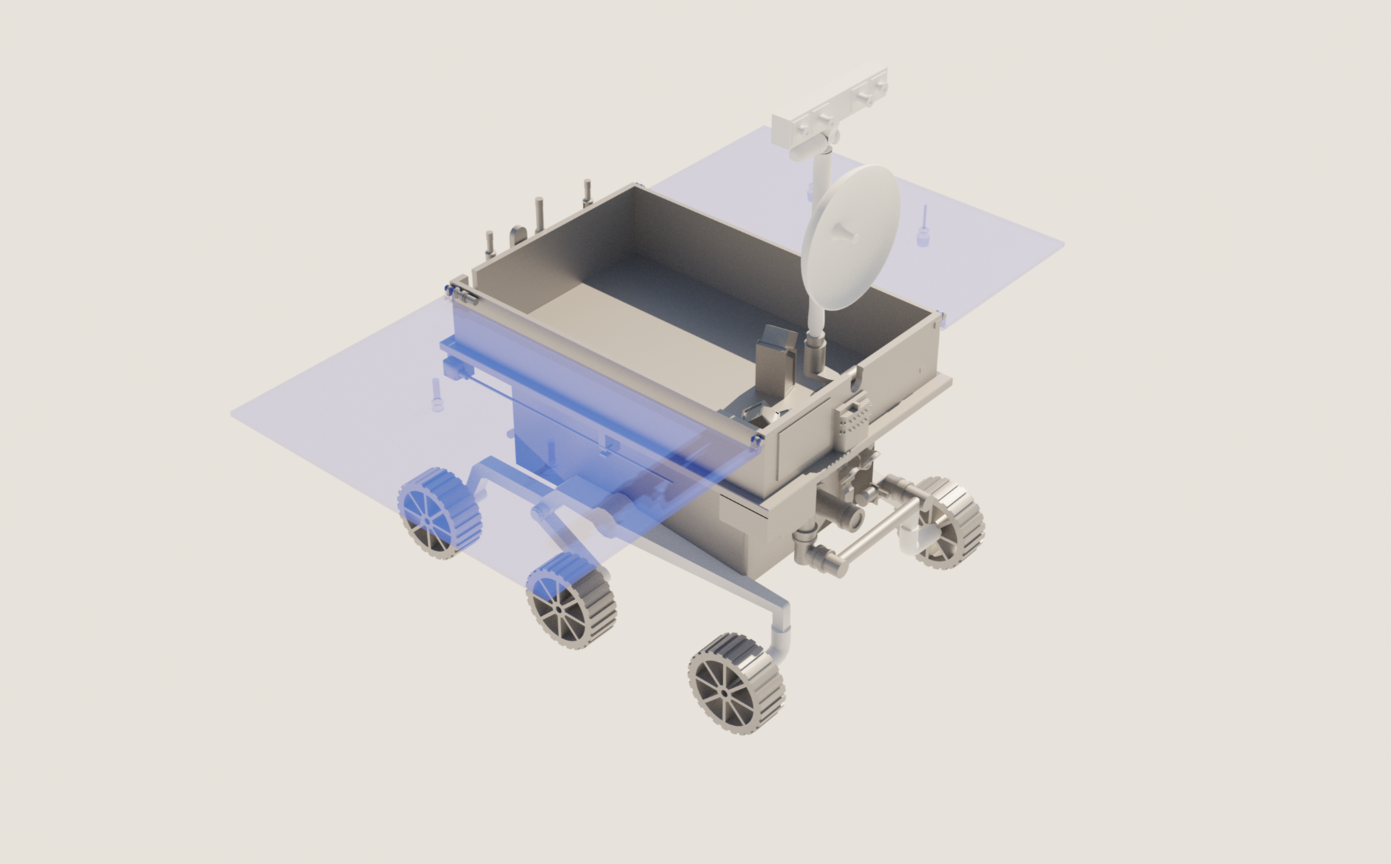}
	\end{subfigure}
	\begin{subfigure}{0.49\textwidth}
	\centering
	\includegraphics[width=3.2in]{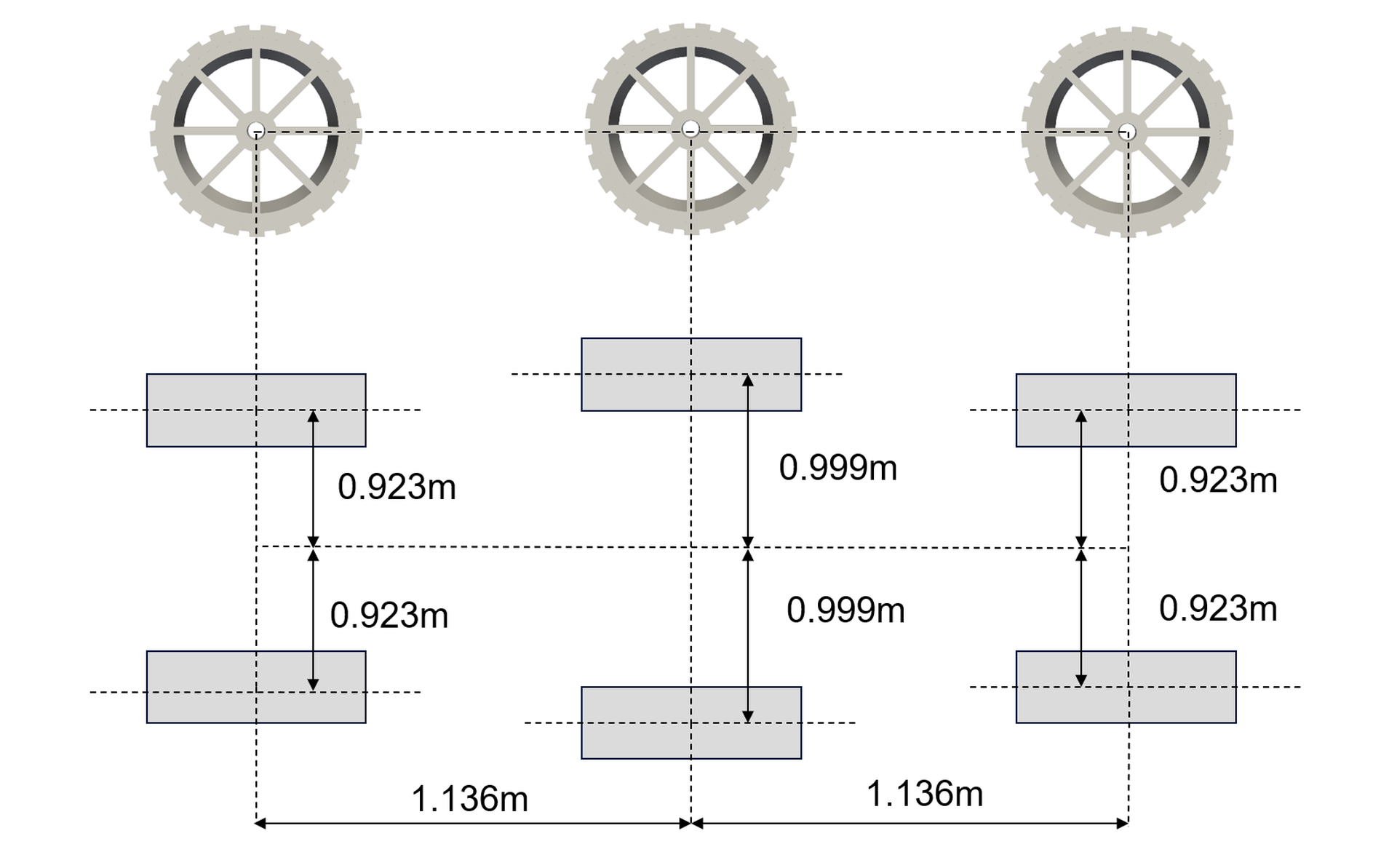}
	\end{subfigure}
	\caption{Setup of the Yutu rover replica used in the simulations.} 
	\label{fig:wheel_position}
\end{figure}

\begin{figure}[htp]
	\centering
	\includegraphics[width=6in]{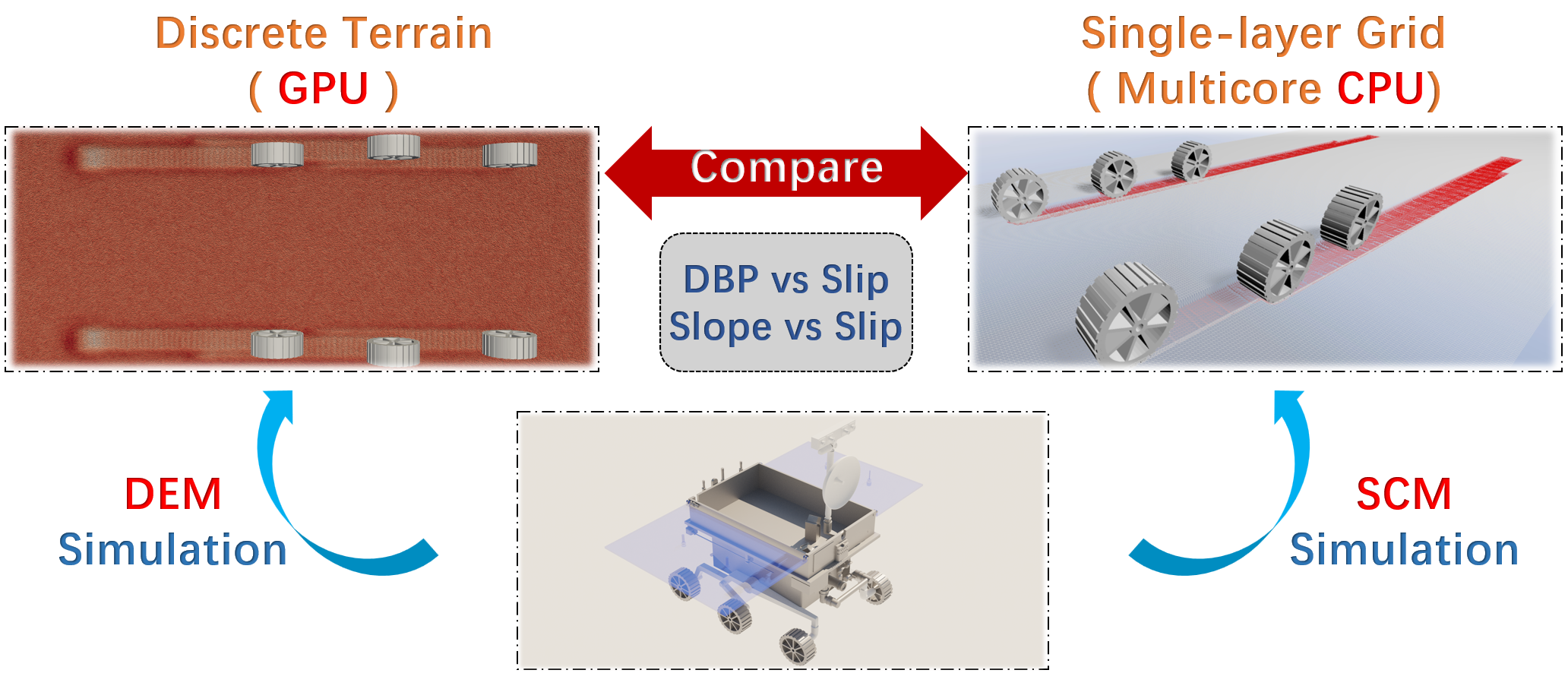}
	\caption{Full rover simulation in DEM and SCM.}
	\label{fig:SCM_DEM_Rover}
\end{figure}

\begin{figure}[htp]
	\centering
	\begin{subfigure}{0.49\textwidth}
		\centering
		\includegraphics[width=3.2in]{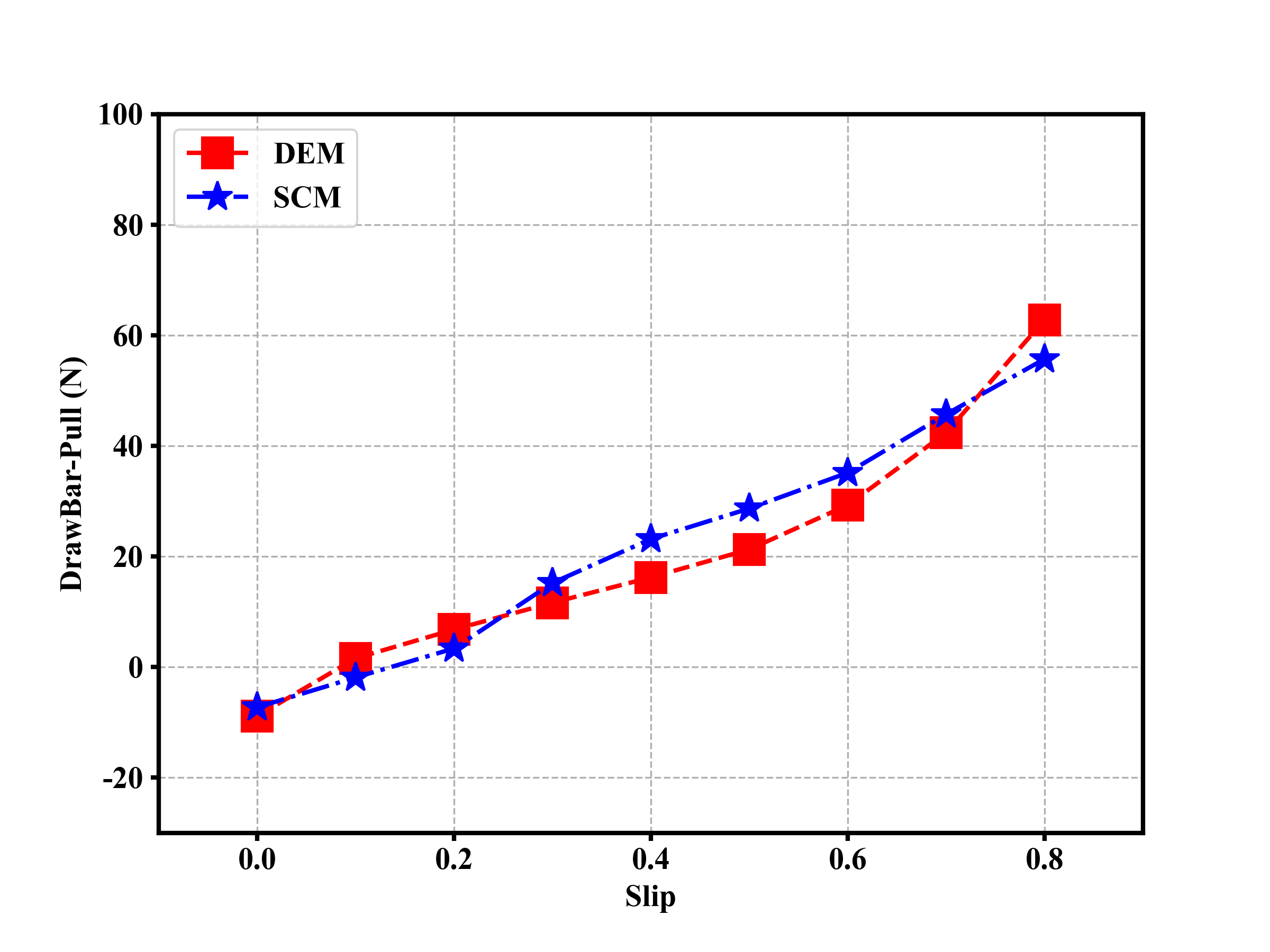}
		\caption{DBP force.}
	\end{subfigure}
	\begin{subfigure}{0.49\textwidth}
		\centering
		\includegraphics[width=3.2in]{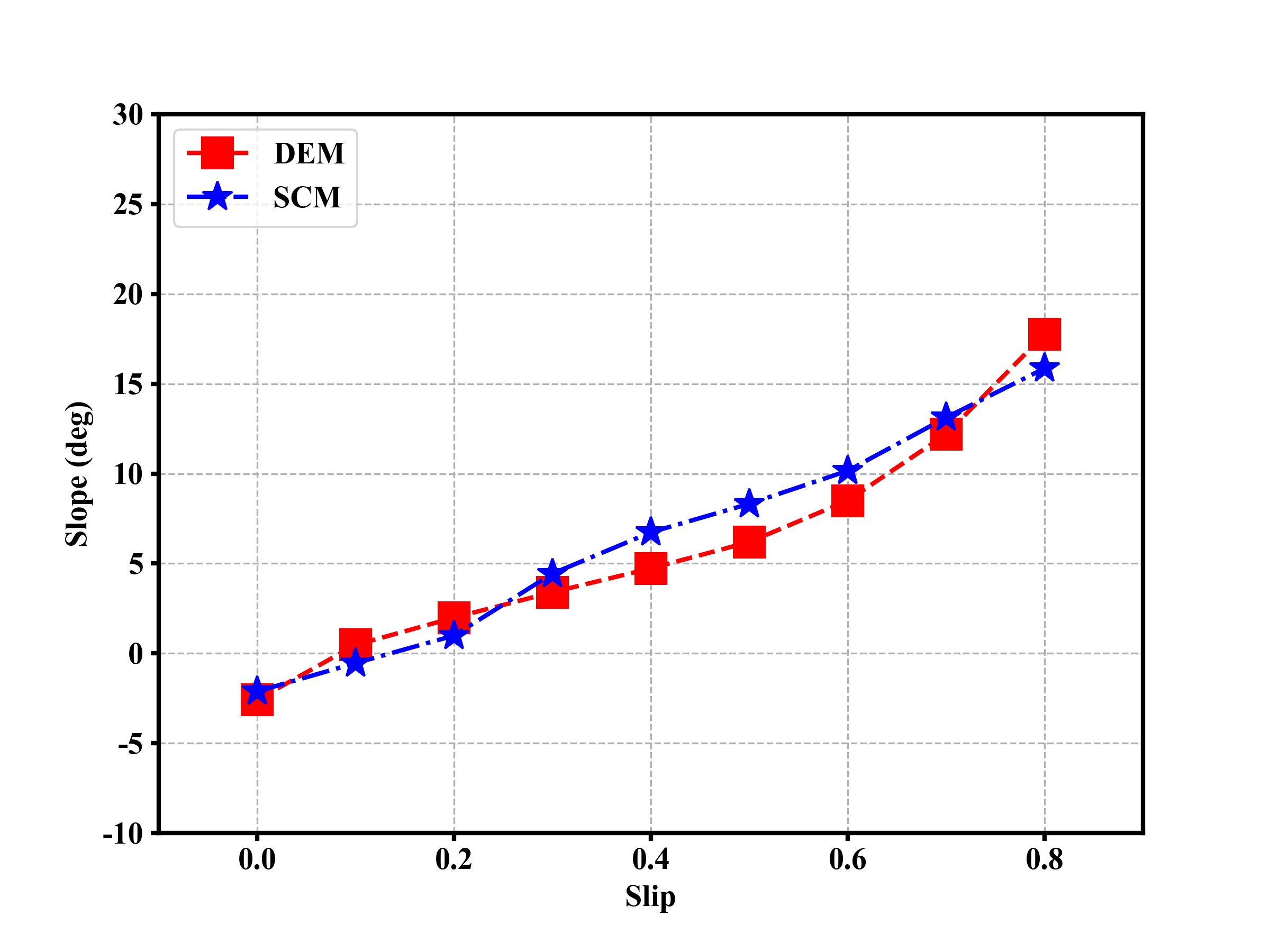}
		\caption{Terrain slope.} 
	\end{subfigure}
	\caption{Comparison of full rover simulation using DEM and SCM.} 
	\label{fig:Rover_DEM_SCM}
\end{figure}

The full vehicle considered is the Yutu rover \cite{dingScienceRobotics2022}, which has six wheels with grousers, see Fig.~\ref{fig:wheel_position}. As shown in the figure, the distance between the wheel axles is 1.136 m. The front and rear wheels are offset by 0.923 m from the rover centerline, while the middle wheels are offset by 0.999 m. The total mass of the rover was approximately 120 kg. The rover was controlled to move at a constant translational velocity and different angular velocities of the wheels. The translational velocity of the rover was fixed at 1 m/s, and the slip between each wheel and soil varied from 0 to 0.8. The angular velocity of the wheels was changed according to the target slip ratio as in $\omega = \frac{v}{r(1-s)}$. Like in the previous subsection, the full rover simulations were run with both SCM and DEM terramechanics and the corresponding results were subsequently compared; the methodology is summarized in Fig.~\ref{fig:SCM_DEM_Rover}. Figure~\ref{fig:Rover_DEM_SCM} offers a quantitative comparison in terms of DBP vs. slip, and slope vs. slip, which indicate that the DEM and SCM results are in agreement. In the DEM simulation, the soil bin was of size 14 m $\times$ 2.5 m $\times$ 0.375 m and it contained approximately \num{15000000} particles. To simulate the 10 s rover test, the SCM simulation took 20 seconds, while the the DEM simulation took \num{10000} seconds.

\begin{figure}[htp]
	\centering
	\begin{subfigure}{0.49\textwidth}
		\centering
		\includegraphics[width=3.2in]{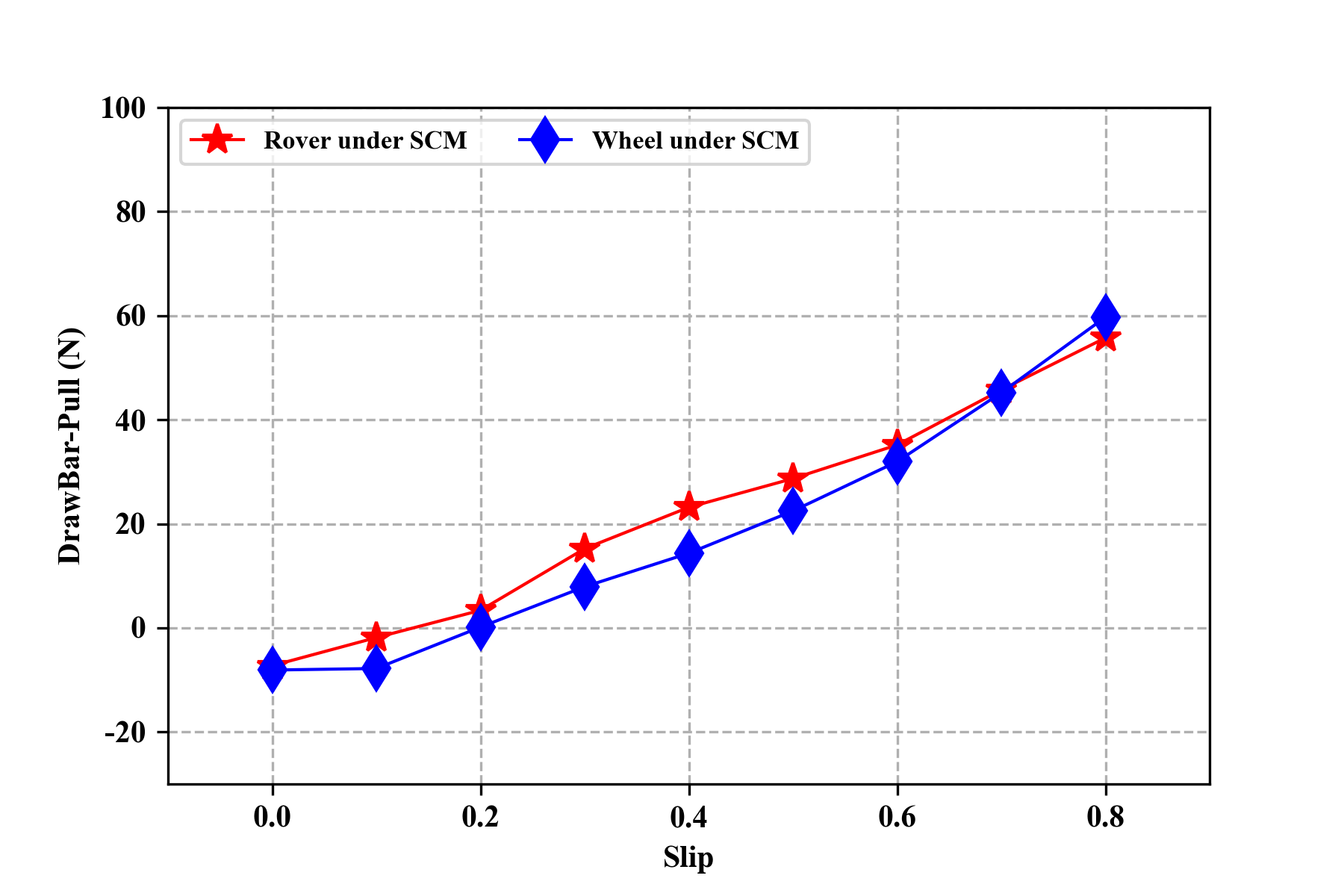}
		\caption{DBP force.}
	\end{subfigure}
	\begin{subfigure}{0.49\textwidth}
		\centering
		\includegraphics[width=3.2in]{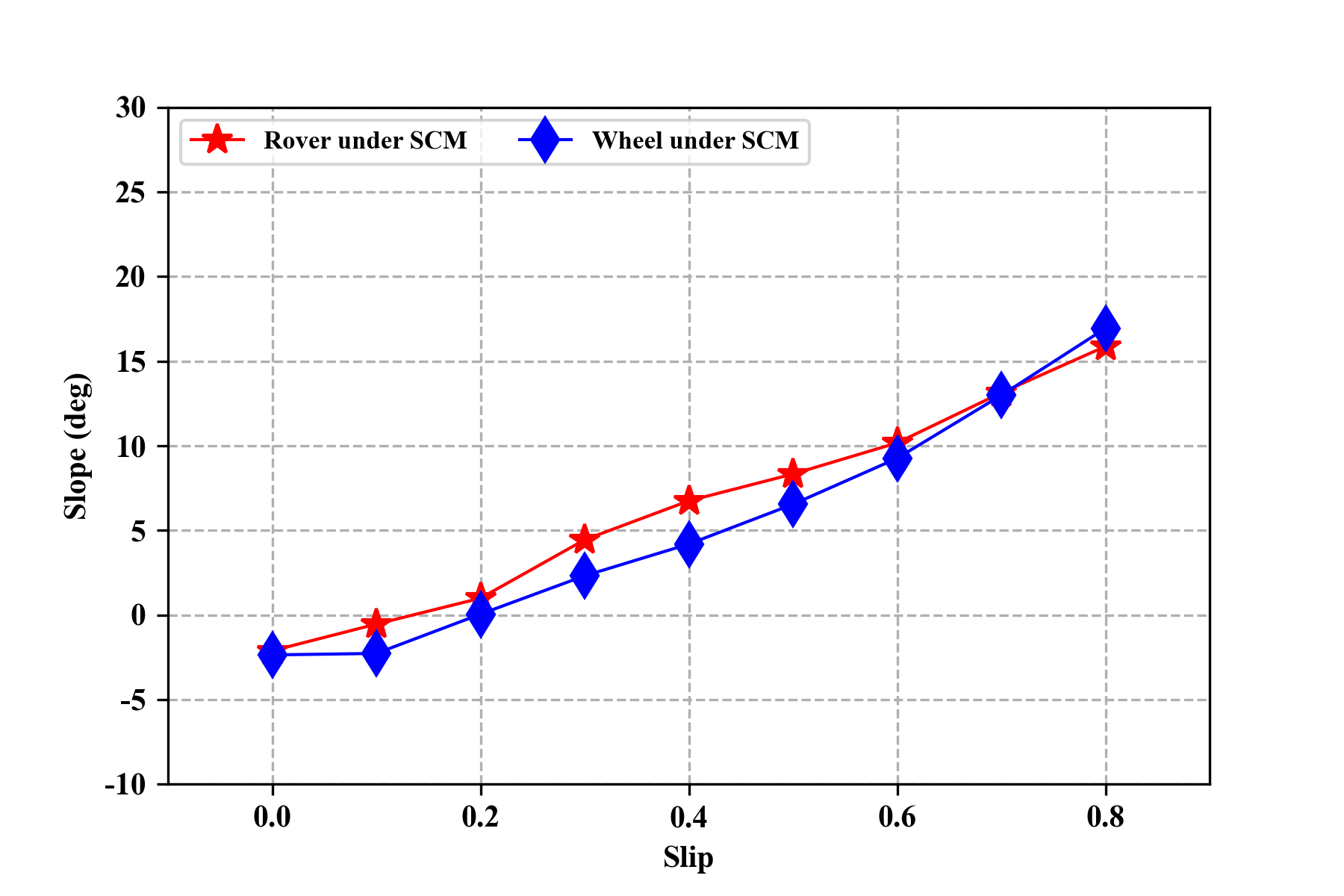}
		\caption{Terrain slope.} 
	\end{subfigure}
	\caption{Comparison of full rover simulation and single wheel simulation using SCM.} 
	\label{fig:Rover_singlewheel_SCM}
\end{figure}

\begin{figure}[htp]
	\centering
	\begin{subfigure}{0.49\textwidth}
		\centering
		\includegraphics[width=3.2in]{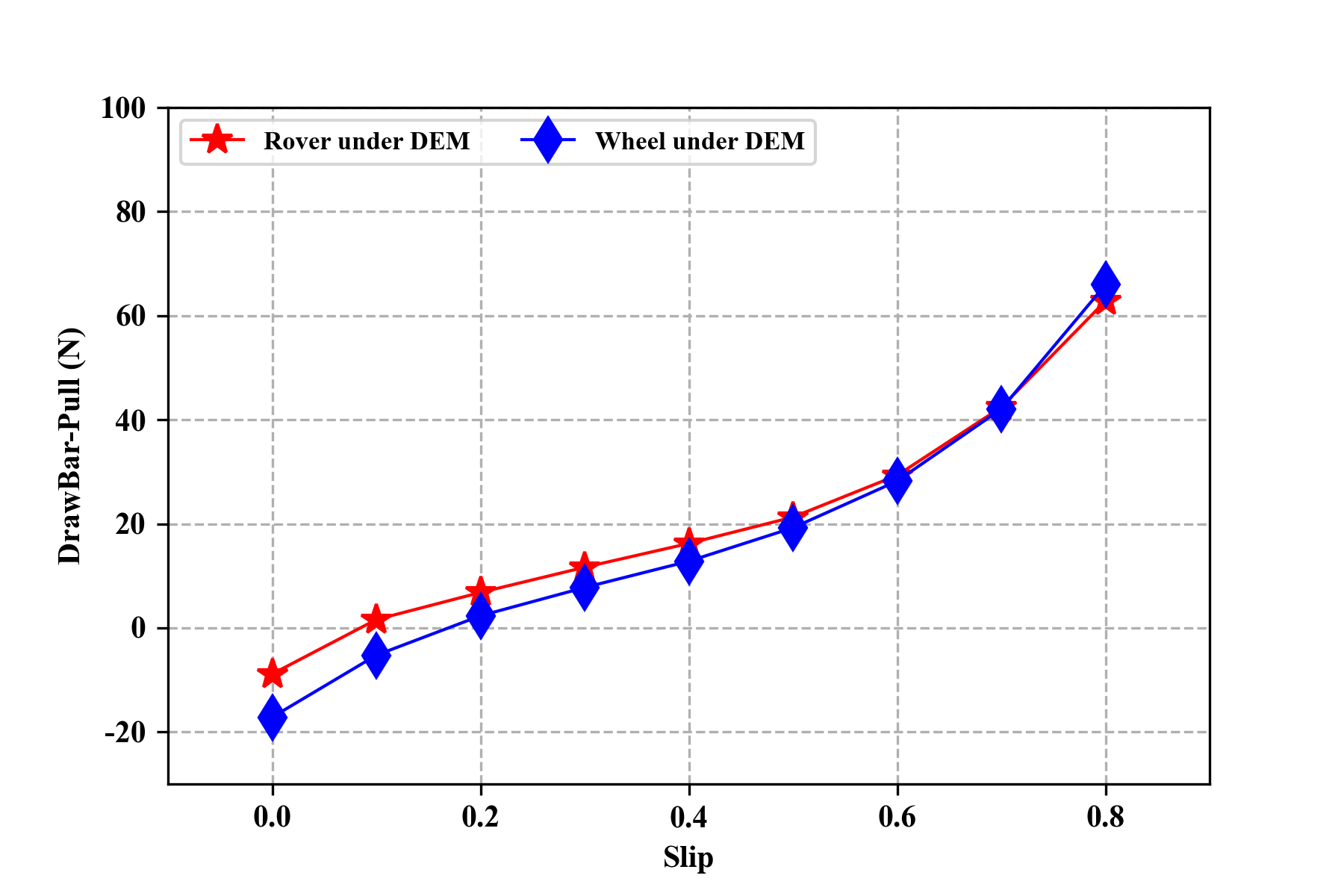}
		\caption{DBP force.}
	\end{subfigure}
	\begin{subfigure}{0.49\textwidth}
		\centering
		\includegraphics[width=3.2in]{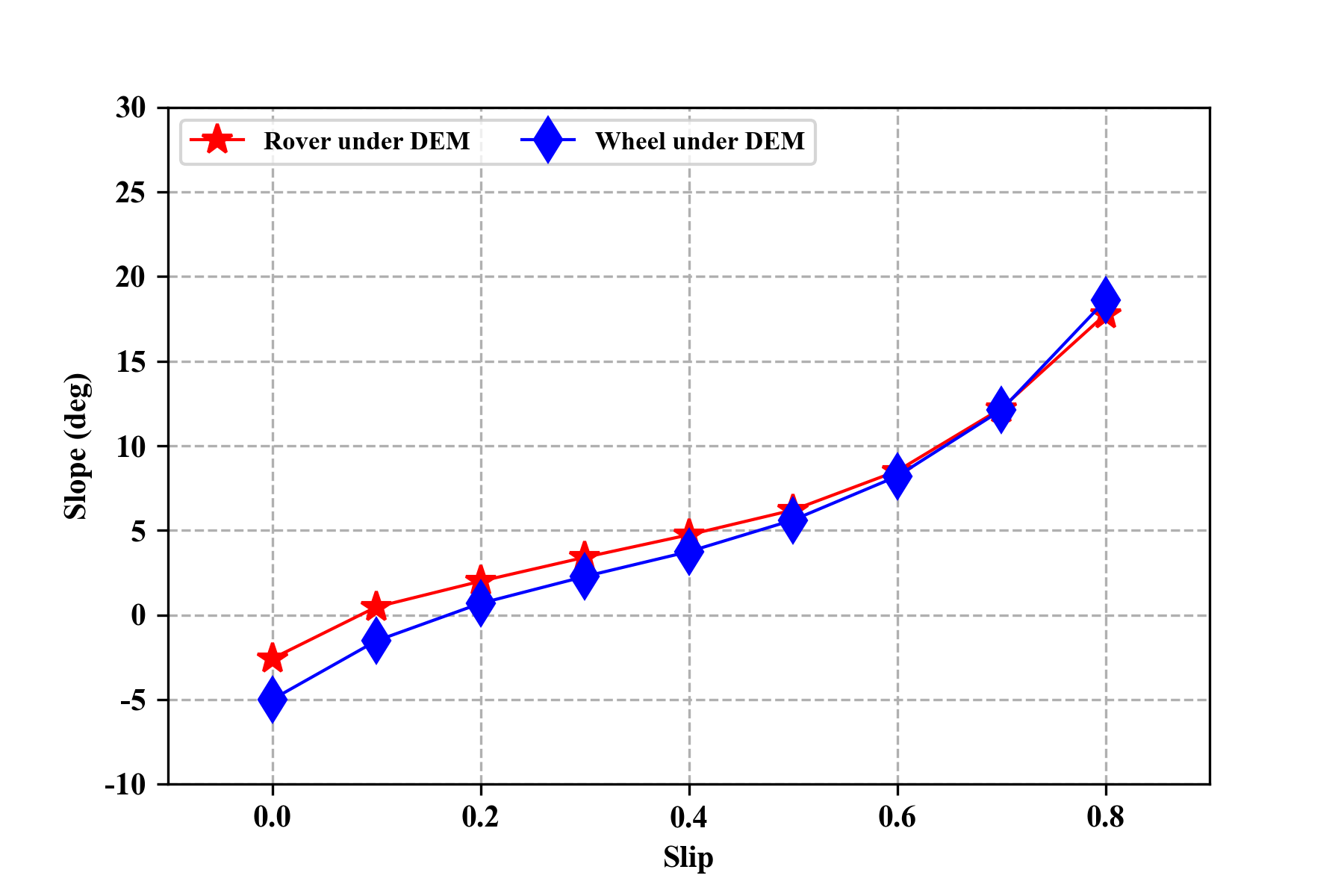}
		\caption{Terrain slope.} 
	\end{subfigure}
	\caption{Comparison of full rover simulation and single wheel simulation using DEM.} 
	\label{fig:Rover_singlewheel_DEM}
\end{figure}

\begin{figure}[htp]
	\centering
	\includegraphics[width=5in]{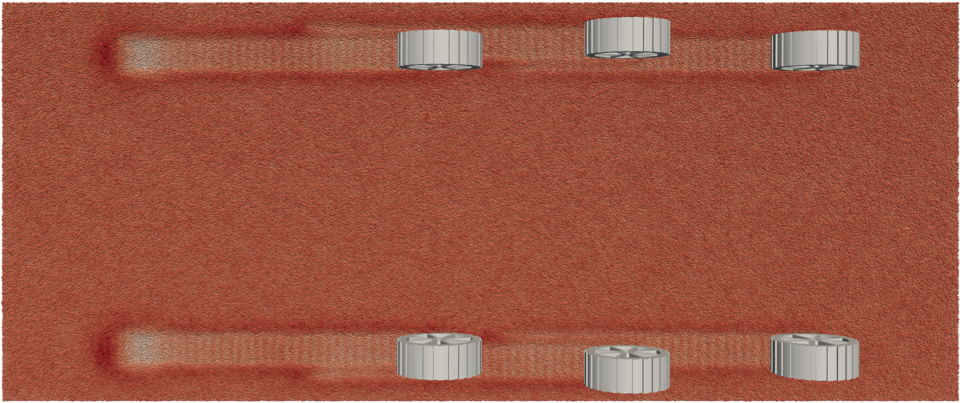}
	\caption{Screenshot of the full rover simulation. Animation of the full rover simulation is provided in the supplementary materials.} 
	\label{fig:Rover_DEM_Screen}
\end{figure}

\begin{figure}[htp]
	\centering
	\begin{subfigure}{0.99\textwidth}
		\centering
		\includegraphics[width=4in]{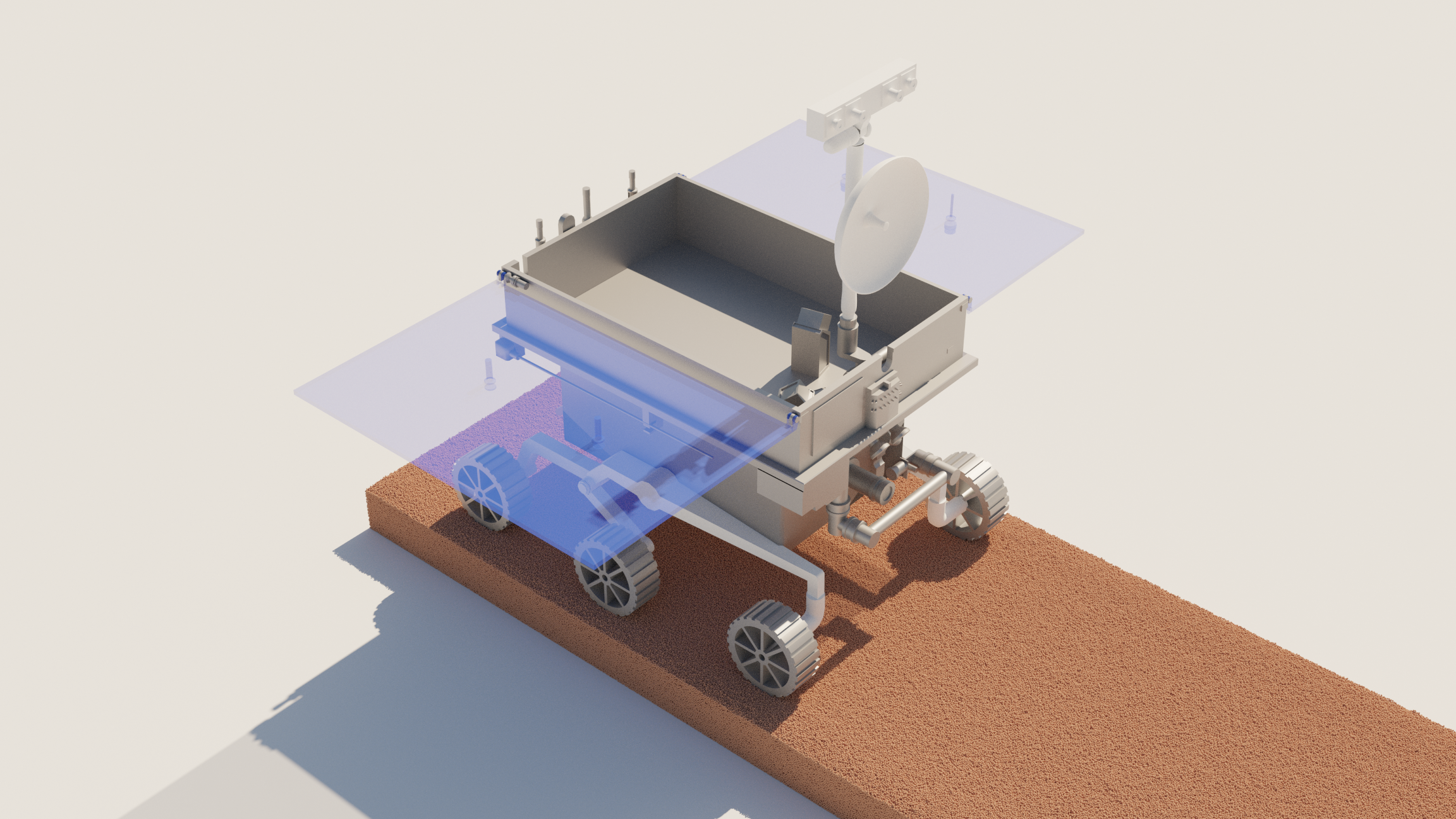} \\
		\caption{Time = 0 s.}
	\end{subfigure}
	\begin{subfigure}{0.99\textwidth}
		\centering
		\includegraphics[width=4in]{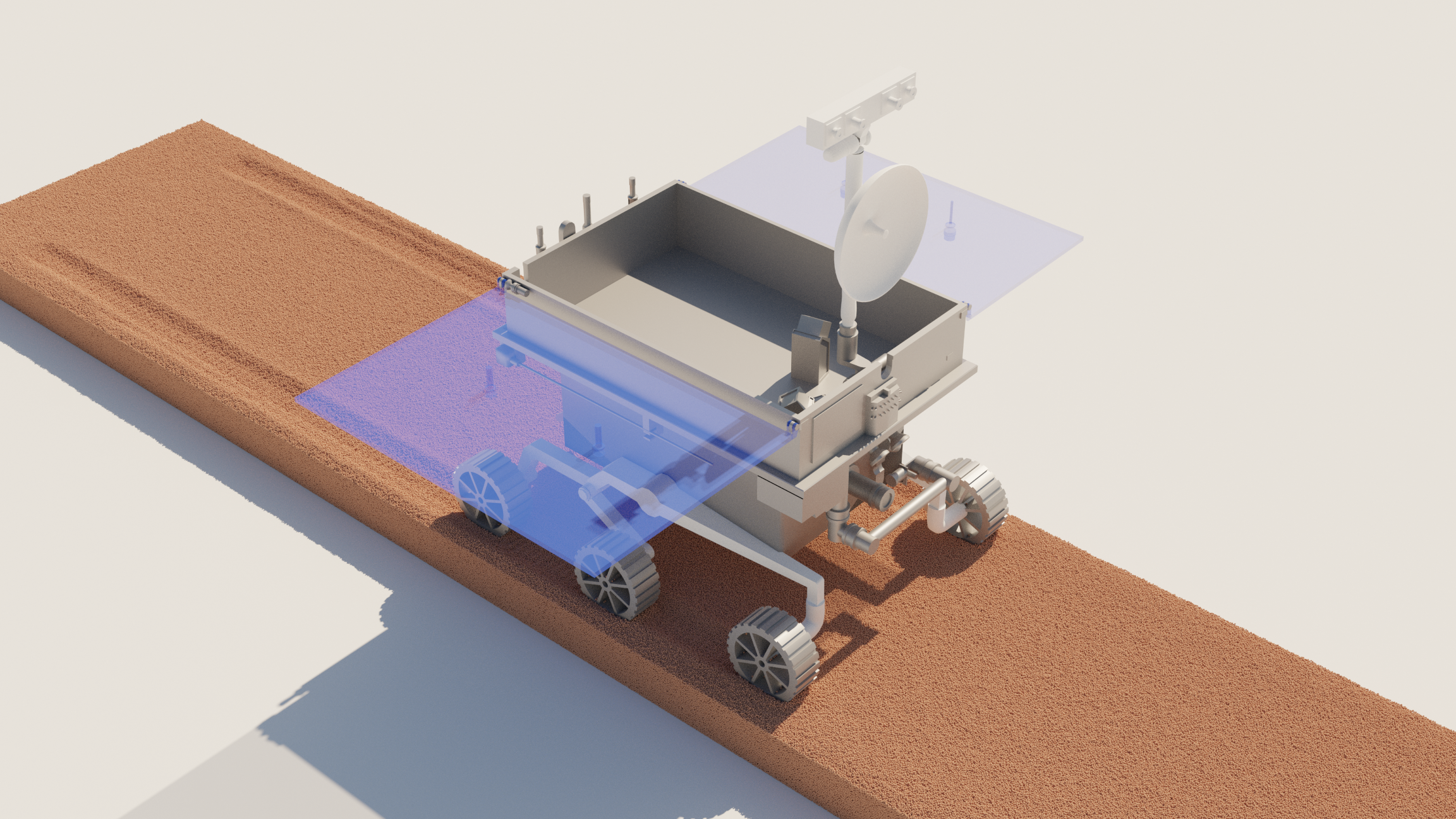} \\
		\caption{Time = 10 s.}
	\end{subfigure}
	\begin{subfigure}{0.99\textwidth}
		\centering
		\includegraphics[width=4in]{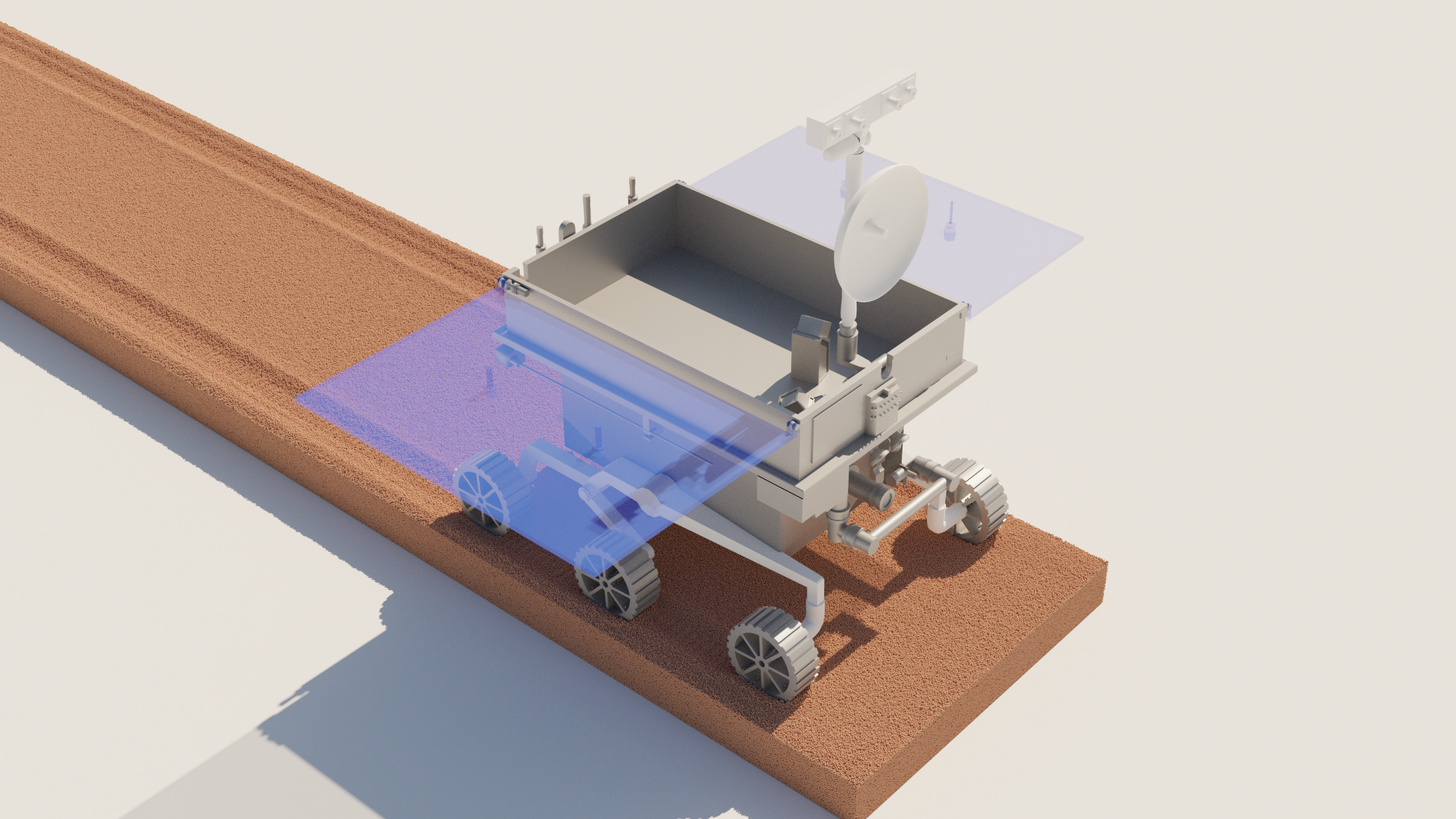}
		\caption{Time = 20 s.} 
	\end{subfigure}
	\caption{Screenshots of the full rover simulation at t = 0 s, 10 s, 20 s.} 
	\label{fig:Rover_DEM_Screen_t}
\end{figure}

Since the DBP force was measured on the rover chassis, which had six wheels connected to it, the rover's DBP force should be roughly six times larger than that noted in the single rover wheel test. Indeed, Fig.~\ref{fig:Rover_singlewheel_SCM} shows this to be the case, indicating that for this rover, a single wheel test is indicative of full vehicle behavior. This is notable since a single wheel simulation experiment is about six times shorter than a full rover experiment. Figure~\ref{fig:Rover_singlewheel_DEM} reports results of a similar comparison between the full rover and a single rover wheel using DEM terramechanics. A screenshot of the full rover's simulation on DEM terrain is shown in Fig.~\ref{fig:Rover_DEM_Screen}. Rover simulation snapshots taken at t = 0 s, 10 s, and 20 s are shown in Fig.~\ref{fig:Rover_DEM_Screen_t}. It is noted that even though the rear wheels moved on terrain which was disturbed by the front wheels, the vehicle DBP response was not significantly affected, otherwise one couldn't see agreement shown in Figs.~\ref{fig:Rover_singlewheel_SCM} and \ref{fig:Rover_singlewheel_DEM}.

\FloatBarrier

\section{Conclusions and directions of future work}
\label{sec:conclusion}
The tests required to calibrate SCM terramechanics for off-road vehicle simulation are costly and cumbersome since they involve a non-standardized bevameter test. This paper advances the idea of using a multibody dynamics virtual bevameter test that employs high-fidelity terramechanics, i.e., DEM-based, to calibrate the lower-fidelity SCM terramechanics model. We use the DEM terramechanics to generate ground truth data that is subsequently used in a Bayesian calibration framework to produce values for the six main parameters associated with SCM terramechanics: $K_c$, $K_\phi$, $n$, $c$, $\varphi$, and $K_s$. It was noted that upon calibration, SCM terramechanics produces results that are on par with the ones produced by DEM terramechanics for single wheel and full rover off-road mobility experiments. This is significant because, for example, a ten-second simulation of a vehicle operating on deformable DEM terrain with \num{15000000} elements required approximately three hours to run on a GPU card, while the SCM counterpart simulation took approximately 20 seconds to produce results of similar quality. Thus, for problems where efficiency is paramount, SCM terramechanics can produce good results at a fraction of the cost. SCM is not without issues, see, for instance \cite{rodriguezHighSpeedGranMatMobility2019}, but properly calibrated it can be a versatile modeling approach suitable for control, planning, and perception studies. For other engineering inquires, e.g., grouser pattern design, or rover topology design, where high fidelity in terramechanics is essential, one should rely on DEM or CRM terramechanics. These methods are physics-based and thus present fewer challenges related to selecting model parameters.

This contribution also provided a broad overview of two terramechanics approaches and showed that SCM and DEM can produce results that agree with each other for non-demanding experimental regimens. The similar behavior was demonstrated for single wheel and full rover experiments; for the single wheel, three wheel geometries were used in the DEM vs. SCM comparison.

In terms of future work, it remains to quantify the degree to which high-fidelity DEM terramechanics captures the physics of the actual bevameter test. To the best of our knowledge, this hasn't been reported in the literature yet.

\FloatBarrier
	
\section*{Acknowledgment}
Support for this work was provided by National Natural Science Foundation of China under grant 12302050. The work was also sponsored by the Oceanic Interdisciplinary Program of Shanghai Jiao Tong University (project number SL2023MS001).
	
\bibliographystyle{unsrt}
\bibliography{main}

\section{Appendix}\label{sec:appendix}
This appendix contains a collection of figures that support the results presented in the main part of the contribution. Details associated with the tests conducted to generate these images are provided in the caption on a case-by-case basis.

\begin{figure}[htp]
	\centering
	\begin{subfigure}[t]{0.49\textwidth}
		\centering
		\includegraphics[width=1.5in]{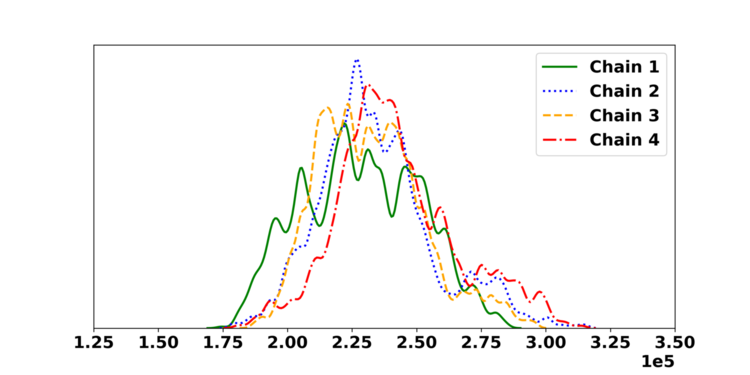}
		\includegraphics[width=1.5in]{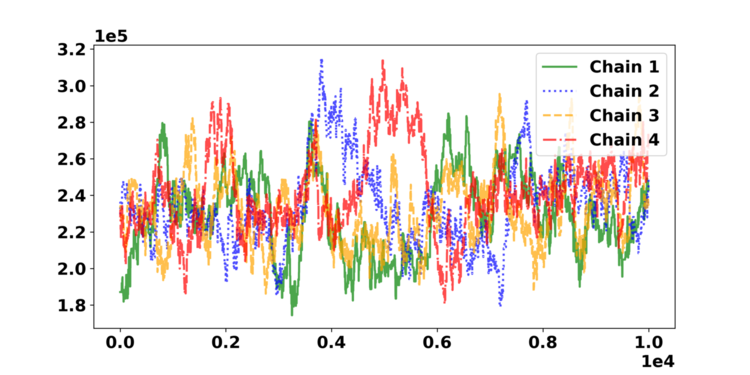}
		\caption{\num{10000} samples.}
	\end{subfigure}
	\begin{subfigure}[t]{0.49\textwidth}
		\centering
		\includegraphics[width=1.5in]{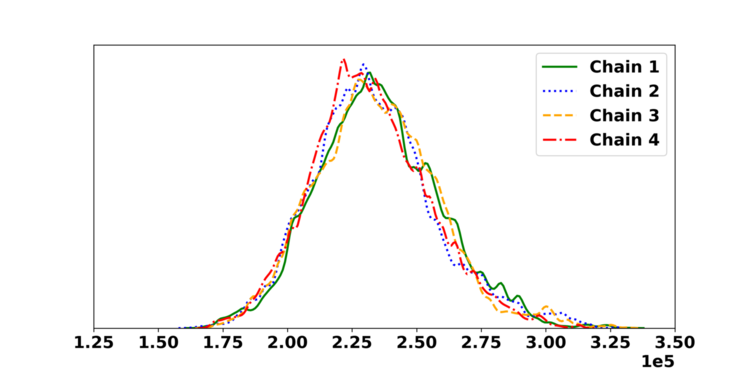}
		\includegraphics[width=1.5in]{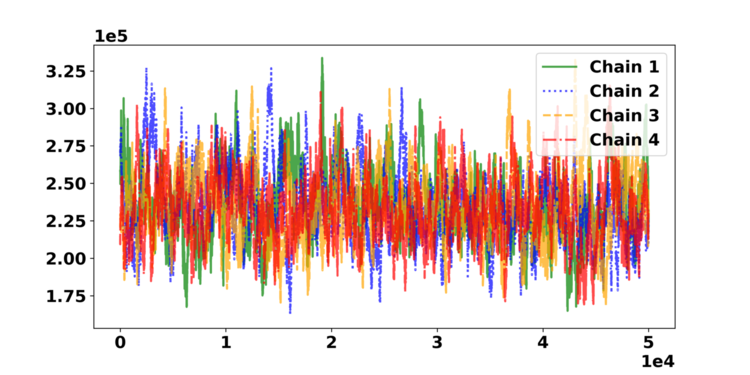}
		\caption{\num{50000} samples.}
	\end{subfigure}
	\begin{subfigure}[t]{0.49\textwidth}
		\centering
		\includegraphics[width=1.5in]{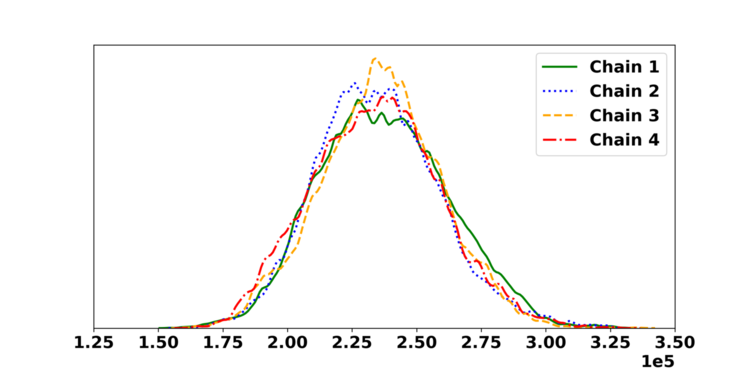}
		\includegraphics[width=1.5in]{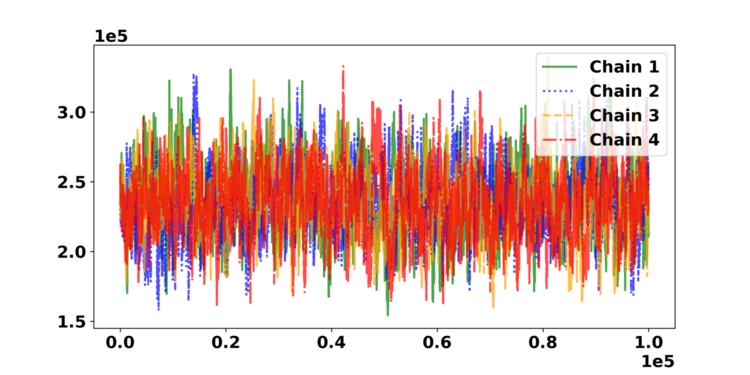}
		\caption{\num{100000} samples.}
	\end{subfigure}
	\begin{subfigure}[t]{0.49\textwidth}
		\centering
		\includegraphics[width=1.5in]{Post_500000_samples_K_phi.png}
		\includegraphics[width=1.5in]{Trace_500000_samples_K_phi.png}
		\caption{\num{500000} samples.}
	\end{subfigure}
	\caption{Posterior probability distribution and trace plot for parameter $K_{\phi}$ in four different Metropolis-Hastings chains.}
	\label{fig:trace_posterior_kphi_4chain}
\end{figure}

\begin{figure}[htp]
	\centering
	\begin{subfigure}{0.49\textwidth}
		\centering
		\includegraphics[width=3in]{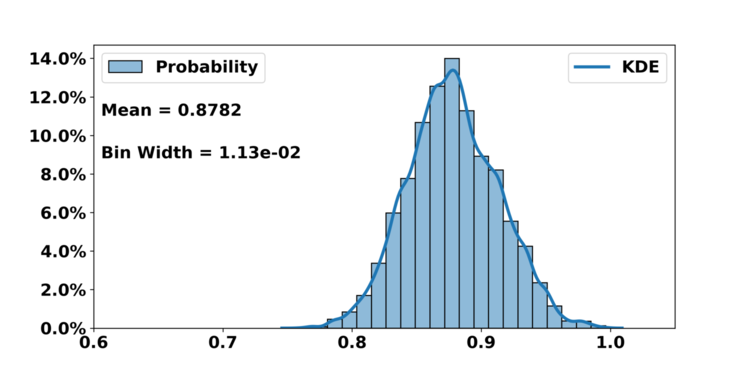}
		\caption{\num{10000} samples.}
	\end{subfigure}
	\begin{subfigure}{0.49\textwidth}
		\centering
		\includegraphics[width=3in]{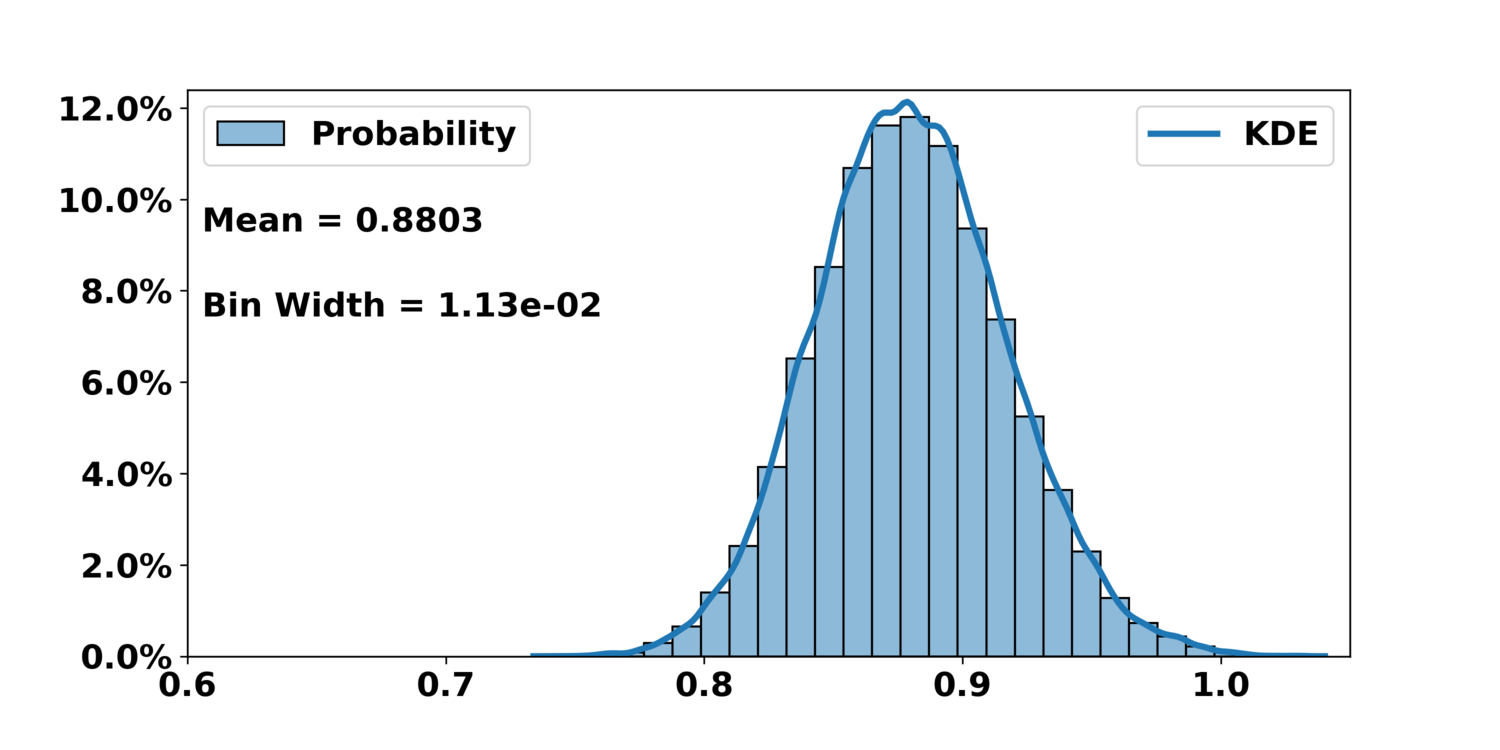}
		\caption{\num{50000} samples.} 
	\end{subfigure}
	\begin{subfigure}{0.49\textwidth}
		\centering
		\includegraphics[width=3in]{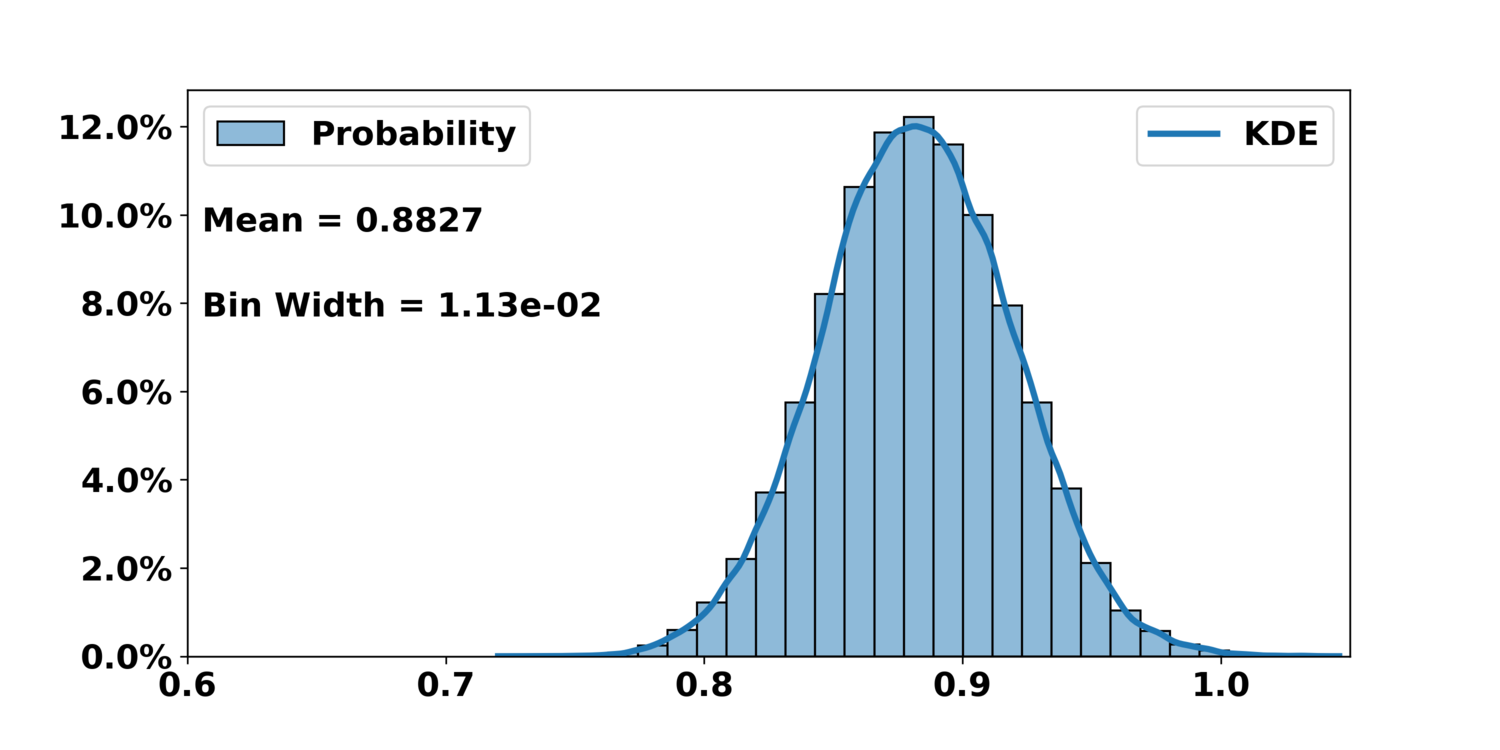}
		\caption{\num{100000} samples.}
	\end{subfigure}
	\begin{subfigure}{0.49\textwidth}
		\centering
		\includegraphics[width=3in]{Post_Mean_500000_samples_n.png}
		\caption{\num{500000} samples.} 
	\end{subfigure}
	\caption{Averaged posterior probability distribution and its kernel density estimation (KDE) for the parameter $n$ across four chains with different sample sizes.} 
	\label{fig:posterior_n}
\end{figure}

\begin{figure}[htp]
	\centering
	\begin{subfigure}[t]{0.49\textwidth}
		\centering
		\includegraphics[width=1.5in]{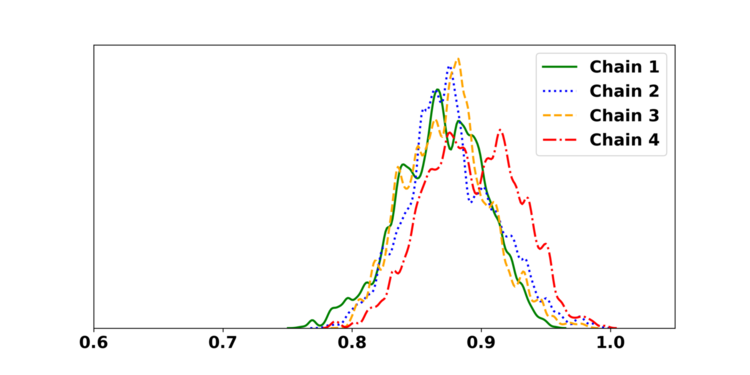}
		\includegraphics[width=1.5in]{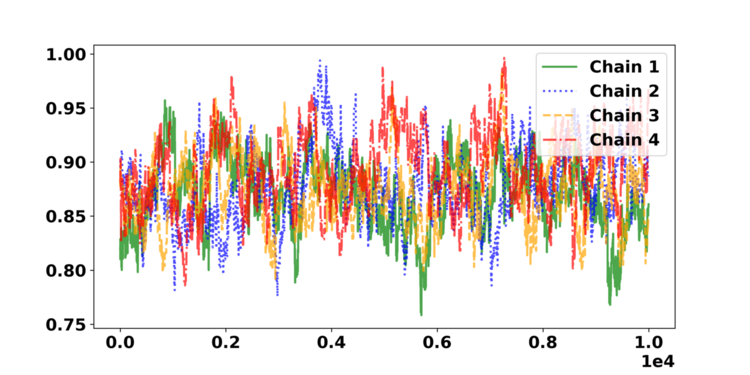}
		\caption{\num{10000} samples.}
	\end{subfigure}
	\begin{subfigure}[t]{0.49\textwidth}
		\centering
		\includegraphics[width=1.5in]{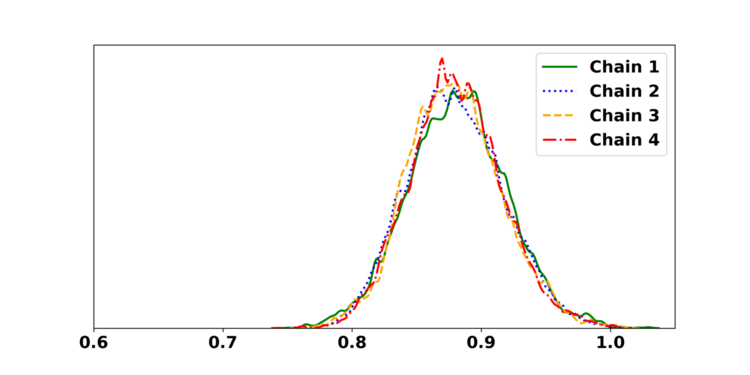}
		\includegraphics[width=1.5in]{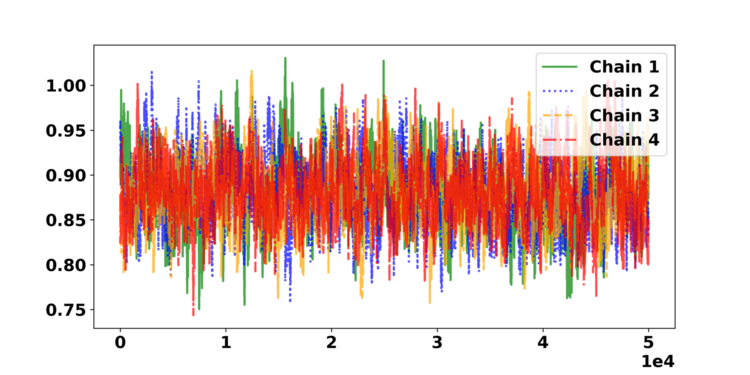}
		\caption{\num{50000} samples.}
	\end{subfigure}
	\begin{subfigure}[t]{0.49\textwidth}
		\centering
		\includegraphics[width=1.5in]{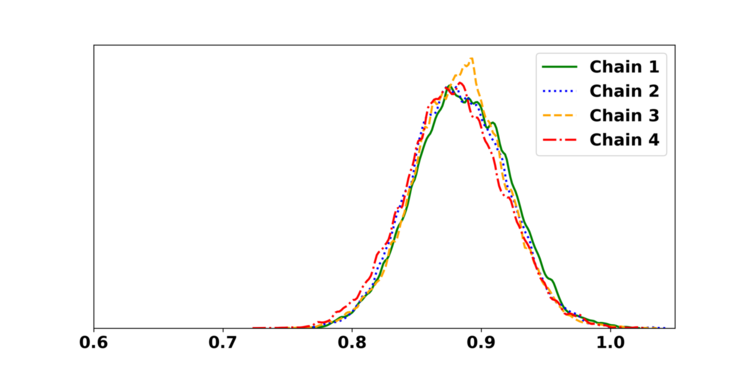}
		\includegraphics[width=1.5in]{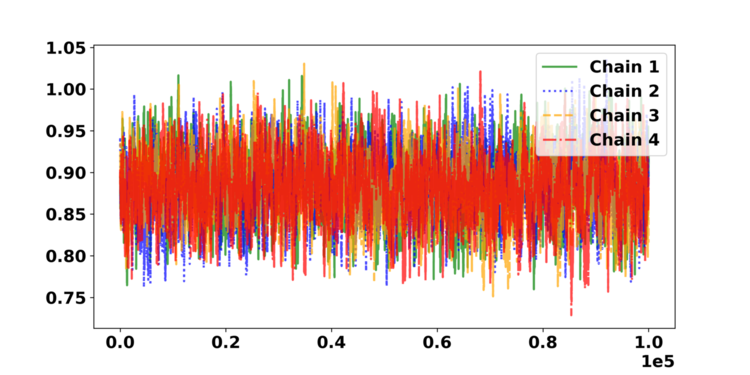}
		\caption{\num{100000} samples.}
	\end{subfigure}
	\begin{subfigure}[t]{0.49\textwidth}
		\centering
		\includegraphics[width=1.5in]{Post_500000_samples_n.png}
		\includegraphics[width=1.5in]{Trace_500000_samples_n.png}
		\caption{\num{500000} samples.}
	\end{subfigure}
	\caption{Posterior probability distribution and trace plot for parameter $n$ in four different Metropolis-Hastings chains.}
	\label{fig:trace_posterior_n_4chain}
\end{figure}

\begin{figure}[htp]
	\centering
	\begin{subfigure}[t]{0.49\textwidth}
		\centering
		\includegraphics[width=1.5in]{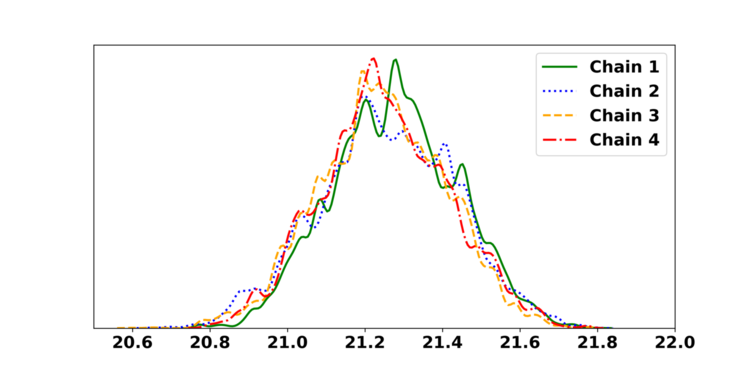}
		\includegraphics[width=1.5in]{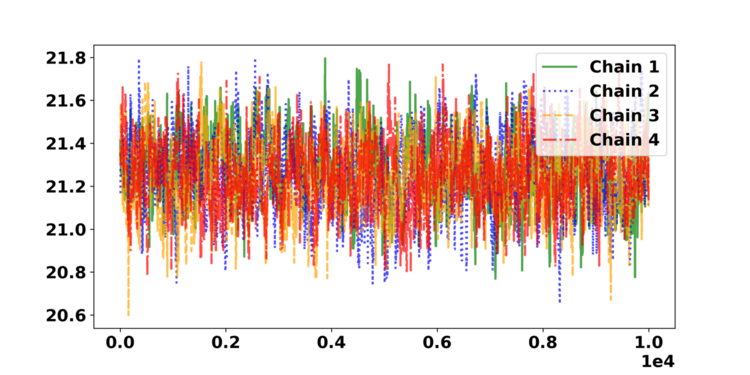}
		\caption{\num{10000} samples.}
	\end{subfigure}
	\begin{subfigure}[t]{0.49\textwidth}
		\centering
		\includegraphics[width=1.5in]{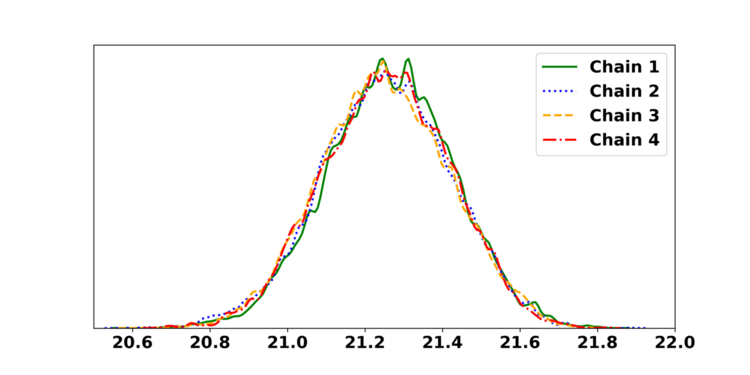}
		\includegraphics[width=1.5in]{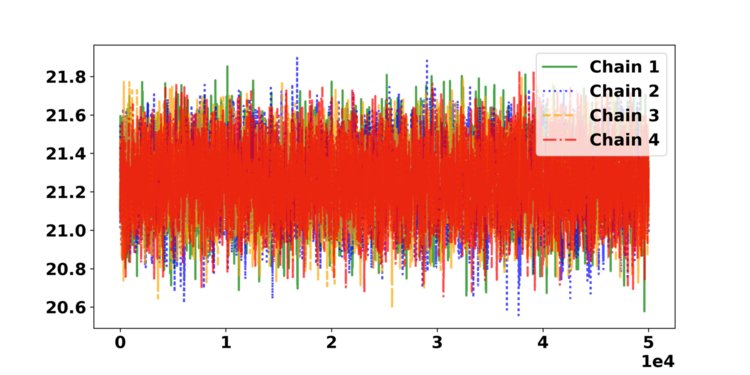}
		\caption{\num{50000} samples.}
	\end{subfigure}
	\begin{subfigure}[t]{0.49\textwidth}
		\centering
		\includegraphics[width=1.5in]{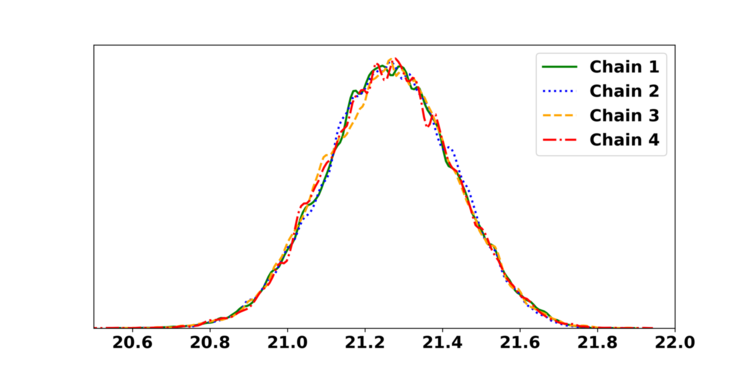}
		\includegraphics[width=1.5in]{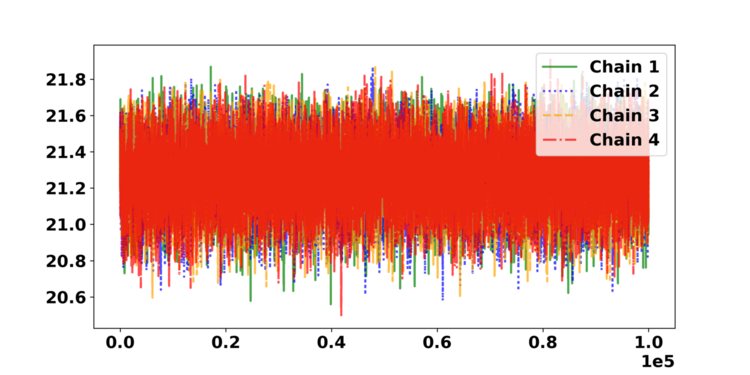}
		\caption{\num{100000} samples.}
	\end{subfigure}
	\begin{subfigure}[t]{0.49\textwidth}
		\centering
		\includegraphics[width=1.5in]{Post_500000_samples_phi.png}
		\includegraphics[width=1.5in]{Trace_500000_samples_phi.png}
		\caption{\num{500000} samples.}
	\end{subfigure}
	\caption{Posterior probability distribution and trace plot for parameter $\varphi$ in four different Metropolis-Hastings chains.}
	\label{fig:trace_posterior_phi_4chain}
\end{figure}

\begin{figure}[htp]
	\centering
	\begin{subfigure}{0.49\textwidth}
		\centering
		\includegraphics[width=3in]{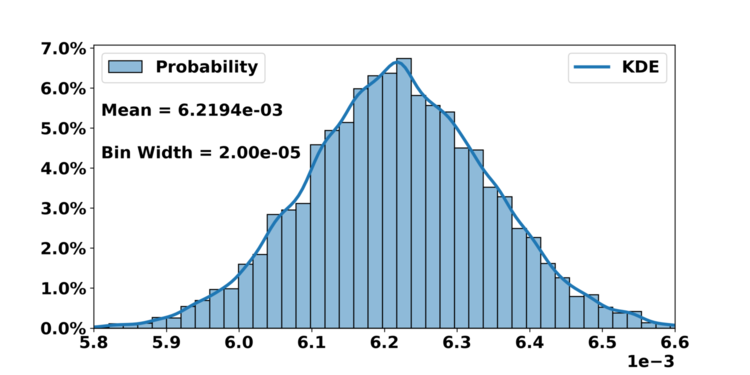}
		\caption{\num{10000} samples.}
	\end{subfigure}
	\begin{subfigure}{0.49\textwidth}
		\centering
		\includegraphics[width=3in]{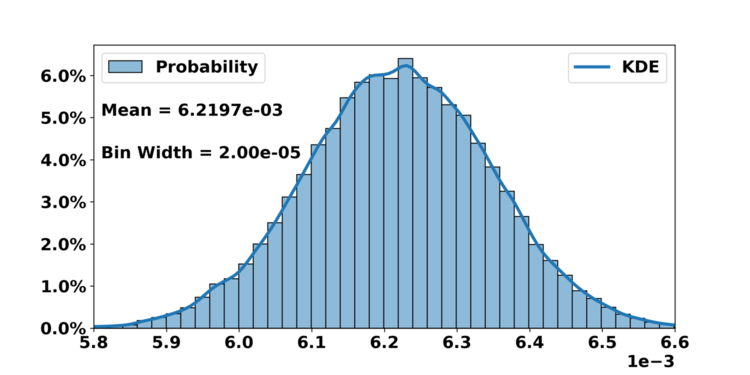}
		\caption{\num{50000} samples.} 
	\end{subfigure}
	\begin{subfigure}{0.49\textwidth}
		\centering
		\includegraphics[width=3in]{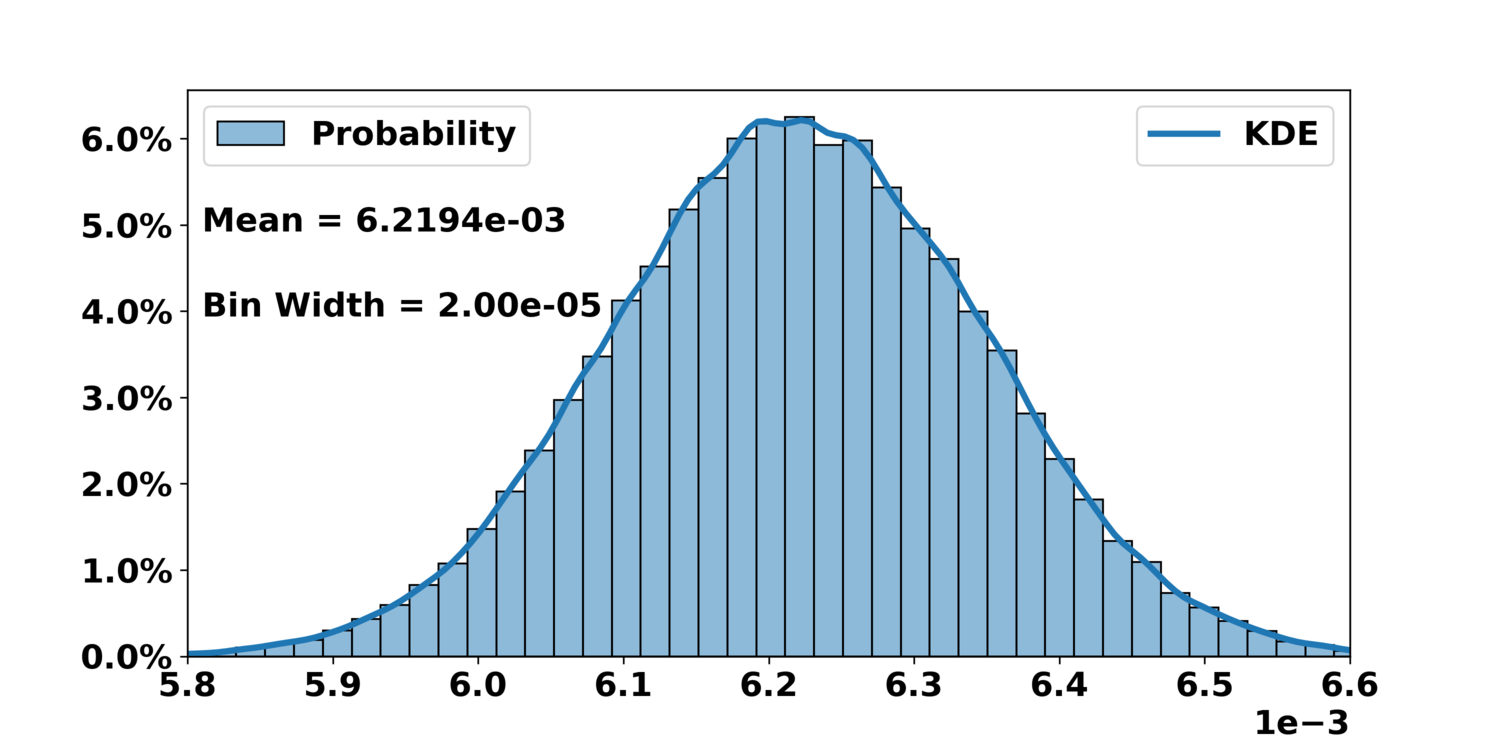}
		\caption{\num{100000} samples.}
	\end{subfigure}
	\begin{subfigure}{0.49\textwidth}
		\centering
		\includegraphics[width=3in]{Post_Mean_500000_samples_K_s.png}
		\caption{\num{500000} samples.} 
	\end{subfigure}
	\caption{Averaged posterior probability distribution and its kernel density estimation (KDE) for the parameter $K_s$ across four chains with different sample sizes.} 
	\label{fig:posterior_ks}
\end{figure}

\begin{figure}[htp]
	\centering
	\begin{subfigure}[t]{0.49\textwidth}
		\centering
		\includegraphics[width=1.5in]{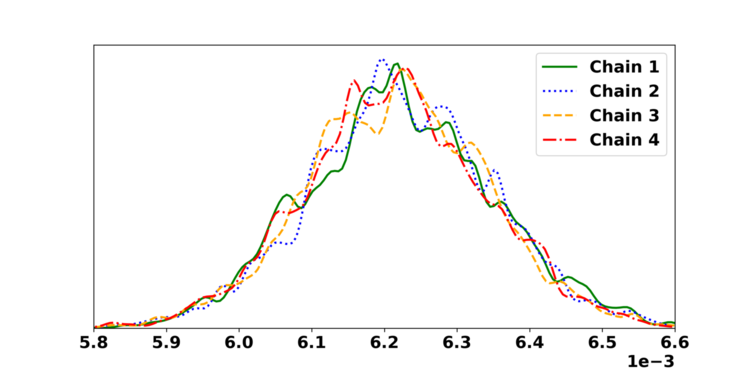}
		\includegraphics[width=1.5in]{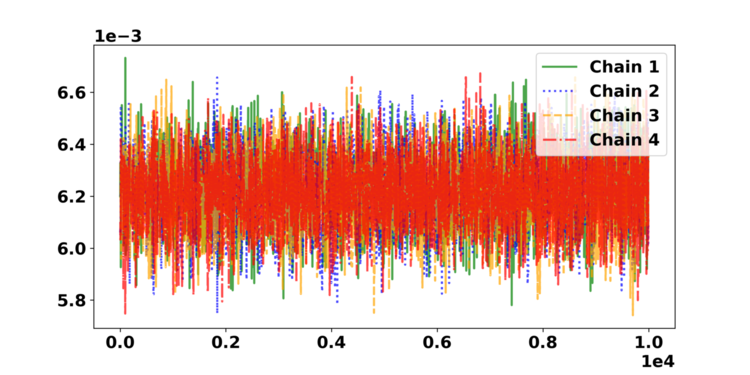}
		\caption{\num{10000} samples.}
	\end{subfigure}
	\begin{subfigure}[t]{0.49\textwidth}
		\centering
		\includegraphics[width=1.5in]{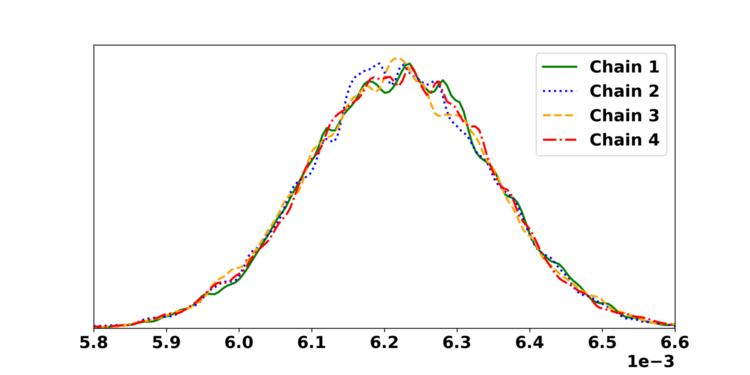}
		\includegraphics[width=1.5in]{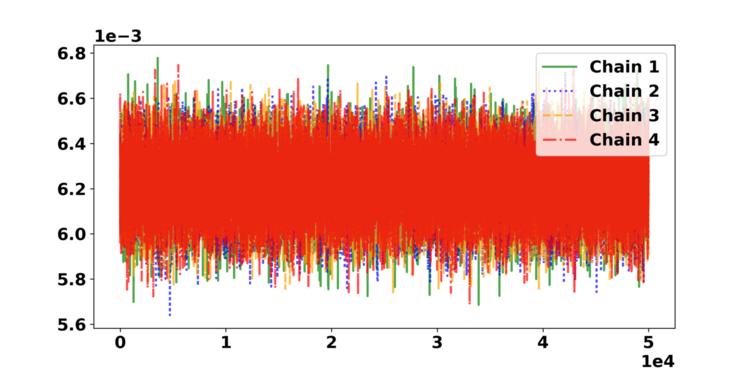}
		\caption{\num{50000} samples.}
	\end{subfigure}
	\begin{subfigure}[t]{0.49\textwidth}
		\centering
		\includegraphics[width=1.5in]{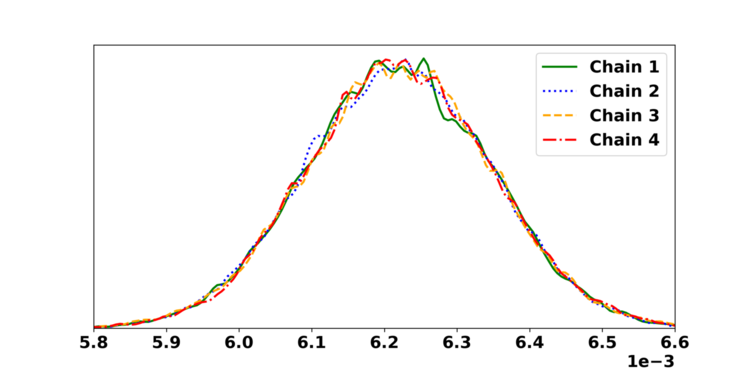}
		\includegraphics[width=1.5in]{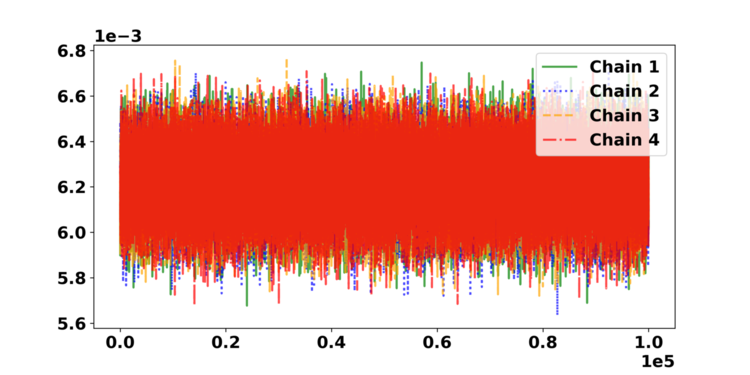}
		\caption{\num{100000} samples.}
	\end{subfigure}
	\begin{subfigure}[t]{0.49\textwidth}
		\centering
		\includegraphics[width=1.5in]{Post_500000_samples_K_s.png}
		\includegraphics[width=1.5in]{Trace_500000_samples_K_s.png}
		\caption{\num{500000} samples.}
	\end{subfigure}
	\caption{Posterior probability distribution and trace plot for parameter $K_s$ in four different Metropolis-Hastings chains.}
	\label{fig:trace_posterior_ks_4chain}
\end{figure}

\end{document}